\documentclass{article}

\usepackage{arxiv}
\usepackage[utf8]{inputenc}
\usepackage[T1]{fontenc}    
\usepackage{hyperref}      
\usepackage{url}            
\usepackage{booktabs}      
\usepackage{amsfonts}      
\usepackage{nicefrac}       
\usepackage{microtype}     
\usepackage{lipsum}		
\usepackage{amsmath}
\usepackage{amssymb}
\usepackage{graphicx}
\usepackage{siunitx}
\DeclareSIUnit\angstrom{\text{\AA}}

\usepackage{doi}
\usepackage[normalem]{ulem}
\usepackage[backend=bibtex,natbib=true,maxnames=3,style=authoryear]{biblatex}
\title{Addressing type Ia supernova color variability with a linear spectral template}

\author{ \href{https://orcid.org/0000-0002-3646-8073}{\includegraphics[scale=0.06]{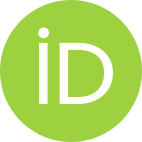}\hspace{1mm}C\'assia S. Nascimento}\\
	Instituto de F\'isica, Universidade Federal do Rio de Janeiro, 21941-972, Rio de Janeiro, RJ, Brazil\\
	\texttt{cassia@pos.if.ufrj.br} \\
	\And
	\href{https://orcid.org/0009-0001-4409-6820}{\includegraphics[scale=0.06]{orcid.png}\hspace{1mm}Jo\~ao Paulo C. França} \\
    Centro Brasileiro de Pesquisas Físicas, Rua Xavier Sigaud st. 150, 22290-180, Rio de Janeiro, Brazil\\
	Instituto de F\'isica, Universidade Federal do Rio de Janeiro, 21941-972, Rio de Janeiro, RJ, Brazil\\
	\texttt{joaofranca@cbpf.br} \\
	\And
	\href{https://orcid.org/0000-0003-1339-2106}{\includegraphics[scale=0.06]{orcid.png}\hspace{1mm}Ribamar R. R. Reis} \\
	Instituto de F\'isica, Universidade Federal do Rio de Janeiro, 21941-972, Rio de Janeiro, RJ, Brazil\\
    Observat\'orio do Valongo, Universidade Federal do Rio de Janeiro, 20080-090, Rio de Janeiro, RJ, Brazil\\
	\texttt{ribamar@if.ufrj.br} \\
}

\renewcommand{\headeright}{}

\begin{document}
\maketitle

\begin{abstract}
Type Ia Supernovae (SNeIa) provided the first evidence of an accelerated expansion of the universe and remain a valuable probe to cosmology. 
They are deemed standardizable candles due to the observed correlations between their luminosity and photometric quantities. 
This characteristic can be exploited to estimate cosmological distances after accounting for the observed variations. 
There is however a remaining dispersion unaccounted for in the current state-of-the-art standardization methods. 
In an attempt to explore this issue, we propose a simple linear 3-component rest-frame flux description for a light-curve fitter. 
Since SNIa intrinsic color index variations are expected to be time-dependent, our description builds upon the mathematical expression of the well-known Spectral Adaptive Light Curve Template 2 (SALT2) for rest-frame flux, whilst we drop the exponential factor and add an extra model component with time and wavelength dependencies. 
The model components are obtained by performing either Principal Component Analysis (PCA) or Factor Analysis (FA) onto a representative training set. 
The constraining power of the model dubbed Pure Expansion Template for Supernovae (PETS) is evaluated and we found compatible results with SALT2 for $\Omega_{m0}$ and $\Omega_{\Lambda 0}$ within $68\%$ uncertainty between the two models, with PETS' fit parameters exhibiting non-negligible linear correlations with SALT2' parameters. 
For both PCA and FA model versions we verified that the first component mainly describes color index variations, proving it is a dominant effect on SNIa spectra. 
The model nuisance parameter which multiplies the color index variation-like fit parameter shows evolution with redshift in an initial binned cosmology analysis. 
This behavior can be due to selection effects and should be further investigated with higher redshift SNeIa samples. 
Overall, our model shows promise, as there are still a few aspects to be refined; however, it still falls short in reducing the unaccounted dispersion. 
\end{abstract}


\section{Introduction}

\label{sec:introduction}

The correlations between light curves' decay rate and color index with luminosity are the main photometric features of Type Ia Supernovae (SNeIa) (\citet{rust1974, pskovskii1977,phillips1993absolute,hamuy1995}). 
By combining both effects, \citet{riess1996, tripp1998, guy2005} built standardization procedures allowing SNeIa to be identified as standard candles.
As a result, \citet{riess1998observational, perlmutter1999measurements} were able to constrain dark energy and mass densities using a few dozen SNeIa with redshift up to $0.7$, providing the first evidence of an accelerated expansion of the universe.
Since then, SNIa was established as a valuable probe to cosmology while the pioneering standardization methods were refined and different standardization approaches were proposed (\citet{Jha_2007, Guy2007, Guy2010, Burns_2011,  Betoule_2014, Rubin_2015, saunders2018snemo,  Pierel_2018, sugar2020, Kenworthy_2021, rubin2023union}). 

The total dispersion estimated for SNIa apparent magnitude after correction is composed of statistic and systematic contributions of the same order of magnitude. 
The latter is commonly linked to flux calibrations and the standardization method itself.
Even the state-of-the-art light-curve fitting model SALT2 (Spectral Adaptive Light-Curve Template 2) by \citet{Guy2010} exhibits an unaccounted dispersion in SNIa apparent magnitude after the standardization process.

The statistical dispersion contribution is expected to be reduced with the next-generation surveys, which intend to supply a major growth in available data. 
For example, considering the Rubin Observatory (LSST) (\citet{Ivezic_2019}) and Nancy Grace Telescope (WFIRST) (\citet{spergel_2015}) projects, we expect to detect around $300.000$ supernovae with redshift up to $z=3$ (\citet{Rose_2021}). 
Such growth in SNIa sample size corresponds to a net increase of two orders of magnitude, thus triggering an attention shift to diminish the systematic dispersion contribution. 

Therefore, in recent years we witnessed the development of new modeling techniques, such as SNEMO (SuperNova Empirical MOdels) (\citet{saunders2018snemo}) and SUGAR (SUpernova Generator And Reconstructor) (\citet{sugar2020}), that employed unsupervised machine learning decomposition methods on reconstructed Spectral Energy Distributions (SEDs) for a set of SNeIa spectra. 

Knowing that the SNIa intrinsic color index, which correlates with luminosity, is expected to be time-dependent and following the aforementioned modeling approaches, in this paper, we drop the exponential factor in the usual rest-frame flux description proposed by SALT2 and add an extra model component with time and wavelength dependencies. 
We extensively explore the description of a generic SNIa spectral surface by a linear expansion 3-components model and compare our results with state-of-the-art fitters. 

Our model, hereafter called PETS (Pure Expansion Template for Supernovae), is trained with the Nearby Supernova Factory Data Release 9 spectra (\citet{Aldering_2020}). 
Using this data we construct regularized SEDs by performing 1D Gaussian Process Regressions (GPRs) on monochromatic light-curves. 
Gaussian priors are defined over the GPR kernel parameters alongside a mean function based on the \citet{Hsiao2007} template, ensuring a high-resolution reconstruction.  
Then, PCA-PETS and FA-PETS components are respectively evaluated by performing Principal Component Analysis (PCA) and Factor Analysis (FA) decomposition methods to the regularized reddened SEDs. 
The regularization process generates SEDs data distributed in an even-spaced 2D grid, suitable for both decomposition methods. 
The analysis is performed over reddened data to avoid the common assumption that both intrinsic and host dust contributions can be described by tuning an extinction law. 
Hence, just like the SALT2 model, we do not separate the color index variation sources.

In section \ref{sec:modeldescpriction} we present the PETS model. 
In section \ref{sec:datasetpreprocess} we illustrate the pre-training process of reconstructing the SEDs and defining a relative normalization. 
In section \ref{sec:model_training} we discuss in detail the process of decomposing the spectral surfaces via Principal Component Analysis and Factor Analysis and analyze model reconstruction performance for both training and validation sets. In section \ref{sec:comparison_lcfitters} we compare PETS light curve fitter performance with SALT2 and SNEMO models.
In section \ref{sec:cosmo_paramters_constraints} we discuss the light-curve fitting results for the Pantheon sample (\citet{Scolnic2018}). 
In section \ref{sec:cosmoresults} we briefly introduce the cosmological results for both PCA and FA approaches when considering $\Lambda$CDM cosmology. 
Finally, in section \ref{sec:conclusions} we present the final remarks. 
Further discussions on the initial cosmology evaluation can be found in Appendix \ref{ap:pcacosm_cont} and Appendix \ref{ap:facosm_cont}. 

\section{The pure expansion spectral model}
\label{sec:modeldescpriction}
SNIa are identified as standardizable candles due to the remarkable similarity between their luminosities. 
This observational fact enables the assumption that a general SNIa SED can be described as a linear combination of a descriptive sample of SNIa SEDs.
Thus, decomposition methods that act on correlated data are suitable to reduce our original descriptive set, as they find the samples' main features by employing an orthonormal transformation to an ordered basis ranked by explained variance. 

Here, we utilize two decomposition methods, PCA and FA, which will be discussed in detail in sections \ref{sec:pca} and \ref{sec:fa}. 
Both methods allow us to reduce our original basis dimension while describing a reasonable variance of the original set. 

The most widely used light-curve fitter, SALT2 (\citet{Guy2007,Guy2010}), proposes the following empirical description for the rest-frame flux of a general SNIa: 
\begin{equation}
    \phi_{SALT2}(p, \lambda; {\bf x}) := x_{0,SALT2}[M_{0,SALT2}(p, \lambda)+x_{1,SALT2}M_{1,SALT2}(p, \lambda)+...]\exp(-c CL(\lambda)),
    \label{eq:restframefluxsalt2}
\end{equation}
where $\lambda$ is the wavelength and $p$ is the phase, \textit{i.e.} the number of days since maximum light in B-band. 
$M_{0,SALT2}$, $M_{1,SALT2}$ and $CL$ are the model components obtained from training over a set of light curves and spectra, while $x_{0,SALT2}$, $x_{1,SALT2}$ and $c$ account for the variations observed for different SNIa.

According to \citet{Guy2007}, the terms in brackets in equation~(\ref{eq:restframefluxsalt2}) can be understood as the training sample average SED plus the first term of a PCA decomposition of their training sample. 
However, at the time the model was proposed there was not enough dense and high-quality data to enable the model components evaluation by diagonalizing the sample covariance matrix.
The $M_{1,SALT2}$ component is usually associated with stretch-like effects, while $CL$ describes color index variations without mention of its source.

More recently, the SNEMO model (\citet{saunders2018snemo}) was introduced and trained over recently available dense high-quality data. 
Its rest-frame flux is analogous to SALT2, but with the term in brackets obtained through FA instead of PCA and with the color law term replaced by FM07 extinction law (\citet{Fitzpatrick_2007}). 

Besides the PCA/FA linear terms, both models rest-frame fluxes include an exponential of a color law. This description is based on the dust extinction effect over observed fluxes. However, to capture the two high variability features in SNIa, luminosity, and stretch, in the first two model components, it is necessary to independently address the reddening effect in spectra. Usually, this process consists of tuning a $R_V$ parameter in an extinction law such as CCM89, \citet{Cardelli1989}, F99, \citet{Fitzpatrick1999} or FM07, \citet{Fitzpatrick_2007}, accounting simultaneously for dust extinction and intrinsic variations in the training spectra.

Here, we take a different approach relying on the SEDs' similarities and knowing SNIa intrinsic color variations are time-dependent. We drop the exponential factor on the usual empirical SNIa rest-frame flux and propose a simple linear expansion model. Thus, we expect the color index variations to be distributed across the PETS components.

Our model describes the rest frame flux of a given SNIa, $\phi(p, \lambda; {\bf x})$, as
\begin{equation}
    \phi_{PETS}(p, \lambda; {\bf x}) := x_{0}[M_{0}(p, \lambda)+x_{1}M_{1}(p, \lambda) + x_{2}M_{2}(p, \lambda)+...].
    \label{eq:restframeflux}
\end{equation}
Once again, $\lambda$ is the wavelength and $p$ is the phase.
The parameter $x_0$ controls the rest frame flux amplitude while the set of surfaces $M_i(p, \lambda)$ consists of SEDs components that describe the observed dispersion of these objects weighted by the remaining free parameters $x_i$.

Using equation~(\ref{eq:restframeflux}) we can obtain the photometric flux model (m) in observer frame, to compare with observed data, $F_Y^{(m)}(p(1+z))$, see \citet{Kessler2009}, by integrating the spectroscopic flux, $\phi(p, \lambda; {\bf x})$, through a given bandpass $Y$ with observer frame transmission function $T_Y(\lambda)$:
\begin{equation}
    F_Y^{(m)}(p(1+z)) = (1+z)\int_0^\infty d\lambda[\lambda \phi(p, \lambda)T_Y(\lambda(1+z))]\,,
    \label{eq:photoflux}
\end{equation}
where $z$ is the SNIa redshift. 

Once the $M_i$ surfaces are obtained, we can construct a light-curve fitter using equation~(\ref{eq:photoflux}) to fit photometric data, thus correcting for SNIa apparent magnitude dispersion.

\section{The data set and pre-processing steps}
\label{sec:datasetpreprocess}
We train our template surfaces  $M_i$, introduced in equation~(\ref{eq:restframeflux}), on Nearby Supernova Factory (SNfactory) Data Release 9 (for more information see \citet{Aldering_2020}). This sample used for model training and validation consists of 2474 spectra from 171 spectroscopically confirmed SNeIa with redshift ranging from $0.01$ to $0.08$. The spectra are already shifted to a common rest-frame ($z=0$), corrected for Milky Way dust extinction following \citet{schlegel1998maps,Cardelli1989} (see \citet{childress2013host, rigault2020strong} for more details) and the observer-frame B-band date of maximum light from a SALT2 analysis is already subtracted. For more information about these procedures see \citet{saunders2018snemo}.

This data set consists of high-quality selected supernovae that have at least five spectra, of those at least one is prior to the date of maximum light, and at least four are between 10 and 35 days. All dates are measured relative to the observer-frame B-band maximum light estimate, which defines the rest-frame phase as $p=(t-t_{B,max})/(1+z)$. This higher number of spectroscopic data than is usually available allows us to construct a SED for each supernova, in a similar fashion as seen in \citet{saunders2018snemo}. These SEDs are surfaces of specific flux in units of erg/s/$\textnormal{cm}^2$/\AA \,(multiplied by an arbitrary factor) as a function of phase, in units of days, and rest-frame wavelength, in units of \AA. To extract the components that best explain the variability in our data set, we begin by evaluating the SEDs in a regular grid.  

The majority of the spectra data from our training sample range from 3305 \AA\, to 8586 \AA\, in wavelength and from -15 to 50 days in phase. Hence, we chose the model boundaries and mesh grid size as a regular grid from -10 to 50 days, with 1-day bins, and from 3310 \AA\, to 8580 \AA, with 10 \AA\, bins. This choice ensures good data coverage while preventing the loss of bandpasses when fitting photometric data. Our model then includes the optical region and a small portion of near-infrared, where SNIa luminosities usually show less dispersion.

\subsection{Pre Processing: Applying Gaussian Process Regression}
\label{sec:gpr} 
We interpolate each spectra and its uncertainties and evaluate them onto a regular wavelength grid. Then, for each supernova, we position the regularized spectral data in a three-dimensional space of flux density per wavelength per phase and to completely reconstruct each supernova SED we perform 1D Gaussian Process Regressions (GPRs) on phase direction for each different wavelength value.

Gaussian processes (GP) are supervised learning methods used in regression and classification problems. For a more detailed discussion see \citet{Rasmussen2005}. Here we seek one different function that explains the data for each monochromatic light curve. Each of these functions will be depicted by a one-dimensional GP with mean $m(\textbf{p})=\mathbb{E}[f(\textbf{p})]$ and covariance function $ K(\textbf{p},\textbf{p}')=\mathbb{E}[(f(\textbf{p})-m(\textbf{p}))(f(\textbf{p}')-m(\textbf{p}'))]$,
\begin{equation}
    f(\textbf{p})\sim \mathcal{GP}(m(\textbf{p}),K(\textbf{p},\textbf{p}')).
\end{equation}

The input values in the regression, $y_i$, are called the target values and they are, apart from a Gaussian noise $\epsilon_i$, equal to the function we wish to model via a GP,
\begin{equation}
    y_i=f(p_i)+\epsilon_i, \quad \epsilon_i\sim \mathcal{N}(0,\sigma_{n,i}^2).
\end{equation}
In our case, we perform a heteroscedastic regression since each input point has a different noise value.

We start from the assumption that the observed target values, $y_i$, and the function $f$ evaluated at the test points, $f_*$, can be described by the same multivariate Gaussian distribution. This allows us to construct the joint distribution,
\begin{equation}
    \begin{bmatrix}
    \textbf{y}\\\textbf{f}_*
    \end{bmatrix} \sim 
    \mathcal{N} \left(\textbf{m}(P), 
    \begin{bmatrix}
    K(P,P)+\sigma_n^2 & K(P,P_*)\\
    K(P_*,P) & K(P_*, P_*)
    \end{bmatrix}
    \right),
    \label{eq:jointdist}
\end{equation}
where the multivariate distribution covariance matrix, $\boldsymbol{\Sigma}$, was depicted in terms of the kernel function evaluated at each pair of points, whether from the train points vector, $P$, or test points vector, $P_*$. Lastly, $\sigma_n^2$ is a diagonal matrix carrying the target values variances. This distribution is completely defined when specifying a mean vector, \textbf{m}(P), and covariance matrix, $\boldsymbol{\Sigma}$.

We can then obtain the conditional distribution that provides the predictions for the function evaluated at the test points given the test points itself, the training points, and the corresponding observed target values, yielding the predictive mean: $\Bar{\textbf{f}}_*=\textbf{m}(P_*)+K(P_*,P)[K(P,P)+\sigma_n^2]^{-1}(\textbf{y}-\textbf{m}(P))$, and predictive covariance: $\textnormal{cov}(\textbf{f}_*)=K(P_*,P_*)-K(P_*,P)[K(P,P)+\sigma_n^2]^{-1} K(P,P_*)$.

We employ the Matérn Kernel for our GPRs. It includes an additional parameter, $\nu$, to regulate the smoothness of the function, when compared to the Radial Basis Function Kernel (RBF), which describes the covariance between two points as an exponential decline with their squared distances and a set of two parameters that tune the variance and the correlation range. In the limit that $\nu \rightarrow \infty$, we recover RBF. The Matérn kernel function is expressed as
\begin{equation}
    k(p_i, p_j) = \frac{\sigma_m^2}{\Gamma(\nu) 2^{\nu-1}} \left[\frac{\sqrt{2 \nu} (p_i - p_j)^2}{\Delta l} \right]^{\nu}K_\nu \left[\frac{\sqrt{2 \nu}}{\Delta l} (p_i - p_j)^2\right],
    \label{eq:matern}
\end{equation}
where $\sigma_m^2$ is the kernel variance and $\Delta l$ is the length scale parameter that controls the correlation range. $K_\nu$ is the modified Bessel functions and $\Gamma(\nu)$ is the Gamma function for a given $\nu$. Here we are especially interested in $\nu=5/2$ to keep a reasonable smoothness.

We apply this GPR formalism to our data via the Python library \citet{gpy2014}. A crucial step when performing the GPR is using the mean information, $\textbf{m}(P)$, otherwise, spectra without pre-maximum data will lead to a wrong behavior at lower phases (here the maximum date is the one specific to each monochromatic light-curve which does not coincide with B-band date of maximum light). This issue affects several supernovae and directly influences the SEDs reconstruction primarily at the region of low wavelengths and before maximum B-band light, profoundly impacting the forthcoming feature extraction. To avoid this incorrect regression we adopted \citet{Hsiao2007} SNIa template as the aforementioned mean. For each SNIa, we select the corresponding monochromatic light curve from the Hsiao template and fit the template to the data by constructing a straightforward $\chi^2$ model allowing a light curve amplitude parameter to vary. Once this normalization is performed, we feed the template as a mean function for GPy and proceed with the analysis.

\begin{figure}
\centering 
\includegraphics[width=0.5\columnwidth]{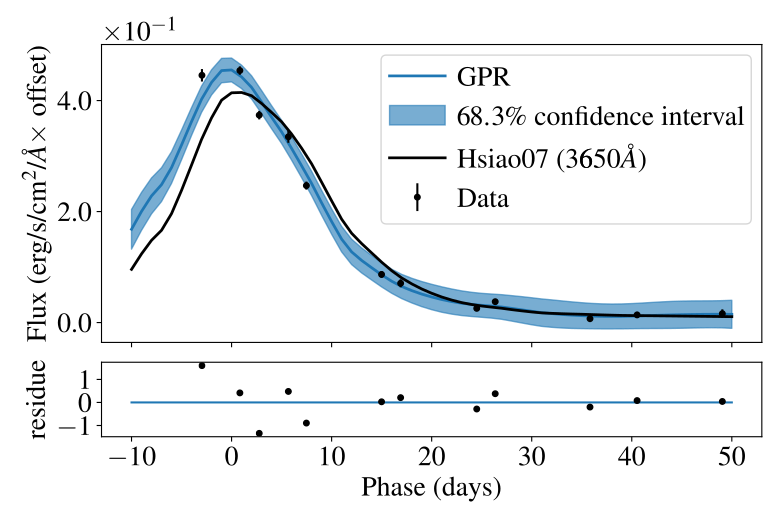} \caption{Gaussian process regression for "Test\_SN14" supernova at the constant wavelength plane of $\lambda=3650$\AA. The black error bars are the interpolated DR9 SNfactory spectra data and in blue we have the predicted GPR curve. The solid black curve represents the template used as the mean.}
\label{fig:gprexample}
\end{figure}

The original data was named by \citet{saunders2018snemo} as either Train or Test supernova followed by an identification number. Thus, the Train and Test flags do not represent our choice of training or validation objects, it just identifies a specific supernova. From the initial sample of 171 supernovas, five were excluded due to the poor quality of their reconstructed SEDs. They had too few spectra taken after maximum, and even with the template information they showed a large amount of nonphysical negative flux in this unconstrained region, they are Train\_SN93, Test\_SN15, Test\_SN26, and Train\_SN96. Train\_SN30 was also excluded since it showed several variations from the mean which are not observed in any other object in this sample. 

For each monochromatic light curve, we find a triplet of values by MLE: correlation length, kernel variance, and noise variance. Since assigning non-informative priors to the three parameters was not effective in reconstructing the expected smooth light curve behavior, we started by adding a Gaussian prior to $\Delta l$. We choose 12 days as the Gaussian mean and 20\% of this value as the standard deviation. This choice was based on 2D GPRs performed over our entire set of SEDs. Including only this prior, we verified in many cases the Gaussian noise surpassing the kernel variance, with the latter being essentially zero. A small kernel variance shows a preference for curves excessively close to the mean. Thus, a prior over this parameter was added by calculating the standard deviation of the data points around the mean. This value was squared and used as the prior. Its standard deviation was chosen as 10\% of this value. Still, on many occasions, we verified the optimized solution was exactly equal to the mean with the remainder dispersion totally explained by the Gaussian noise. This problem was solved by adding a Gaussian prior to this parameter, with mean equals 20\% the kernel variance and corresponding 10\% standard deviation.

For the entire set of 90272 GPRs (more than 500 for each SN) we find the following values: $\chi^2$/ndof: 0.8$\pm$ 0.5 , $\Delta l$: 10.7$\pm$1.5, $\sigma_\textnormal{Mátern}$: 0.07$\pm$ 0.04 and $\textnormal{noise}$ 0.03$\pm$0.02. The set of applied priors allowed us to reproduce the smooth light-curves, with a kernel variance prior accommodating the dispersion around the mean and at the same time limiting the amount of white noise allowed. 

\subsection{Pre Processing: Normalization}
\label{sec:norm}
After constructing the regularized SEDs, we normalize them by finding
the mean SED and integrating its reconstructed spectra at the day of maximum light in B-band (p = 0) over the g-filter from Dark Energy Survey, \citet{Abbott_2018}. Every SN SED is then multiplied by a normalization factor such that the integral of their spectra at maximum light over the g-filter equals the mean SED one. This procedure defines the relative normalization between each pair of SEDs. The normalization was seen to affect the model training, and since it is sample dependent, the model training will rely on the assumption the training set SEDs are representative of the SNIa class of objects. 

\section{Model Training}
\label{sec:model_training}
\subsection{Dimensionality reduction of supernovae SEDs with Principal Component Analysis}
\label{sec:pca}
From the set of reconstructed SEDs, we select a random subsample of 150 SN to compose the training set and save the remaining objects for a further validation step. Due to the observed similarities among SNeIa, one may argue most of the features in this sample should be correlated and partially redundant. Following \citet{Guy2007, Kim_2013, Sasdelli_2015, He_2018}, we investigate an orthonormal transformation to an ordered lower dimension basis via a Principal Component decomposition. 

The Principal Component Analysis (PCA), \citet{hotelling1933analysis,pearsonpca}, aims to find the uncorrelated directions that successively maximize the explained variance of the original data. This process translates into diagonalizing the sample covariance matrix, $\textbf{S}=1/(N-1) \sum_{n=1}^N (\textbf{x}^n-\textbf{m})(\textbf{x}^n-\textbf{m})^T$, where $\textbf{x}^n$ is our data set and $\textbf{m}$ is the sample mean vector, ordering the eigenvectors, $\textbf{e}^1, ..., \textbf{e}^D$ (D is the dimension of the original basis), according to the highest eigenvalues, and identifying the eigenvectors as the Principal Components. By analyzing the explained variance in terms of number of components, a dimensionality reduction can be applied, i.e. only a few M first principal components (M<D) should be enough to describe the majority of the training sample variance. Where M is the dimension of the transformed basis. More detailed discussions can be found in \citet{jolliffe2002principal} and \citet{barber2012bayesian}.

This process assumes the training set is representative of the diversity of SNIa and their environments. Additionally, it assumes that there is a lower dimension hyperplane capable of describing the original data with a few first Principal Components. It is important to note PCA does not assume the existence of hidden variables and the Principal Components which are linear combinations of the original basis are not affected by the choice of the new basis dimension and neither are directly associated with any physical interpretation.

To find accurate PCs using Singular Value Decomposition (SVD) the mean of each feature, in our case, each grid element, is required to be zero. Therefore, we remove from each SN the average SED before applying the decomposition method to the training sample. $M_0$ in equation~(\ref{eq:restframeflux}) is the average surface while the remaining components, $M_i$ (with $i>0$), are the principal components.

\begin{figure}
\centering
\includegraphics[width=0.5\columnwidth]{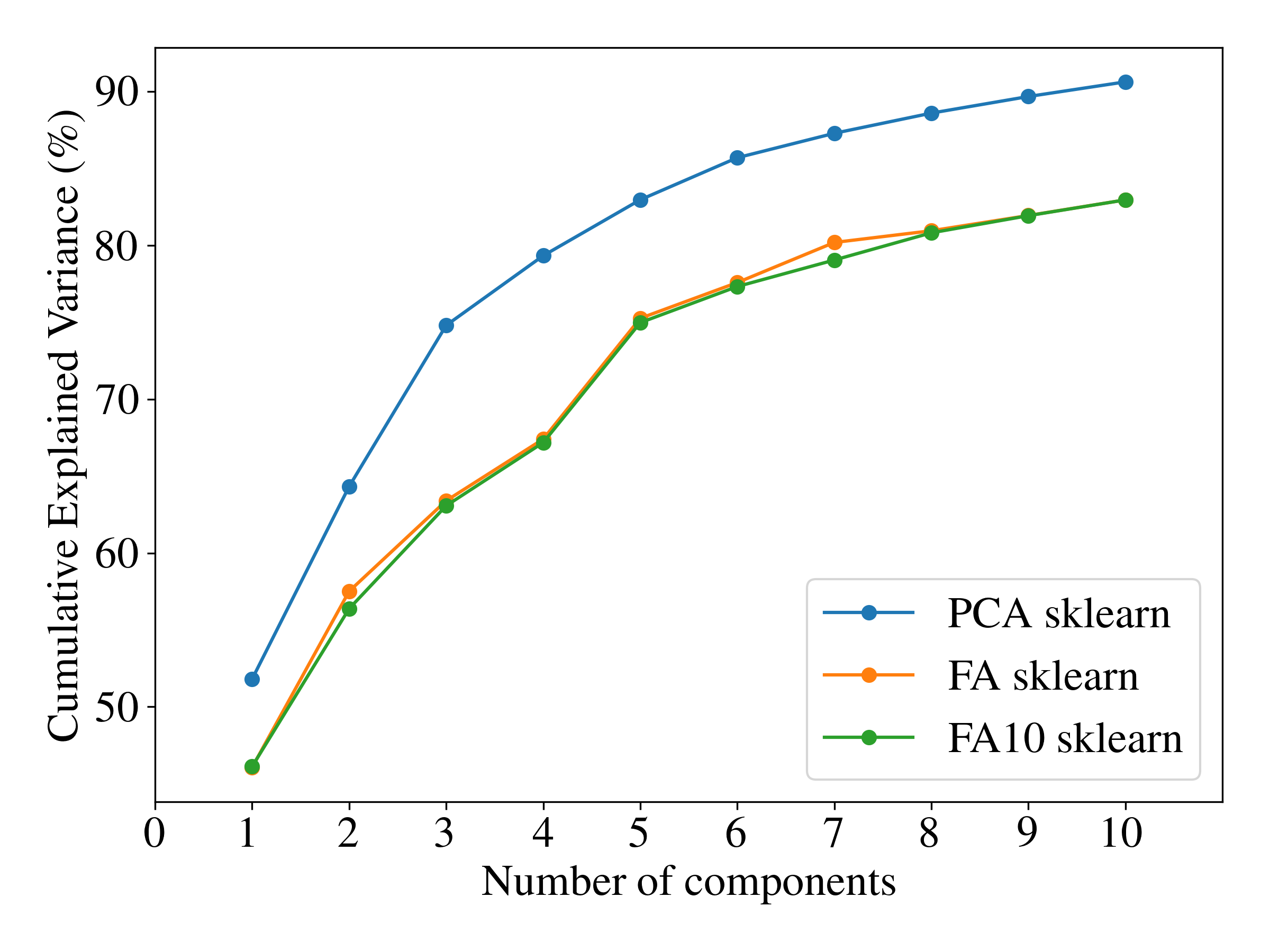} 
\caption{Cumulative explained variance in terms of number of components, both for Principal Component Analysis and Factor Analysis. As expected, first-order templates hold a higher variance. FA10 represents the cumulative explained variance in a Factor Analysis generated with 10 components. FA is the cumulative explained variance encountered for each number of components for a Factor Analysis generated with that specific number of components.} 
\label{fig:cumexpvariance}
\end{figure}

We perform this dimensionality reduction through the Python package Scikit-learn, \citet{scikit-learn}. The input data is a $150 \times 32208$ (150 training SEDs) matrix of the flattened SN SEDs, previously regularized and normalized, placed as rows. Fig.~\ref{fig:cumexpvariance} shows the cumulative explained variance as a function of the number of components kept from the new ordered basis, $N_c$. For PCA $N_c = 2$, we explain $64.3\%$ of the training sample variance, the 90\% explained variance mark is only surpassed for $N_c\geq 10$.

Analyzing the explained variance as a function of $N_c$ can offer insight into the number of components to keep after the orthonormal transformation. We verify the second and third components describe a similar amount of variance. Also, the fourth, fifth, and sixth explain a similar amount of variance. These different numbers of model components were investigated but here for the remainder of this study, we attain to $N_c=2$, maintaining 3 fitting parameters in equity to other light curve fitters.

\begin{figure}
\begin{minipage}[b]{.45\textwidth}
\centering
\includegraphics[width=\columnwidth]{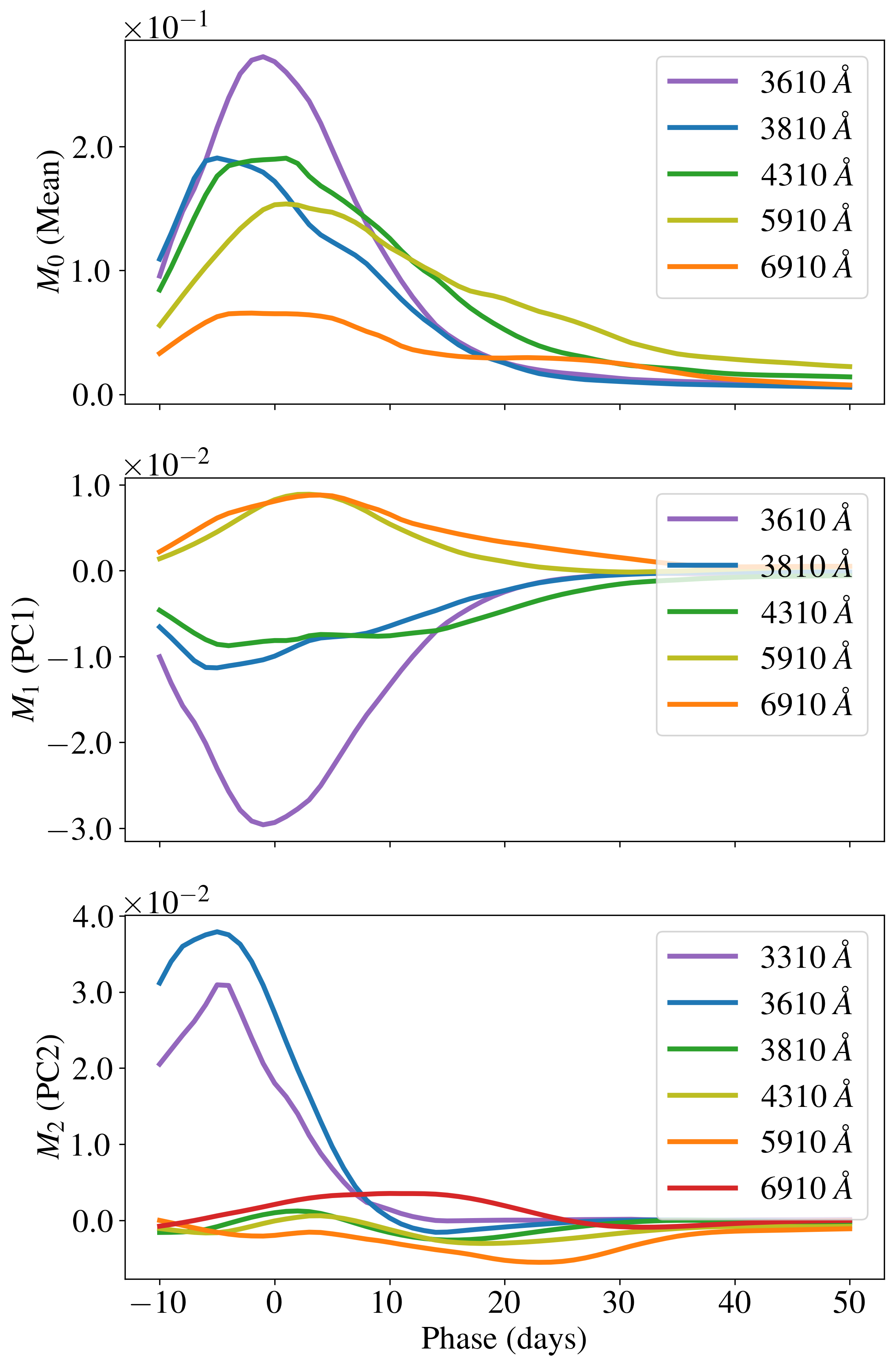} 
\end{minipage}
\hfill
\begin{minipage}[b]{.45\textwidth}
\centering
\includegraphics[width=\columnwidth]{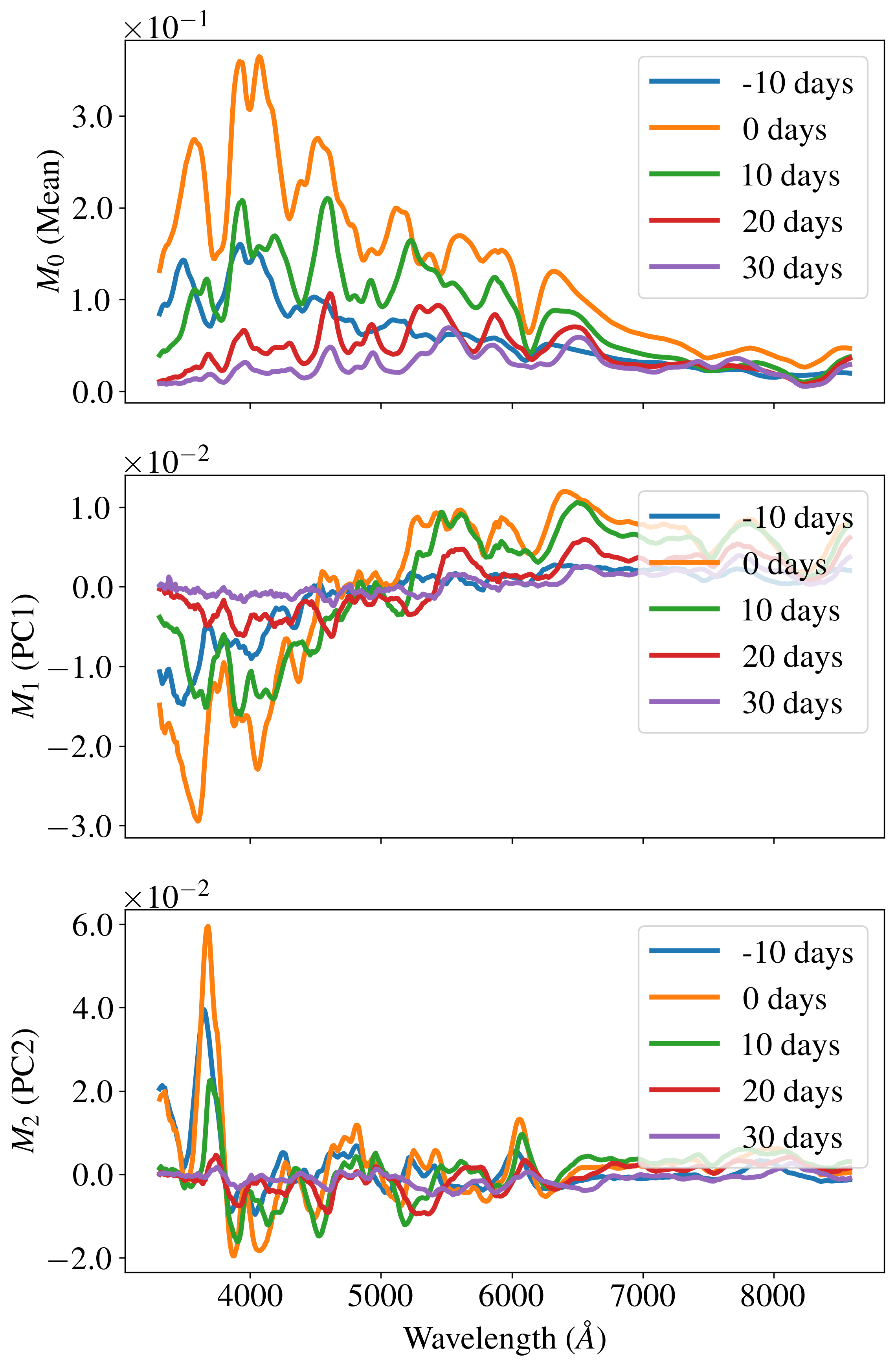} 
\end{minipage}
\caption{PCA-PETS model components in units of flux (erg/s/cm$^2$/\AA $\times$ arbitrary factor), as a function of days (left) and wavelength (right).}
\label{fig:pcacomponents}
\end{figure}

In Fig.~\ref{fig:pcacomponents} we illustrate our PCA-PETS model components. In the left panel, the components are illustrated as a function of days for different wavelengths, while the right panel shows their behavior as a function of wavelength for different days. The first row portrays the average training surface, corresponding to our model component $M_0$, recalling equation~(\ref{eq:restframeflux}). The second row shows the first principal component, which plays the role of $M_1$, and in the third row, we have the second principal component, $M_2$ in our model. $M_1$ is the component that explains the most variance around the average surface, the corrections it adds to the mean are concentrated close to the date of maximum light with the correction changing sign with growing wavelength. $M_2$ corrections also concentrate close to the date of maximum light and rapidly go to zero for higher wavelengths. 

In Fig.~\ref{fig:broadband_pcacomponents} the PCA-PETS model components are shown after integration in CSP filter. $M_1$ corrections are concentrated close to the day of maximum light while $M_2$ main corrections concentrate 10 to 15 days after the day of maximum light.

\subsection{Dimensionality reduction of supernovae SEDs with Factor Analysis}
\label{sec:fa} 
Another method that was implemented for SNIa feature extraction by \citet{saunders2018snemo} and \citet{sugar2020} is Factor Analysis (FA). FA relies on an important underlying assumption that there are latent (i.e. unobserved) variables that can explain our original data, often with a reduced set of components. 

As we discussed previously, the PCs are constructed as linear combinations of the original variables, without any mention of their physical interpretations nor with an explicit model. Nonetheless, for Factor Analysis, the observed quantities, \textbf{x}, are the ones assumed to be linear combinations of the latent variables, \textbf{f}, except for an error term, \textbf{e}, (i.e. $\textbf{x}=\boldsymbol{\Lambda}\textbf{f}+\textbf{e}$), 
with the $\boldsymbol{\Lambda}$ matrix carrying the coefficients. This generative model also requires some assumptions over the distributions and correlations of the data, the latent variables, also known as common factors, and the error terms, also known as specific factors. These three quantities are assumed to have null expected values and the common and specific factors are assumed to be uncorrelated, with themselves and with each other.

Without loss of generality the common and specific factors are considered to be described by Gaussian distributions, $\textbf{f}\sim \mathcal{N}(0,\textbf{I})$ and $\textbf{e}\sim \mathcal{N}(0,\boldsymbol{\Psi})$, respectively. Then the conditional distribution of the observed variables is given by
\begin{equation}
    p(\textbf{x}|\textbf{f})=\mathcal{N}(\textbf{x}|\boldsymbol{\Lambda}\textbf{f}+\boldsymbol{\mu},\boldsymbol{\Psi}).
    \label{eq:faequation}
\end{equation}
Here $\boldsymbol{\mu}$ is an offset and $\boldsymbol{\Psi}$ is a diagonal matrix with different entries, characterizing a heteroscedastic noise. This decomposition consists of an iterative method using maximum likelihood estimation with an SVD approach and it was also implemented using the Python package Scikit-learn, \citet{scikit-learn}. This iterative method solves for the common factors $\textbf{f}$ for a fixed $\boldsymbol{\Psi}$, then with common factors solution a new noise variance is evaluated and this process is repeated until a convergence criterion is met.

It is interesting to note that there is a probabilistic description of PCA that is closely related to FA, Probabilistic Principal Component Analysis, known as PPCA. This also linear-Gaussian framework differs from FA only by defining a homoscedastic, isotropic noise, where $\boldsymbol{\Psi}=\sigma^2\textbf{I}$. In the limit $\sigma^2\rightarrow 0$ it recovers the PCA results, hence the name. For a more detailed discussion on these methods see \citet{bishop2006pattern} and \citet{tipping1999mixtures}.

Overall, as argued by \citet{jolliffe2002principal}, PCA and FA can be understood as explaining different aspects of the sample covariance matrix. While PCA concentrates on explaining the diagonal elements through maximizing explained variance, FA concentrates on the off-diagonal elements by forcing the specific factors to be uncorrelated and leaving the off-diagonal elements to be fully explained by the common factors.

In addition, the solution for $\textbf{f}$ is not unique, arbitrary rotations can be performed over this matrix and some rotations can facilitate recognizing the role of the new variables. We can perform orthogonal and oblique rotations, the latter allowing for correlated factors. These rotations will be further explored in future work, for now, we provide the results for FA in the absence of rotations. In the absence of oblique rotation, the $\textbf{f}$ factors are orthogonal. 

As in PCA, to obtain accurate results it is necessary to center each feature before applying the decomposition algorithm. The average SED is also the $M_0$ component for FA-PETS and it is subtracted from each SN SED before the orthonormal transformation. The input is the same $150 \times 32208$ matrix of training set SEDs.

Fig.~\ref{fig:cumexpvariance} shows the cumulative explained variance by the common factors for Factor Analysis. As previously mentioned, this decomposition method describes the observations as a linear combination of common factors with the addition of specific factors. This decomposition is then performed specifically for a chosen number of hidden variables, i.e. common factors. A 2-component Factor Analysis solution does not generally correspond to a subset of a 10-component Factor Analysis solution. Fig.~\ref{fig:cumexpvariance} exemplifies this difference, "FA10 sklearn" refers to a 10-component FA with corresponding components explained variance. For "FA sklearn" a new FA solution with $N_c$ number of components was fitted for each case and the explained variance was evaluated for the last component. Considering the latter, for $N_c$=2 FA describes $57.5\%$ of the training sample variance. 

\begin{figure}
\begin{minipage}[b]{.45\textwidth}
\centering
\includegraphics[width=\columnwidth]{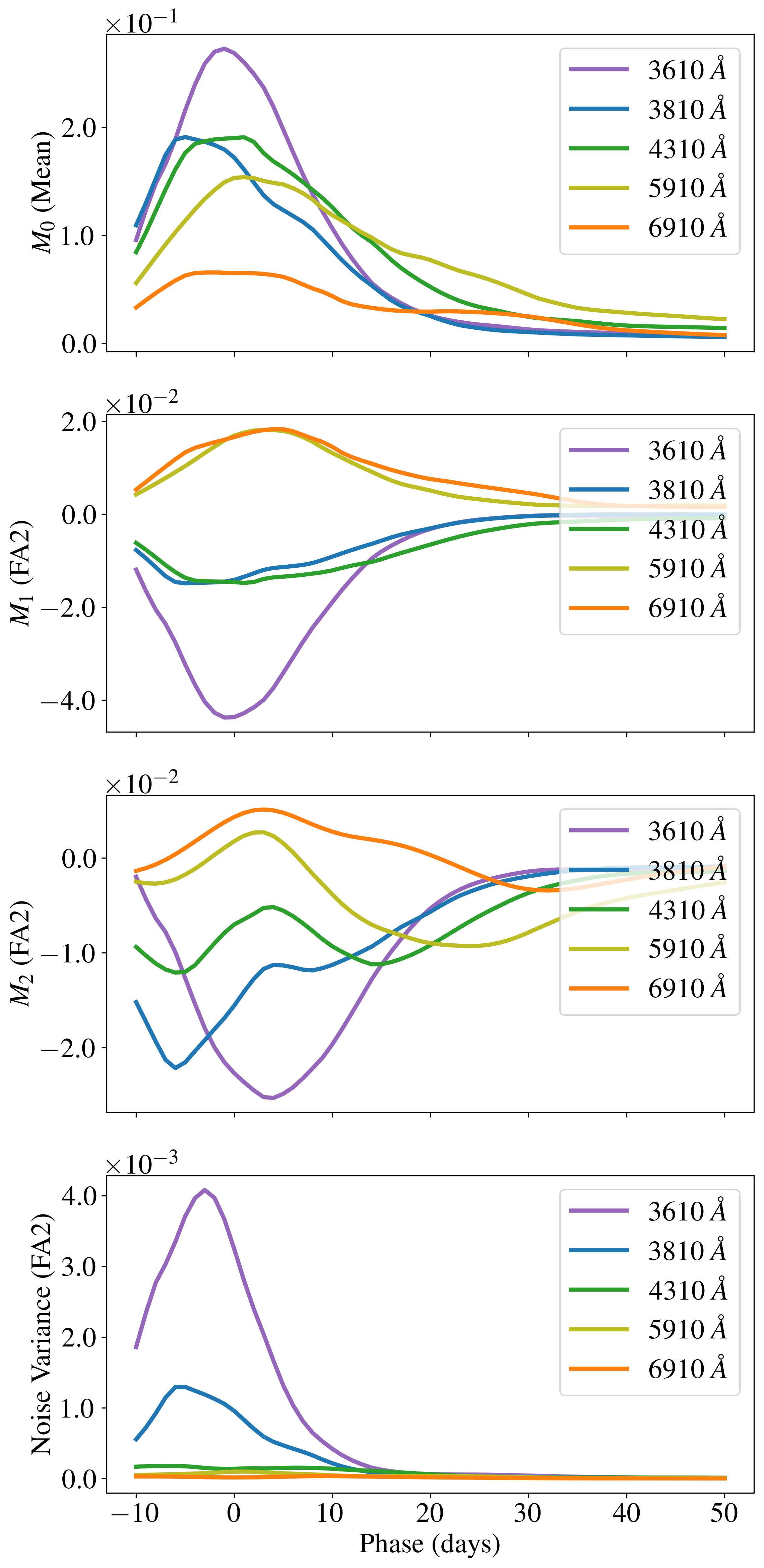} 
\end{minipage}
\hfill
\begin{minipage}[b]{.45\textwidth}
\centering
\includegraphics[width=\columnwidth]{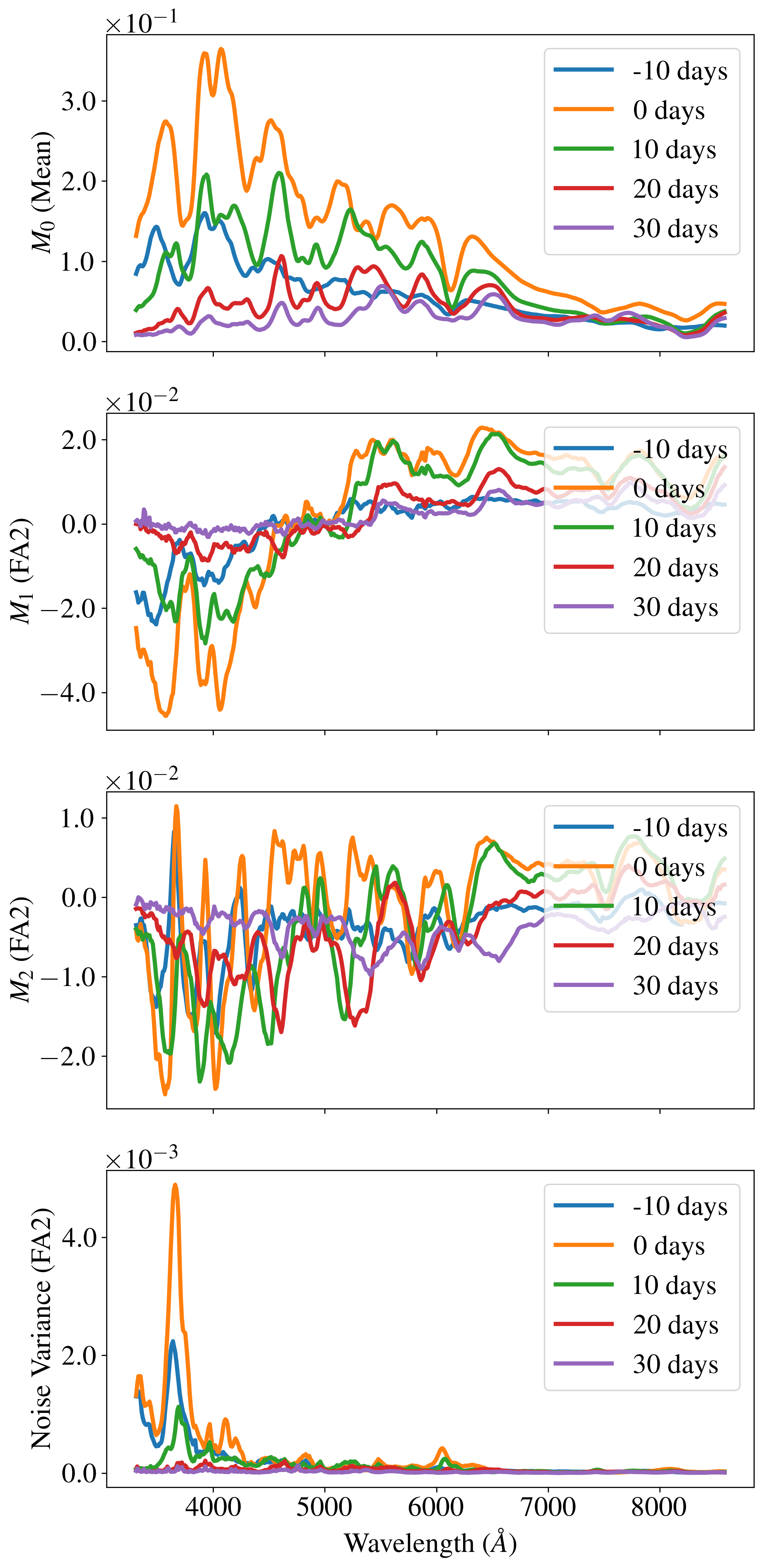} 
\end{minipage}
\caption{FA-PETS model components in units of flux (erg/s/cm$^2$/\AA $\times$ arbitrary factor) as a function of days (left) and wavelength (right).}
\label{fig:facomponents}
\end{figure}

Fig.~\ref{fig:facomponents} shows the common factor solutions obtained for a FA assuming two hidden variables, as a function of days and wavelength. For this model, the SNIa rest-frame flux $M_0$ component is the training average SED. $M_1$ is the first common factor and $M_2$ is the second common factor. In the same figure we also portray $\boldsymbol{\Psi}$ from equation~(\ref{eq:faequation}), the noise variance fitted for FA with $N_c=2$. The initial guess for noise variance was a $1 \times 32208$ vector of the average training SEDs variances obtained from the GPRs. For PPCA, instead of a noise variance vector we would wave a scalar, thus no change in variance with phase and wavelength would be allowed. And in the limit this scalar approaches 0, we recover the PCA solution.

\begin{figure}
\begin{minipage}[c]{.45\textwidth}
\centering
\includegraphics[width=\columnwidth]{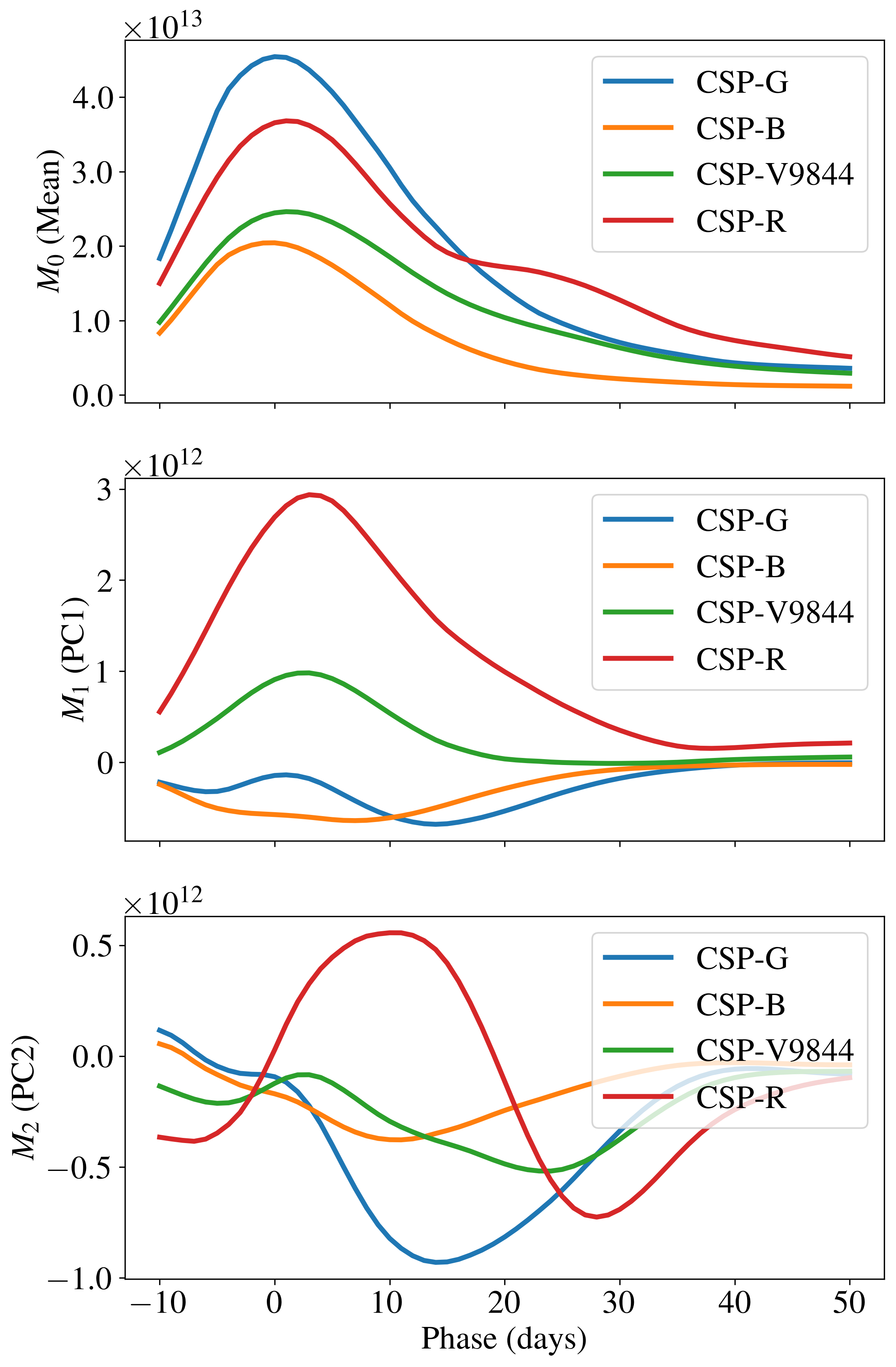} 
\caption{PCA-PETS model components integrated in CSP filters.}
\label{fig:broadband_pcacomponents}
\end{minipage}
\hfill
\begin{minipage}[c]{.45\textwidth}
\centering
\includegraphics[width=\columnwidth]{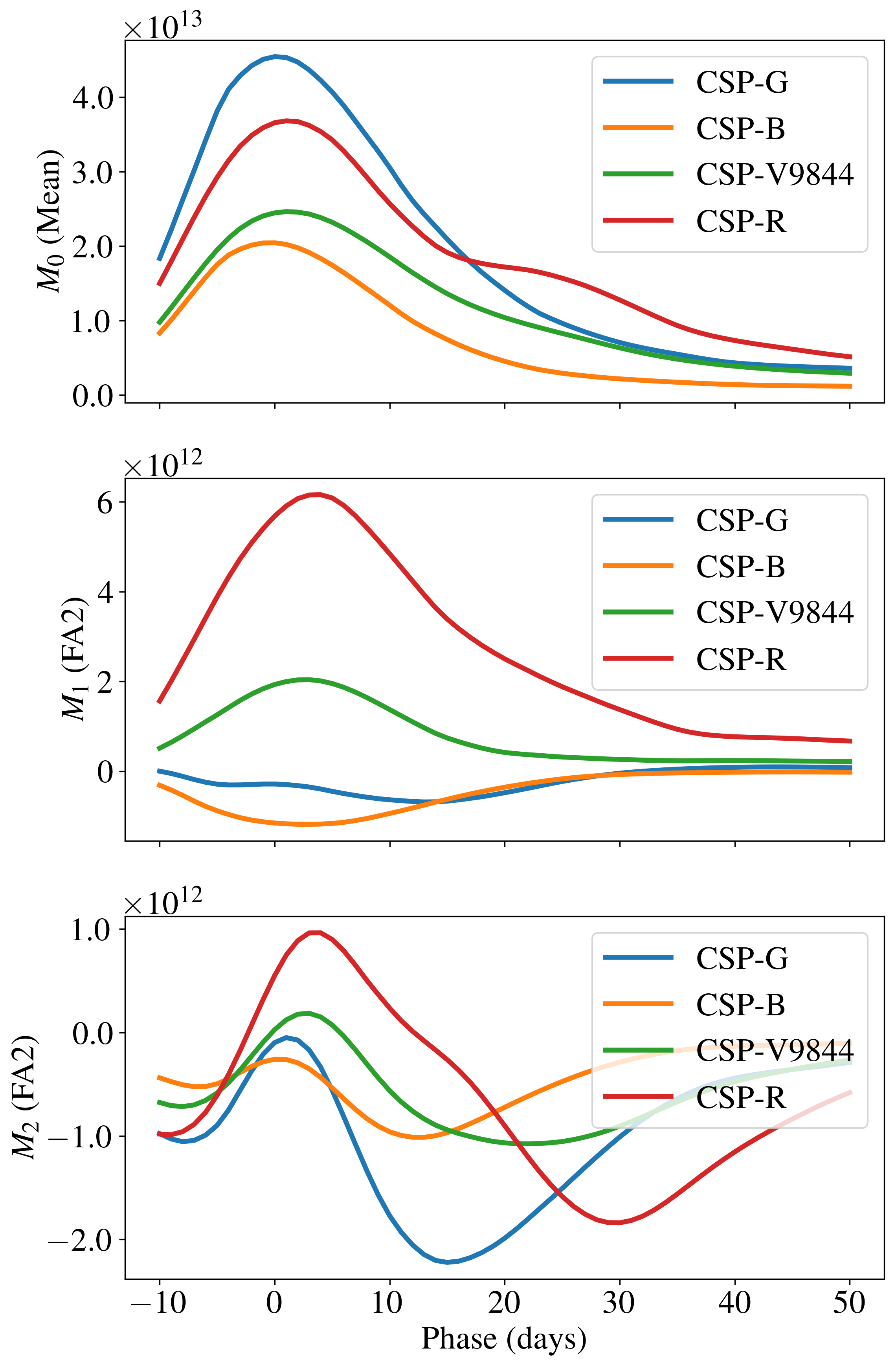}
\caption{FA-PETS model components integrated in CSP filters.}
\label{fig:broadband_facomponents}
\end{minipage}
\end{figure}

As the majority of the emitted flux is concentrated close to the day of maximum light and around $3000$ \AA\, to $6000$ \AA, it is expected that the corrections also concentrate on the same region. As both PCA and FA methods are trained over the same training set, their $M_0$ surface is identical. FA $M_1$ resembles the corresponding PCA component, with corrections changing signs with wavelength growth. FA $M_2$ differs significantly from its PCA counterpart, its corrections concentrate close to the date of maximum light, are mainly negative, and are more significant at higher wavelengths. Lastly, $\boldsymbol{\Psi}$ concentrates in the region of maximum light and around 3000\AA \,to 4000\AA. Fig.~\ref{fig:broadband_facomponents} shows the components integrated in CSP passbands. PCA-$M_1$ and FA $M_1$ differ more significantly in the g-band, i.e. for lower wavelengths. FA $M_2$ shows corrections closer to the day of maximum light with a sharper peak, in comparison to PCA $M_2$.

\subsection{Validation fits}
\label{sec:validationfits} 

The remaining 16 SNIa that did not take part in the feature extraction form a set for model validation. The entire analysis described in this section was performed for both validation and training sets to check for overfitting.

We investigated the performance of PCA and FA decomposition methods when reconstructing the original SEDs after applying the orthonormal transformations, reducing the original basis dimension. This procedure consists of projecting the original SEDs on the new basis and returning to the original one. Information is lost since the lower dimension basis does not entirely spawn the original space. The goodness of fit metric chosen to quantify the reconstruction quality was the $\chi^2/\textnormal{ndof}$, considering uncertainties from the previous GPRs.

\begin{figure}
\begin{minipage}[b]{.45\textwidth}
\centering
\includegraphics[width=\columnwidth]{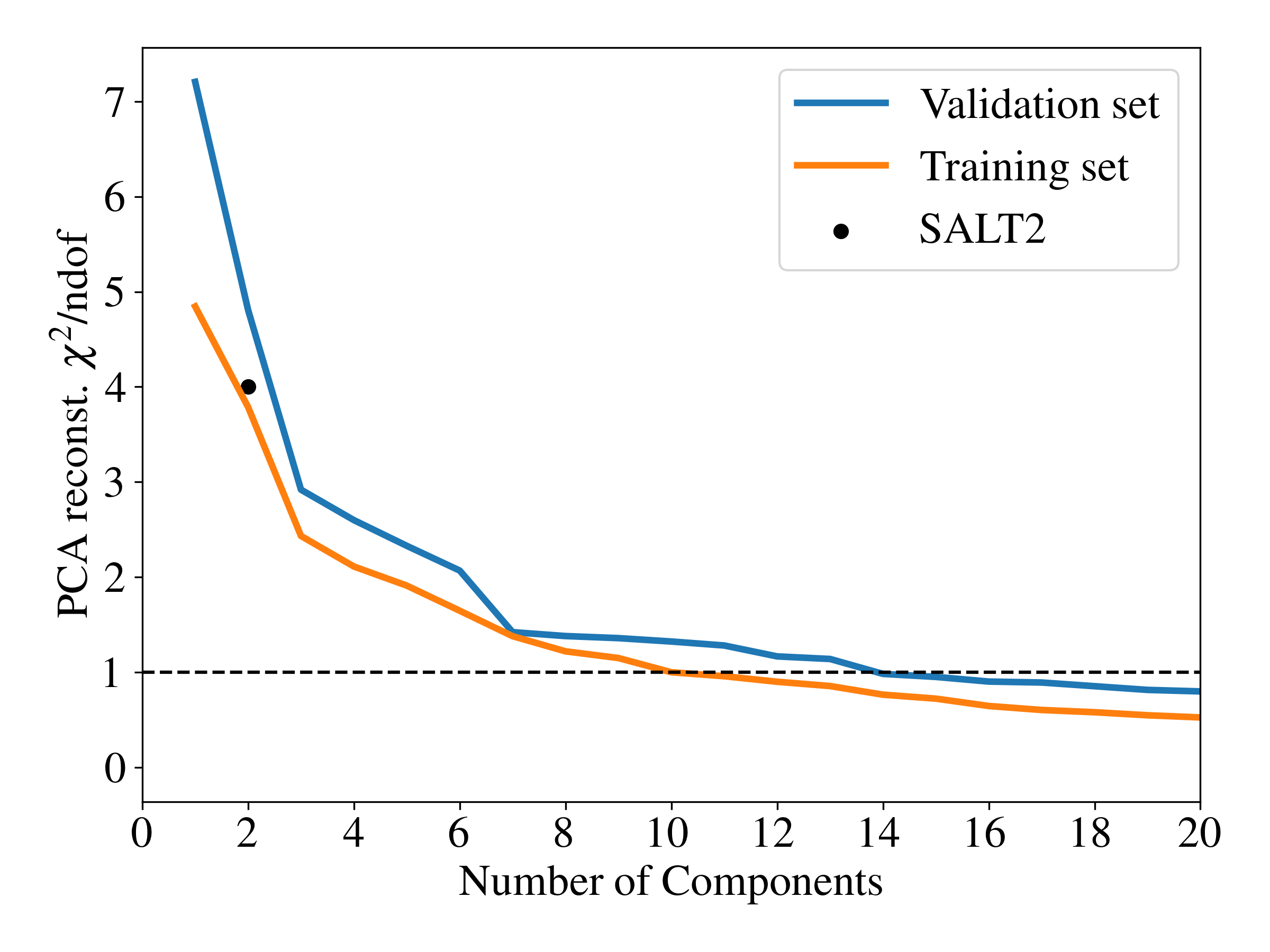} 
\end{minipage}
\hfill
\begin{minipage}[b]{.45\textwidth}
\centering
\includegraphics[width=\columnwidth]{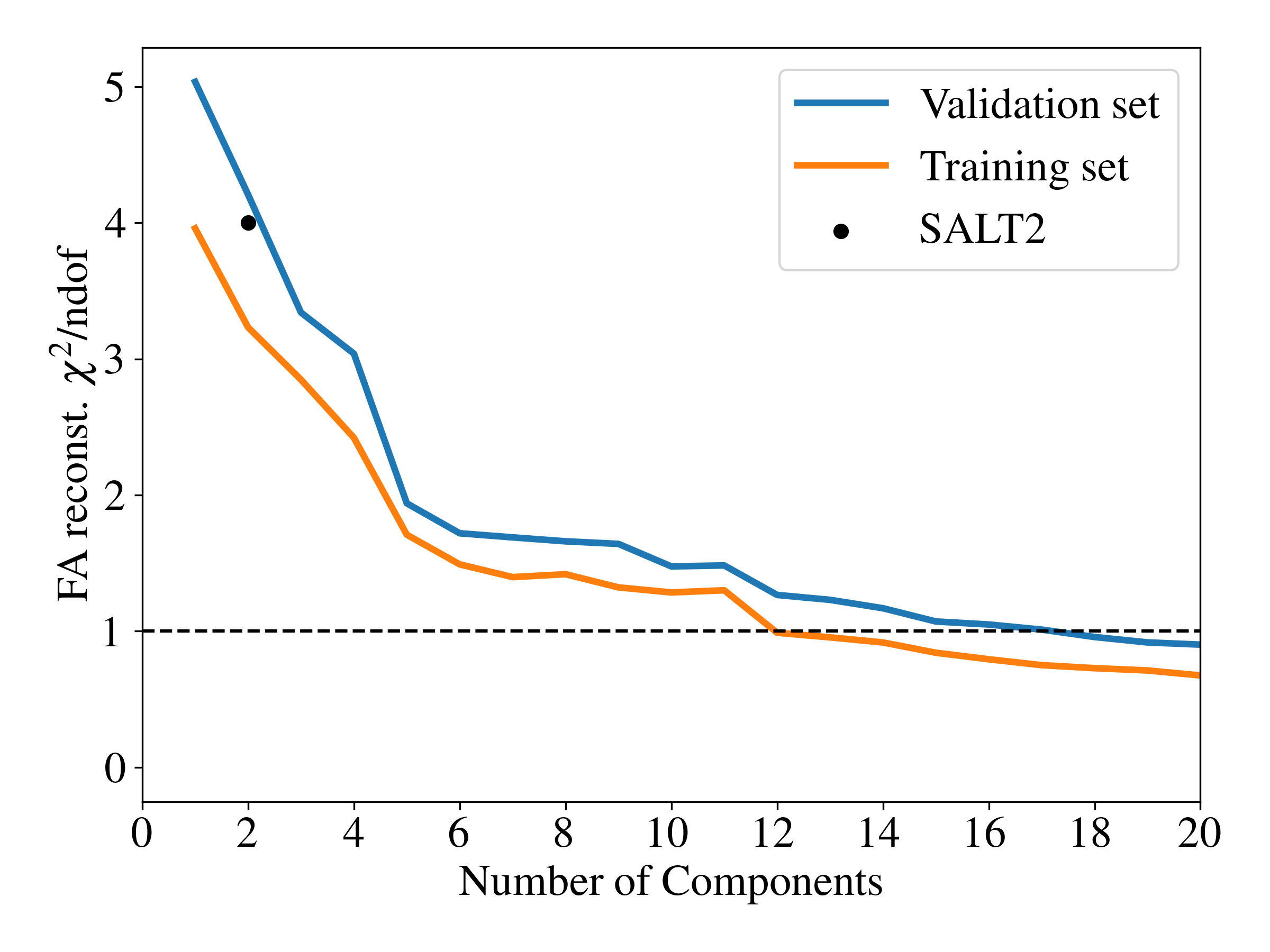} 
\end{minipage}
\caption{$\chi^2$/ndof as a function of the number of components for Principal Component Analysis (left) and Factor Analysis (right) reconstructions. The black dot illustrates the corresponding SALT2 value.}
\label{fig:chi2_val_train_fits}
\end{figure}

The $\chi^2$/ndof is evaluated for each SNIa SED, the average value is obtained for validation and training sets and divided by the number of degrees of freedom (\# of grid points-\# of components). Fig.~ \ref{fig:chi2_val_train_fits} shows the evolution in reconstruction quality when increasing the number of principal components or common factors in the decomposition. As expected, when increasing the dimension of the new basis, more information can be stored. For PCA, the training set discrepancy tends to 0 with the growing number of PCs included in the reconstruction. For $N_c=150$ the new basis will consist of an orthogonal rotation of the original basis and the entire original space is spawned. 
For PCA, around $N_c=14$ we obtain $\chi^2$/ndof below one. 

Concerning FA, for each number of components a new FA was generated and the $\chi^2$/ndof corresponds to keeping all components in each scenario. We can visualize a bottom triangular matrix, where each row corresponds to the number of common factors generated from 1 to 150. The columns correspond to the number of components we decide to keep after the FA. Each element of this matrix has a $\chi^2$/ndof associated and since FA supposes the existence of hidden features, increasing the number of components does not guarantee a continuous drop in the goodness of the reconstruction metric. The curve on the left in Fig.~ \ref{fig:chi2_val_train_fits} corresponds to the main diagonal of this matrix. The transition for $\chi^2$/ndof below one occurs for $N_c=17$. 

In Fig.~ \ref{fig:chi2_val_train_fits} we also show the corresponding average $\chi^2$/ndof if we fitted the SEDs to SALT2' rest frame flux, equation~(\ref{eq:restframefluxsalt2}). We note that SALT2 performs better than our PCA model version and as well as our FA model version when considering our training and validation sets.

\begin{figure}
\centering
\includegraphics[width=0.8\columnwidth]{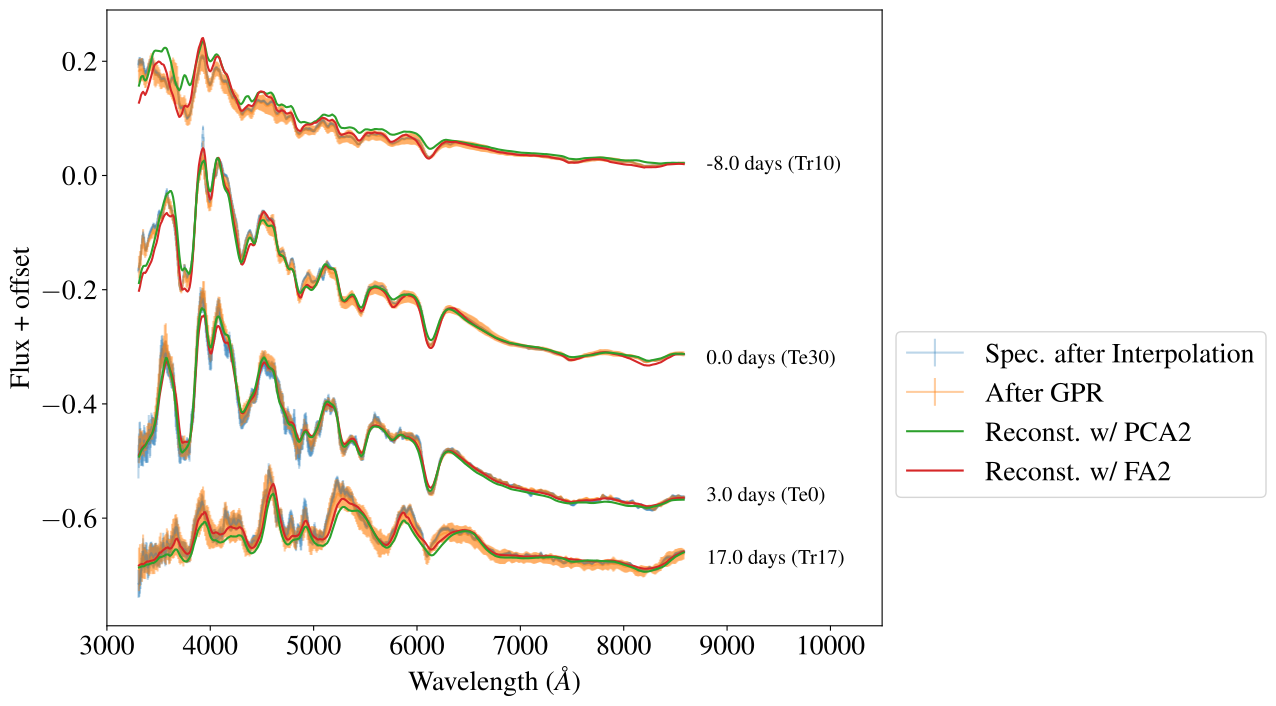} 
\caption{Comparison of "Test\_SN0", "Test\_SN30", "Train\_SN10" and "Train\_SN17"  spectra with its reconstructions. We show the original spectra, the spectra after applying Gaussian process regression to regularize the entire SED, the reconstruction considering only the first two principal components of PCA, and last, the reconstruction considering only the first two common factors of FA.}
\label{fig:spec_recons}
\end{figure}

The methods reconstruction performance for spectra can be seen in Fig.~\ref{fig:spec_recons}. We show arbitrary spectra of "Test\_SN0", "Test\_SN30", "Train\_SN10" and "Train\_SN17". The plot includes the original data, the reconstruction after SED regularization with GPR, and reconstructions after PCA and FA decompositions with two components. The model struggles most in reconstructing data before $p=0$ and for lower wavelengths.

\begin{figure}
\begin{minipage}[b]{.5\textwidth}
\centering
\includegraphics[width=\columnwidth]{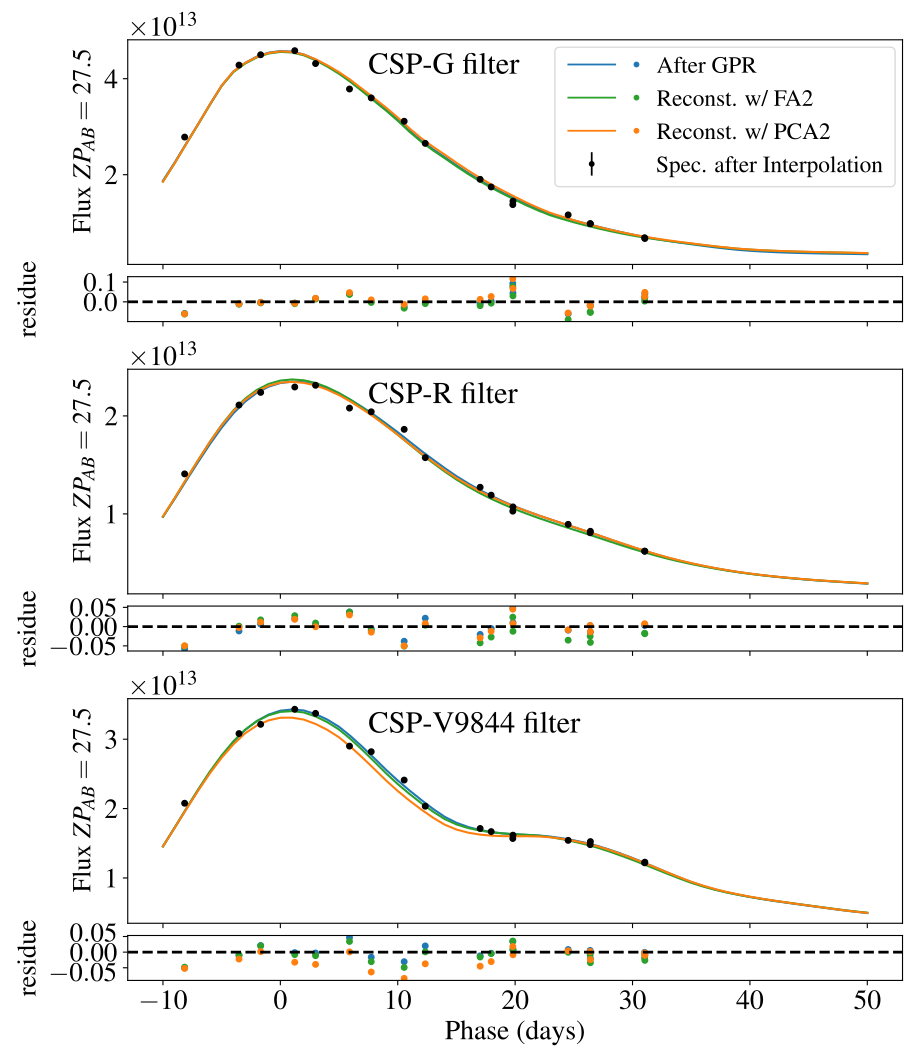} 
\end{minipage}
\hfill
\begin{minipage}[b]{.5\textwidth}
\centering
\includegraphics[width=\columnwidth]{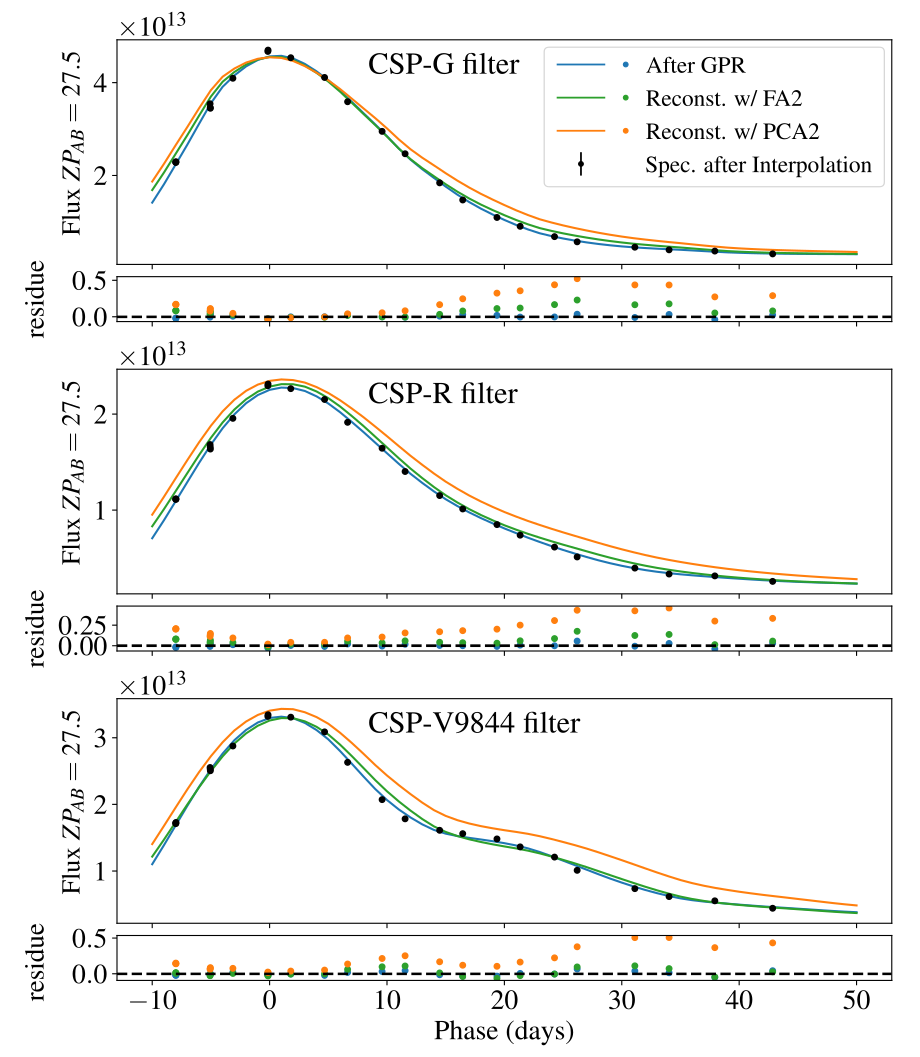} 
\end{minipage}
\caption{Broad-band reconstructions of "Test\_SN0" and "Train\_SN10" SEDs, respectively.}
\label{fig:broadband_recon}
\end{figure}

In Fig.~\ref{fig:broadband_recon} we observe the broadband reconstruction for "Test\_SN0" and "Train\_SN10" SEDs, both from the validation dataset. PETS-FA once again outperforms PCA-PETS, managing to recover more accurately the light-curve shape.

We also fit the light curve models PCA-PETS and FA-PETS over synthetic photometry. For this step, we used the Python package for Supernova Cosmology (\texttt{SNCosmo}), \citet{barbary_kyle_2022_7117347}. With \texttt{SNCosmo} we were able to create a spectral time series function of arbitrary parameters and fit the light curves by generating synthetic photometry from the SEDs of both validation and training sets. To draw maximum information from this data we chose the set of filters used by The Carnegie Supernova Project (CSP), \citet{Krisciunas2017}. Our photometric system consists of filters CSP-g, CSP-r, and CSP-V9844 with AB magnitudes. The first two filters cover a range from about 4000\AA\, to 7000\AA\, but with a lower coverage around 5500\AA. This same region is mainly contained in CSP-V9844. These three filters extract an important amount of information from our original SEDs and capture a good portion of the optical region. The considered models were created by truncating the rest-frame flux expansion, equation (\ref{eq:restframeflux}), after different numbers of terms based on previous discussions of similar amounts of explained variances by neighboring components. Here we include the covariance between measurements taken at the same phase but in different bandpasses and the GPRs flux uncertainties.

\begin{figure}
\begin{minipage}[b]{.49\textwidth}
\centering
\includegraphics[width=\columnwidth]{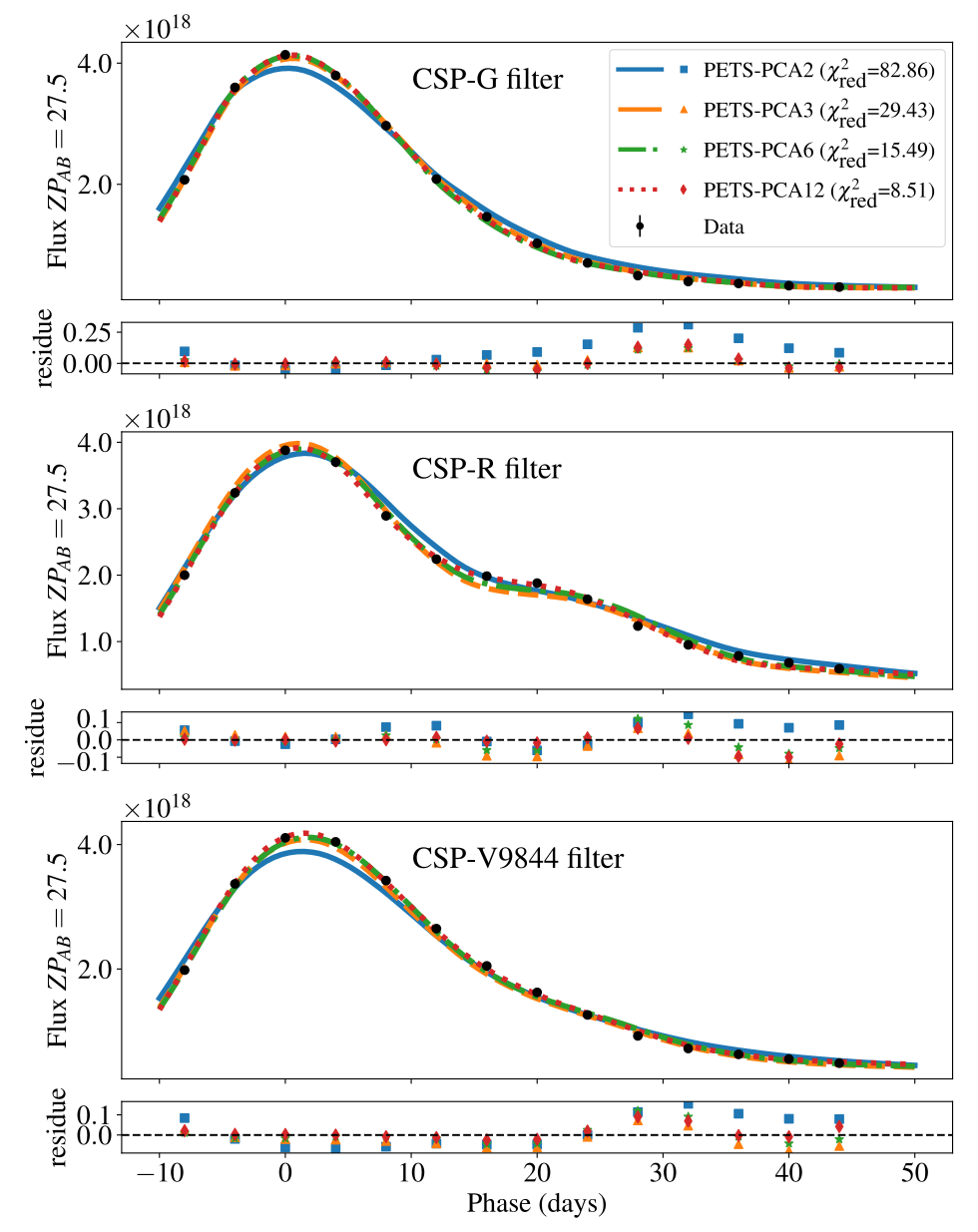} 
\end{minipage}
\hfill
\begin{minipage}[b]{.5\textwidth}
\centering
\includegraphics[width=\columnwidth]{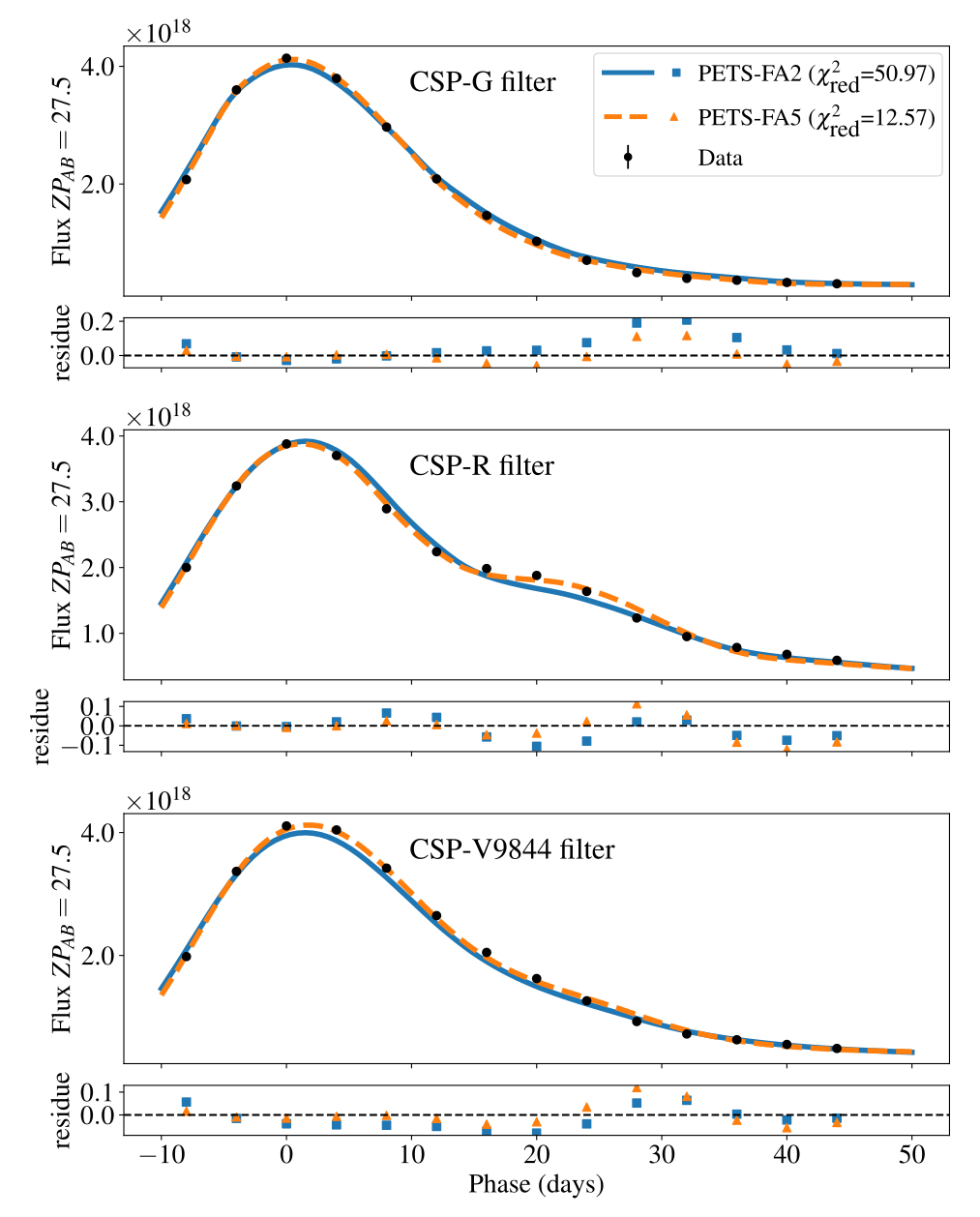} 
\end{minipage}
\caption{Light curve fitting of the representative validation set supernova "Train\_SN78". On the left figure, we have a comparison of fitting results for models constructed with 2, 3, 6, and 12 principal components. In the right figure, we have a comparison of two models, the first constructed with 1 common factor and the second with 5 common factors.}
\label{fig:validationfit}
\end{figure}

Fig.~\ref{fig:validationfit} shows a comparison of different PETS light curve fits for the validation SN "Train\_SN78". As the synthetic photometry underestimates the uncertainties, the residue corresponds to the relative discrepancy.
For PCA-PETS, the 2-component model has difficulty fitting simultaneously every filter at $p=0$. The remaining models perform equally well.
For FA-PETS, the improvement is more subtle. Across the validation and training sets the same behavior was observed, FA-PETS outperforms PCA-PETS. Since the uncertainties are underestimated and PETS currently lacks a model covariance, the $\chi^2/\textnormal{ndof}$ is not a good quality fit ruler and is presented just to enable comparison between different models. No indication of overfitting is observed.

\begin{figure}
\centering
\includegraphics[width=0.8\columnwidth]{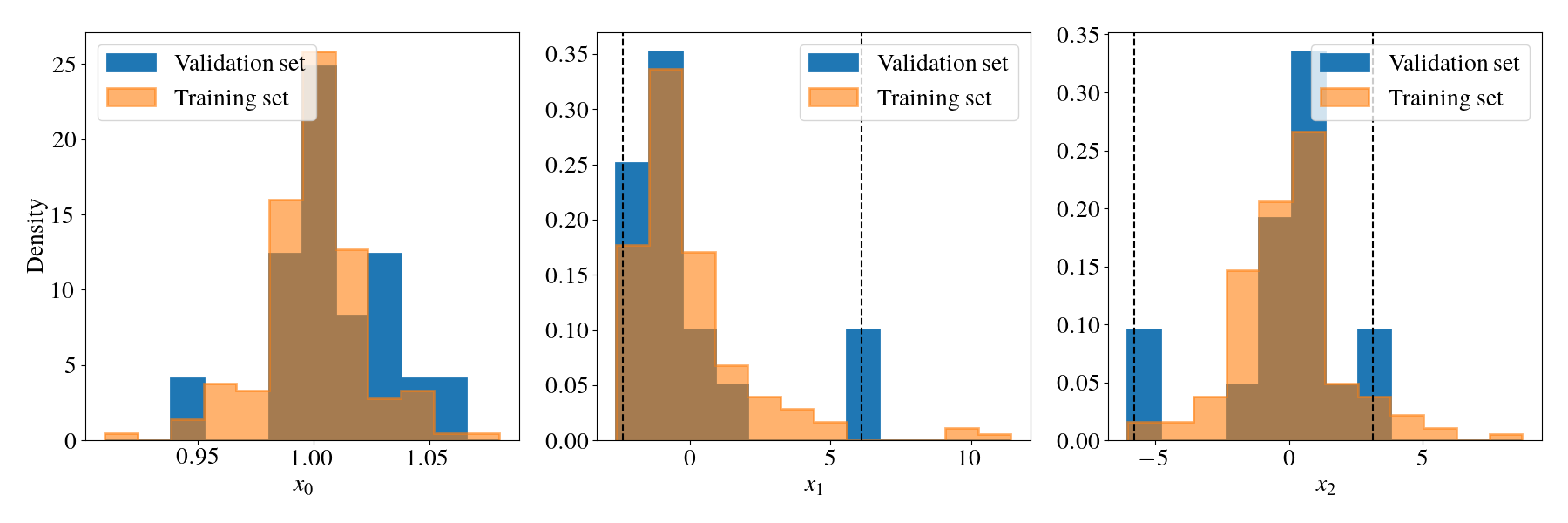} 
\caption{Validation and training sets fitted parameter distributions for PCA-PETS.}
\label{fig:val_pardist_pca}
\end{figure}

\begin{figure}
\centering
\includegraphics[width=0.8\columnwidth]{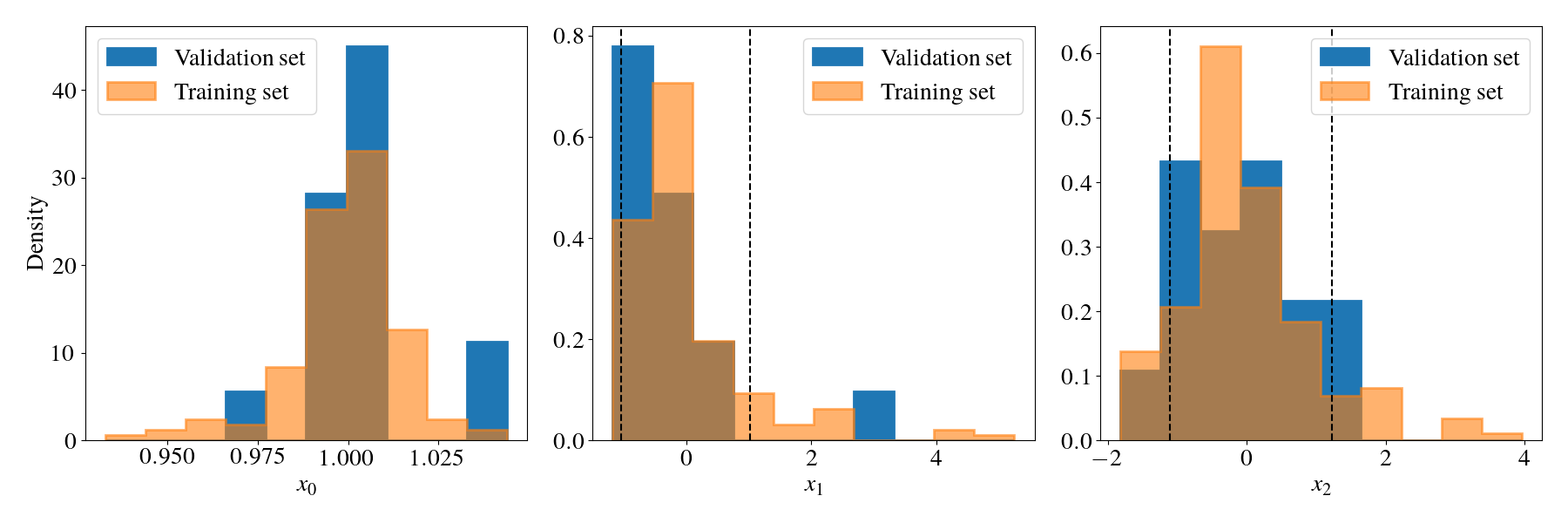} 
\caption{Validation and training sets fitted parameter distributions for FA-PETS.}
\label{fig:val_pardist_fa}
\end{figure}

The fit parameters distributions for validation and training sets can be seen in Fig.~\ref{fig:val_pardist_pca} and Fig.~\ref{fig:val_pardist_fa}, respectively. We can evaluate the Kolmogorov–Smirnov p-value to check if the hypothesis of the validation and test histograms being realizations of the same Gaussian distribution is rejected. We find for each pair of histograms $p>0.05$ and the null hypothesis is not rejected.

\section{Comparisons with SALT2 and SNEMO2}
\label{sec:comparison_lcfitters}

\begin{figure}
\begin{minipage}[b]{.32\textwidth}
\centering
\includegraphics[width=\columnwidth]{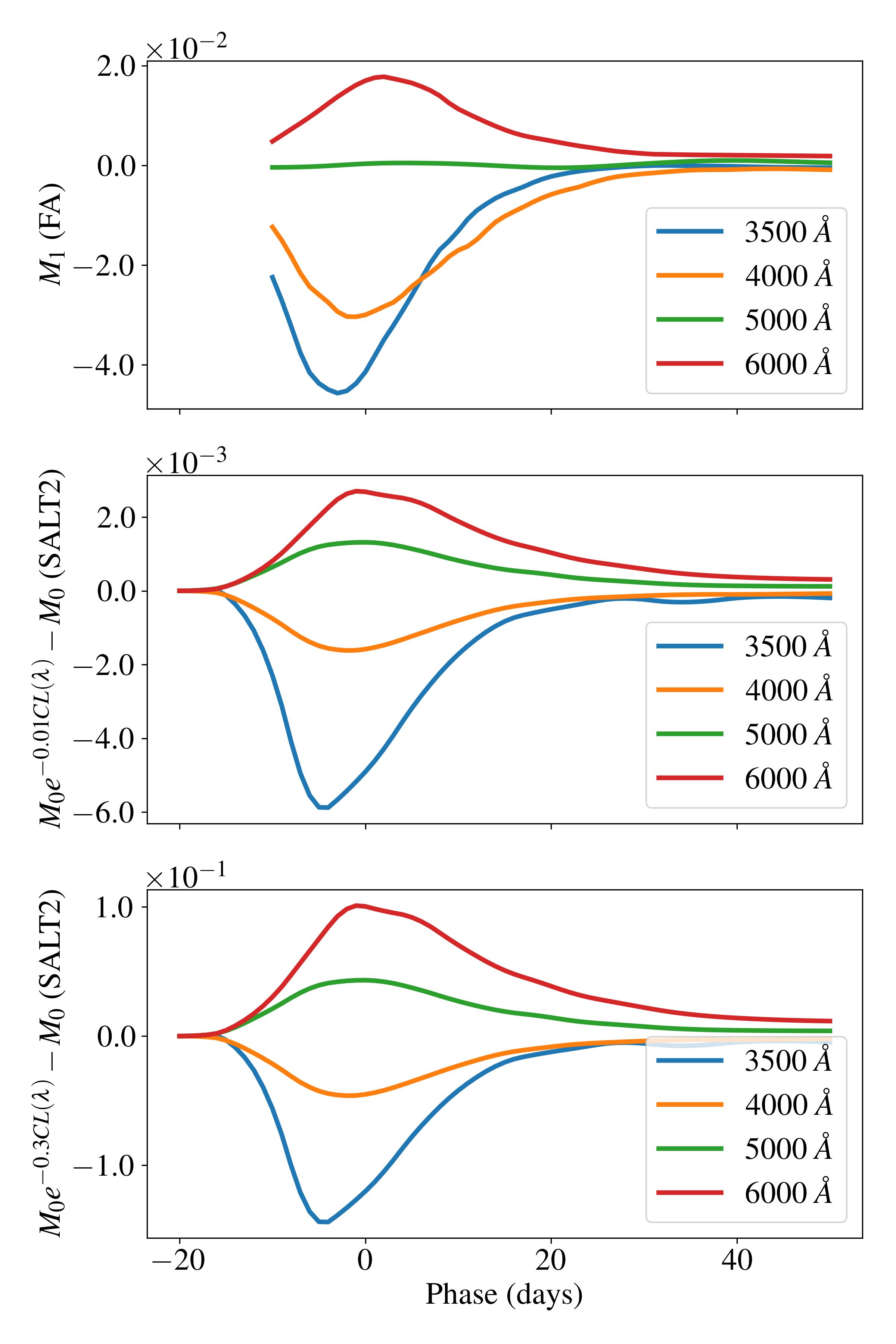} 
\end{minipage}
\begin{minipage}[b]{.32\textwidth}
\centering
\includegraphics[width=\columnwidth]{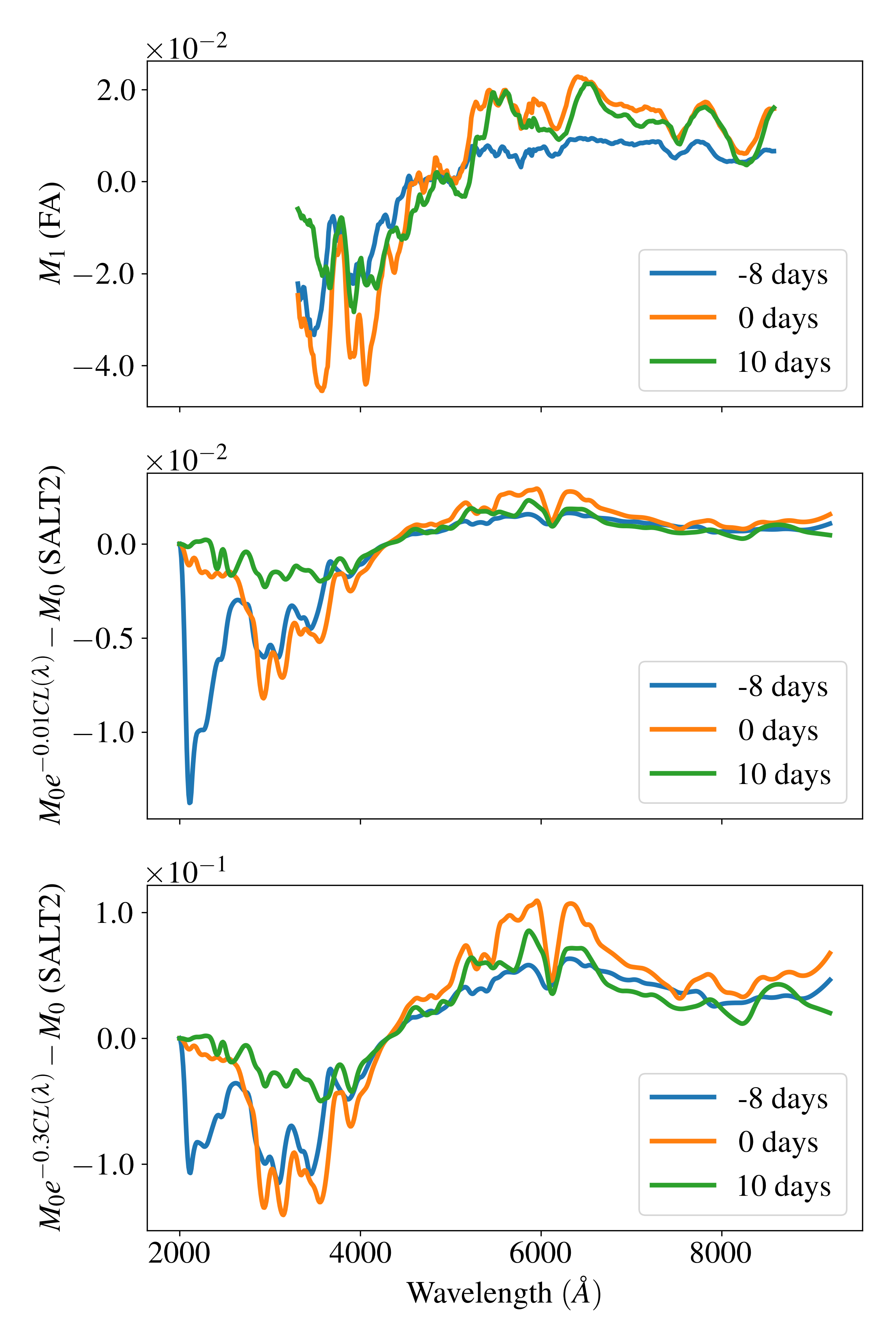} 
\end{minipage}
\begin{minipage}[b]{.32\textwidth}
\centering
\includegraphics[width=\columnwidth]{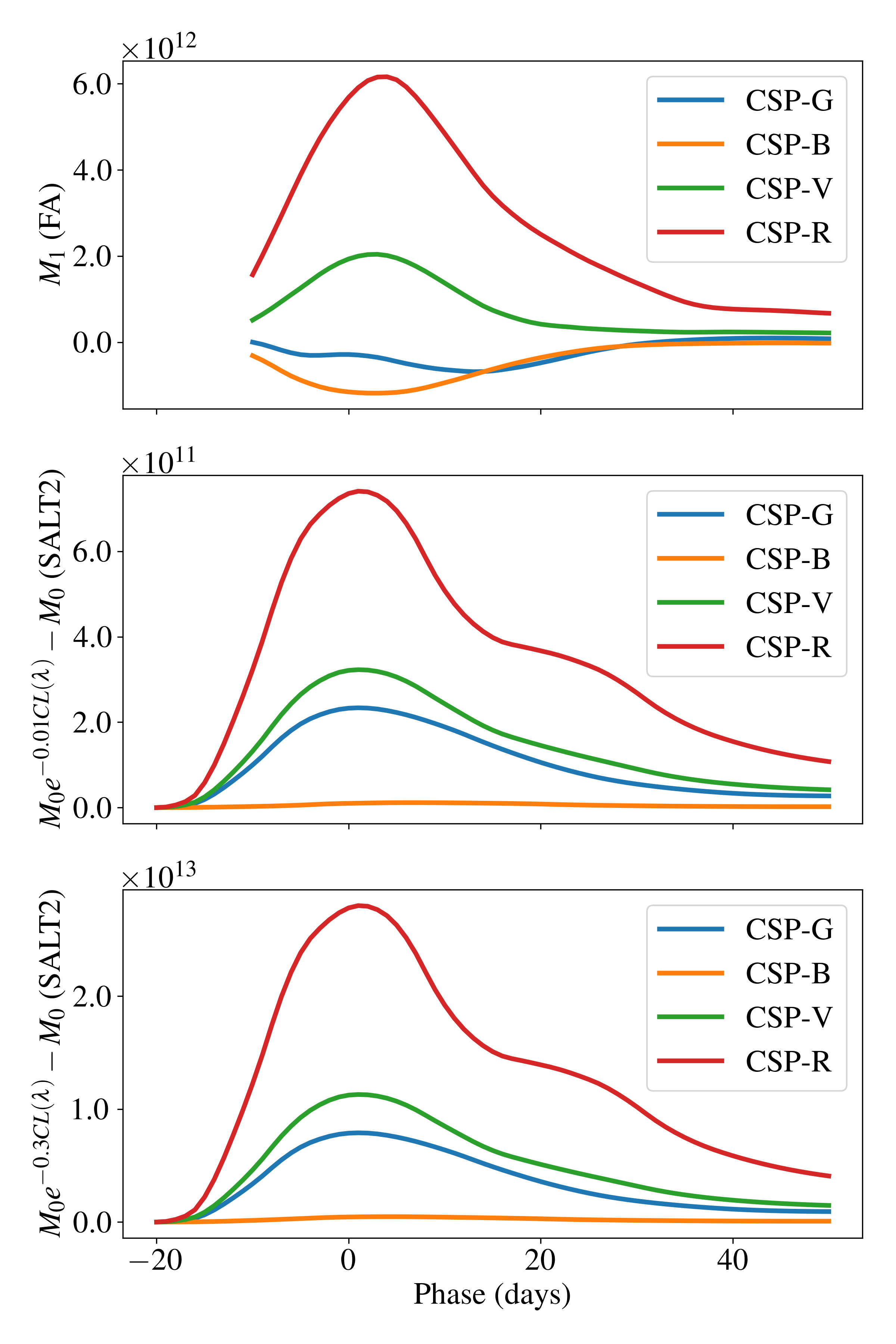} 
\end{minipage}
\caption{Comparison of Factor Analysis $M_1$ component and SALT2 $M_{0,SALT2} e^{-cCL(\lambda)} - M_{0,SALT2}$ color component.}
\label{fig:comparing_M0}
\end{figure}

In Fig.~\ref{fig:comparing_M0} we compare the first common factor obtained by our trained 2-component Factor Analysis with $M_{0,SALT2} e^{-cCL(\lambda)} - M_{0,SALT2}$, which describes the color index variations in the context of the SALT2 light-curve fitter. In the top panels, we have our FA $M_1$ component for different wavelengths, phases, and broad-bands. In the middle and bottom panels, we have the SALT2 color correction component for $c=0.01$ and $c=0.3$, respectively. 

We identify similarities between FA $M_1$ and $M_{0,SALT2} e^{-cCL(\lambda)} - M_{0,SALT2}$, especially in the color-phase section, despite differing in relative amplitudes. The SALT2 sections of constant phase are smoother for lower values of $c$. A similar effect is achieved for FA $M_1$ by tuning the $x_1$ fit parameter. Regarding the broad-band curves, FA $M_1$ allows for negative values, has a sharper peak than the SALT2 counterpart and shows a more rapid decline with phase.  

\begin{figure}
\centering
\includegraphics[width=0.55\columnwidth]{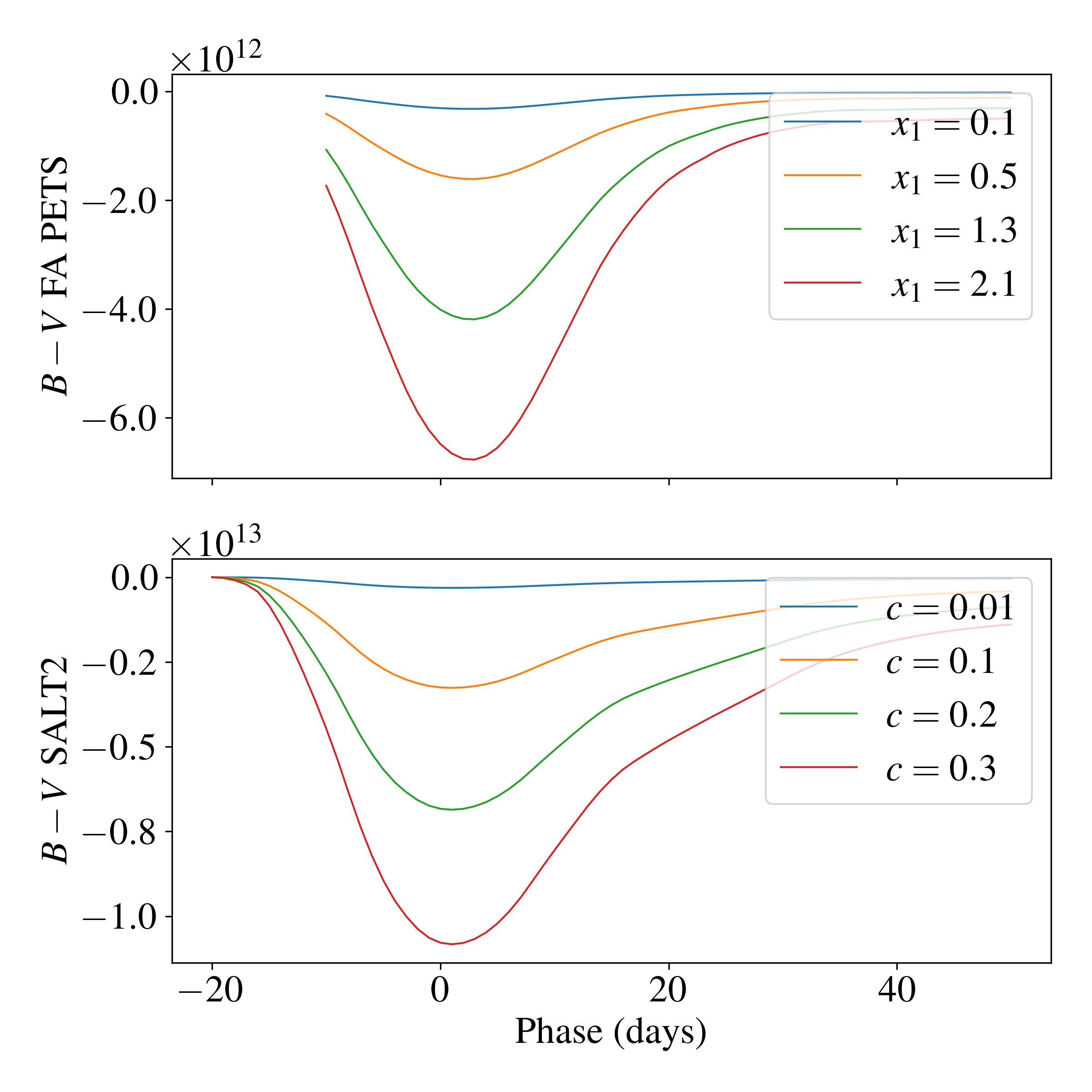} 
\caption{Time evolution of the $B-V$ color for FA-PETS (top) and SALT2 (bottom). Here, the color is obtained from FA $x_1 M_1$ and SALT2 $M_{0,SALT2} e^{-cCL(\lambda)} - M_{0,SALT2}$. The curves correspond to different values of FA $x_1$ and SALT2 $c$.}
\label{fig:comparing_M0_broadband}
\end{figure}

We show in Fig.~\ref{fig:comparing_M0_broadband} the time evolution of the $B-V$ color obtained from FA $x_1M_1$ and SALT2 $M_{0,SALT2} e^{-cCL(\lambda)} - M_{0,SALT2}$ surfaces. We can see that the color information is distributed in a different way among the components for each method and that the FA-PETS $B-V$ color curve is sharper and approaches zero faster. We recall that the color information is expected to be distributed among all PETS components.

\begin{figure}
\begin{minipage}[b]{.5\textwidth}
\centering
\includegraphics[width=\columnwidth]{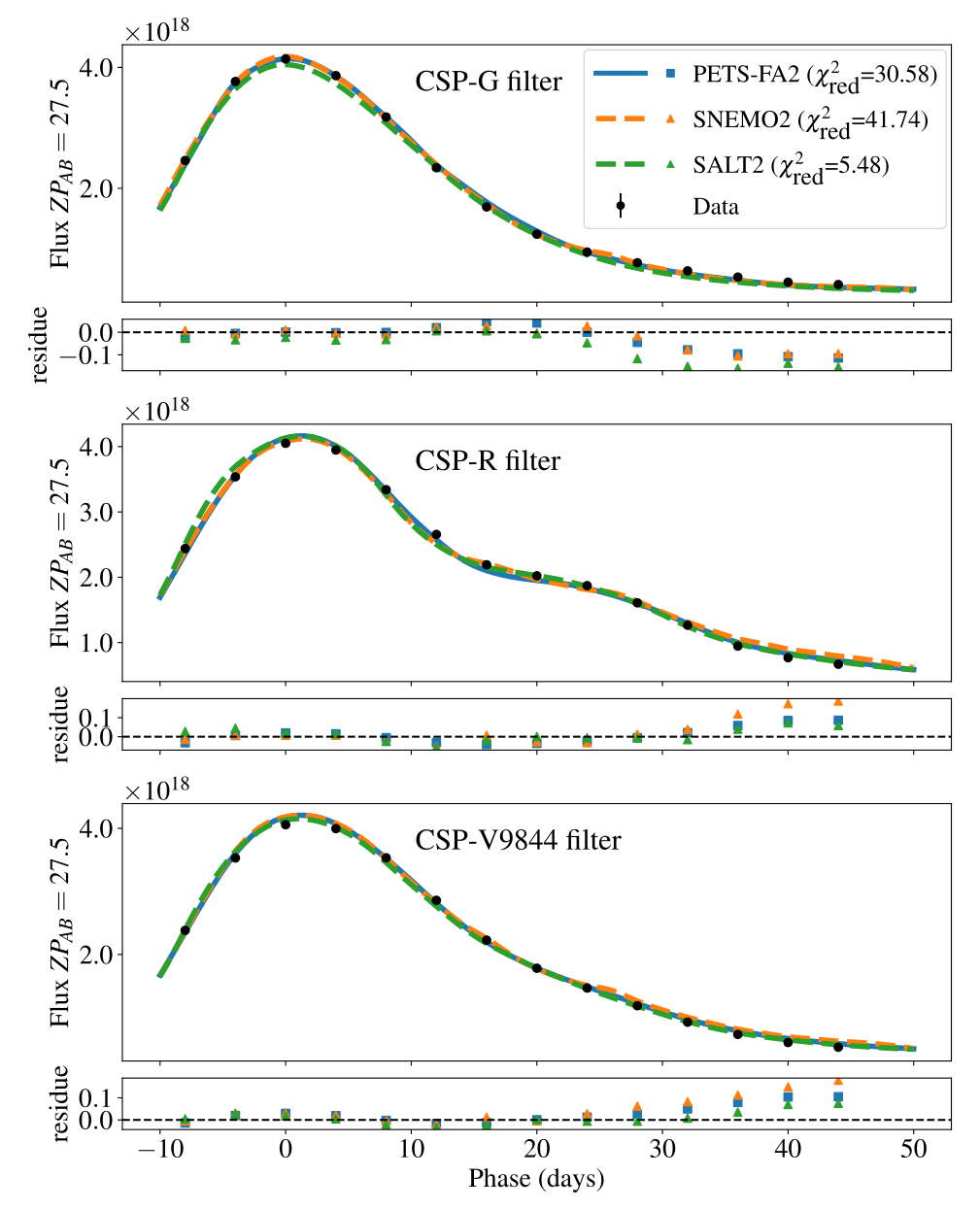} 
\end{minipage}
\begin{minipage}[b]{.5\textwidth}
\centering
\includegraphics[width=\columnwidth]{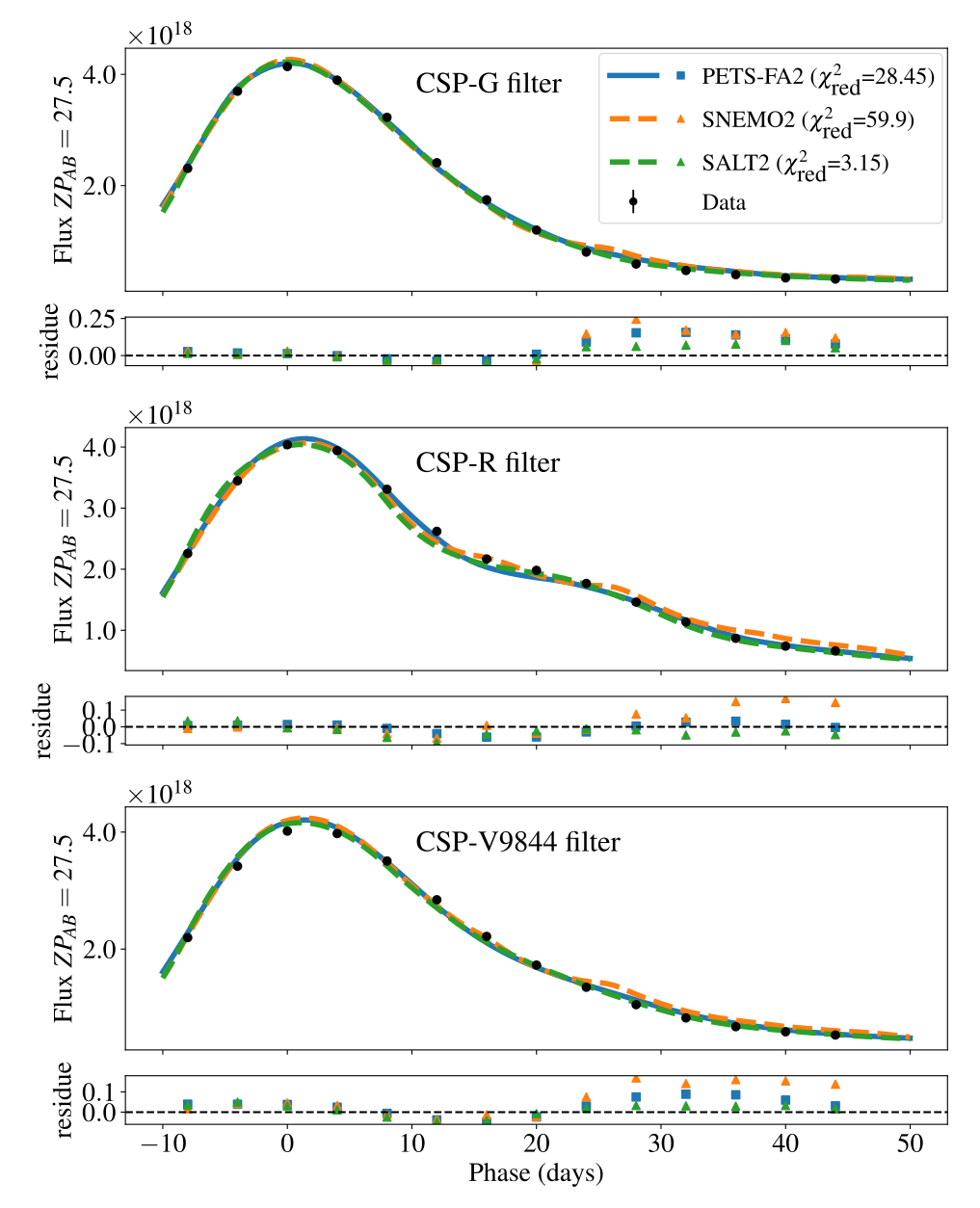} 
\end{minipage}
\caption{Representative PETS-FA, SNEMO2 and SALT2 fits of "Test\_SN8" and "Train\_SN84" synthetic photometry, respectively. "Test\_SN8" was part of the validation/test set for both FA2 and SNEMO2, while "Train\_SN84" was part of the training set for both models.}
\label{fig:fitsnemo}
\end{figure}

\begin{figure}
\begin{minipage}[c]{.3\textwidth}
\centering
\includegraphics[width=\columnwidth]{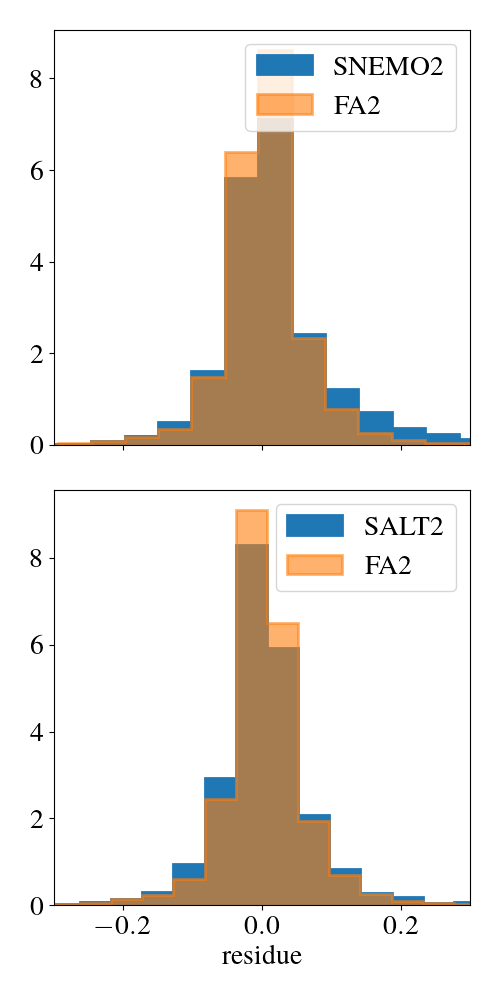} 
\end{minipage}
\hfill
\begin{minipage}[c]{.65\textwidth}
\centering
\includegraphics[width=\columnwidth]{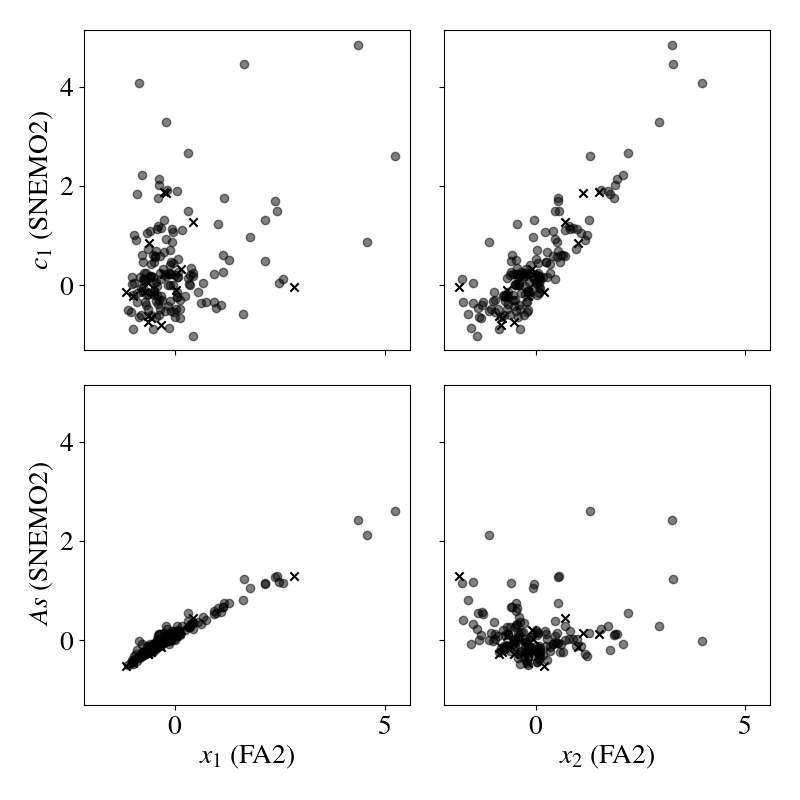} 
\end{minipage}
\caption{On the left are the relative residues for synthetic photometry light curve fitting. On the right are the correlations between SNEMO2 and PETS-FA model parameters for test ("o") and validation ("x") sets.}
\label{fig:corrsnemo}
\end{figure}

In Fig.~\ref{fig:fitsnemo} we have the comparison of light curve fitting for PETS-FA, SNEMO2, and SALT2 for a test set and a validation set SN. The three models are equally good in fitting this set, as supported by the residues shown in Fig.~\ref{fig:corrsnemo} (in this case, our residues are the relative discrepancy since the uncertainties are underestimated). 

A correlation plot of FA-PETS and SNEMO2 fit parameters is portrayed in Fig.~\ref{fig:corrsnemo}. It shows our $x_1$ parameter strongly correlates with SNEMO $As$ which controls the color index variations in their model. Thus, most color information is contained in our FA $M_1$.

\section{Initial cosmology inference with PETS}
\label{sec:cosmo_paramters_constraints}

As our final objective is to apply PETS fit parameters to correct SNIa magnitude and perform cosmology inference, here we evaluate its performance in this scenario in contrast to SALT2. This initial evaluation offers insights into the model potential as a competitive light-curve fitter.

To perform the cosmological analysis, we chose the Pantheon sample from \citet{Scolnic2018}. This sample consists of 1048 spectroscopically confirmed SNe Ia. In the low redshift region, it includes CfA1-CfA4, \citet{
riess1999bvri,jha2006ubvri,hicken2009cfa3,hicken2009improved,hicken2012cfa4} and CSP, \citet{folatelli2009carnegie,contreras2010carnegie,stritzinger2011carnegie}. Populating the intermediate redshift region we have SDDS, \citet{frieman2007sloan,Kessler2009,Sako_2018}, SNLS \citet{conley2010supernova,sullivan2011snls3} and PS1 \citet{rest2014cosmological,scolnic2014systematic}. And in high-z region we have data from HST, \citet{suzuki2012hubble,riess2004type,riess2007new,graur2014type,rodney2014type,riess2018new}. This sample is already cross-calibrated and the light curves were retrieved from \texttt{SNANA}, \citet{kessler2009snana}.

The spectral time series model was constructed in \texttt{SNCosmo} using as input the training SEDs average and the first two components returned by either our Principal Component Analysis or Factor Analysis. We restricted our fitting to the region -10 days to 50 days and 3310\AA\, to 8580\AA\,. We also corrected for extinction by dust in the Milky Way given the color excess, $E(B-V)_{MW}$, following \citet{Fitzpatrick1999}. The fitting process returns for each supernova a value of $t_0$ (date of maximum light in B-band), $x_0$, $x_1$, and $x_2$, with corresponding uncertainties. 

\subsection{Fitting results for PCA-PETS}

From the 1048 SNe in the initial sample, 1023 had a successful fit according to \texttt{SNCosmo}. Since our model does not have a model covariance and our UV cutoff limits the band passes used for fitting high redshift Ia supernovas, we decided to apply quality cuts over the fit parameters based on the distributions percentiles. These quality cuts will reduce the effect of poorly fitted objects on the cosmological analysis. 

We verified that when there are few or no data points before or after the maximum light the model struggles to constrain the date of the maximum and, consequently, in obtaining accurate values for $x_1$ and $x_2$, resulting in bad-quality fits. We encountered additional challenges when trying to determine the date of maximum for the low redshift sample. To address this issue, we used a flat prior $\mathcal{U}(-5, 5)$ around the date of maximum reported in the SNANA files. For the remaining objects we conducted fits for $t_0$, $x_0$, $x_1$, and $x_2$ without using any prior knowledge.

\begin{figure}
\includegraphics[width=\columnwidth]{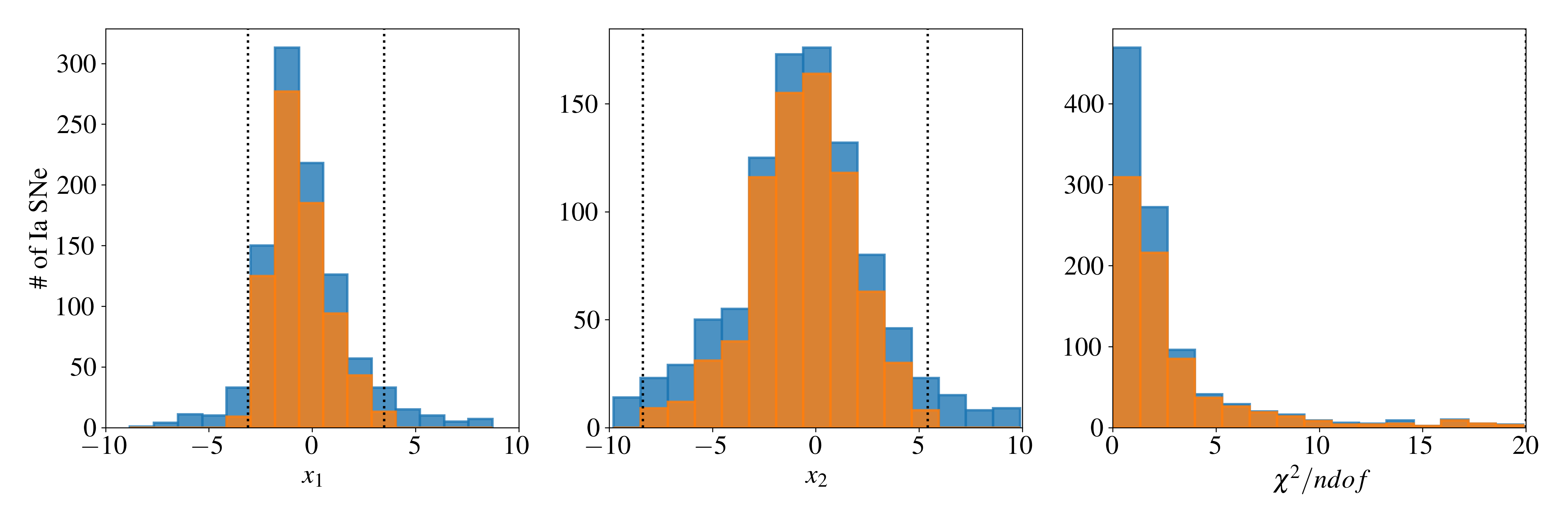} 
\caption{PCA-PETS fit parameter distributions: in blue before the cuts and in orange after the cuts. The dashed line indicates the applied cuts.}
\label{fig:PCA_param_dist}
\end{figure}

The cuts proposed for the PCA-PETS parameters are shown in Table~\ref{tab:cutsexp}, where we can also see how many supernovae are accepted in each step. We first eliminate the outliers by restricting the $x_1$ and $x_2$ distributions to the range $[-10,10]$. Next, in order to provide a sample independent quality cut, we find the 5th and 95th percentiles of $x_1$ and $x_2$ distributions and for the corresponding uncertainties $\sigma_{x_1}$ and $\sigma_{x_2}$, we find the 10th and 90th percentiles of the distributions. These values will define the cuts, portrayed by the dotted lines in Fig.~\ref{fig:PCA_param_dist}. Cuts on the chi-squared are more flexible, fits with $\chi^2/\textnormal{ndof} \leq 20$ are kept for the cosmological analysis. Fig.~\ref{fig:PCA_param_dist} shows in blue the original sample and in orange the reduced sample, after cuts. Most of the eliminated supernovas do not fail in all three cuts, many fits with higher values of $x_1$ and/or $x_2$ have low $\chi^2/\textnormal{ndof}$, leaving us with a reduced sample of 746 SNIa.

\begin{figure}
\centering
\includegraphics[width=0.5\columnwidth]{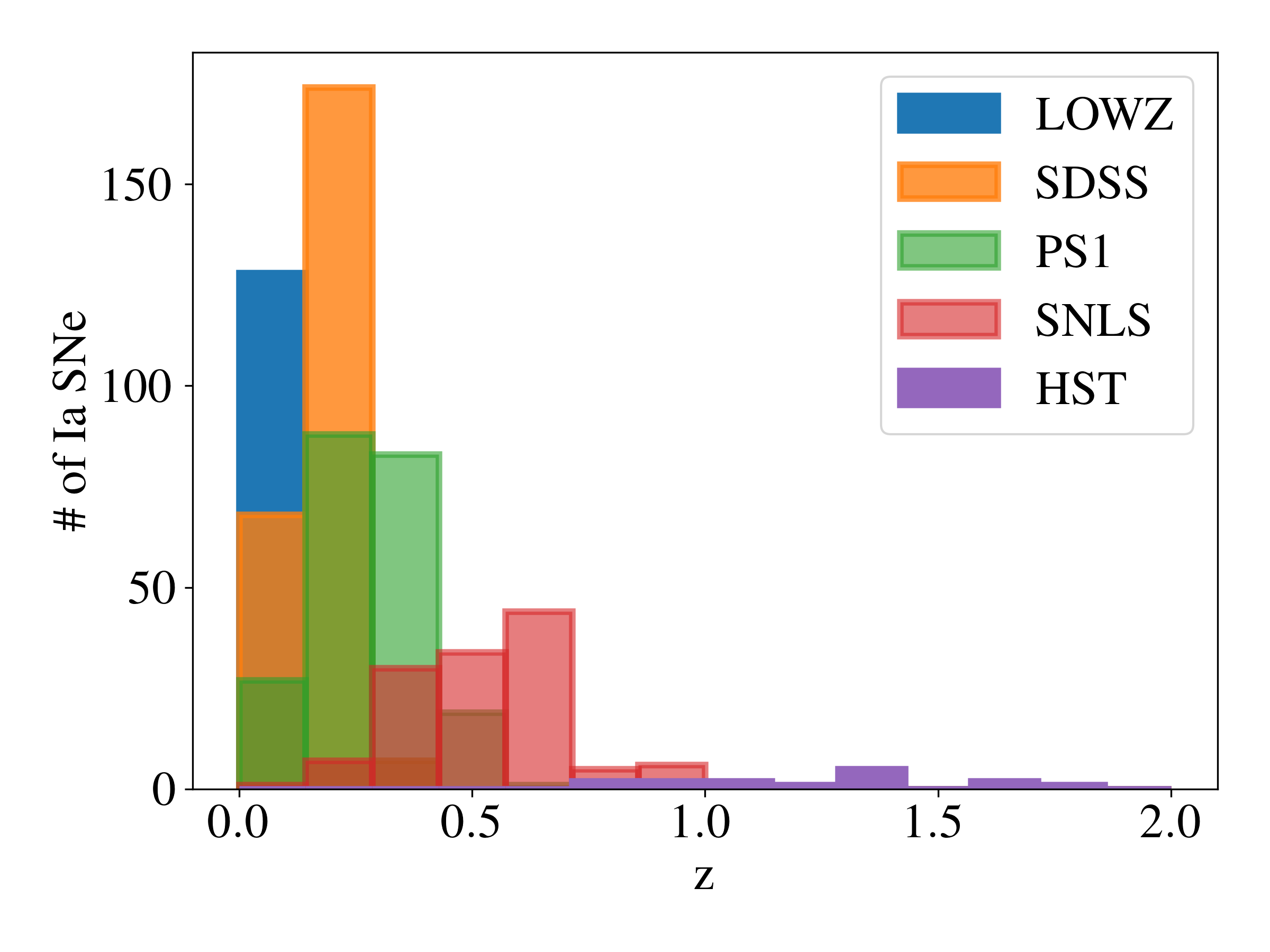} 
\caption{Redshift distribution of the supernovae sample used in the cosmological analysis with PCA-PETS.}
\label{fig:redshiftdist_pca}
\end{figure}

\begin{figure}
\begin{minipage}[b]{.5\textwidth}
\centering
\includegraphics[width=\columnwidth]{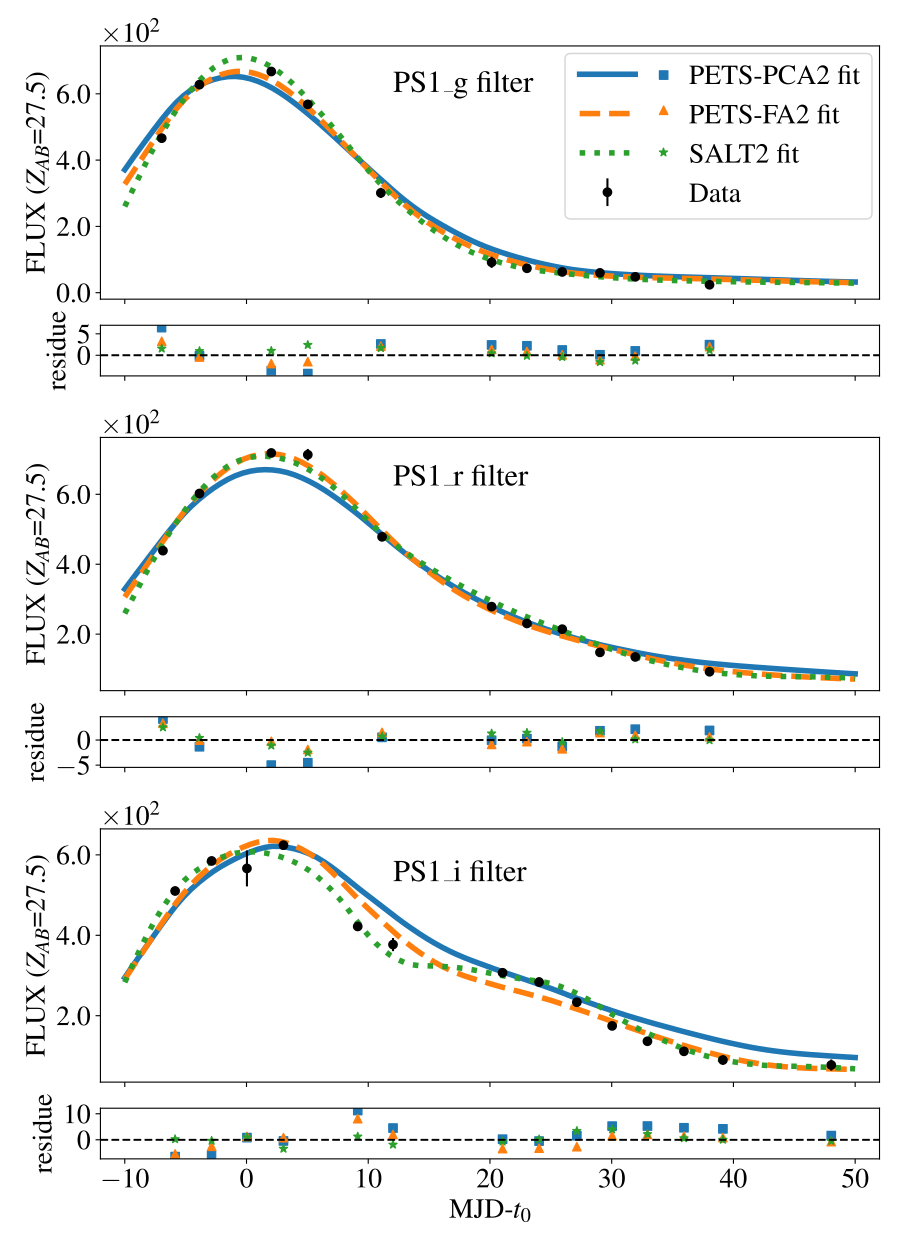} 
\end{minipage}
\hfill
\begin{minipage}[b]{.5\textwidth}
\centering
\includegraphics[width=\columnwidth]{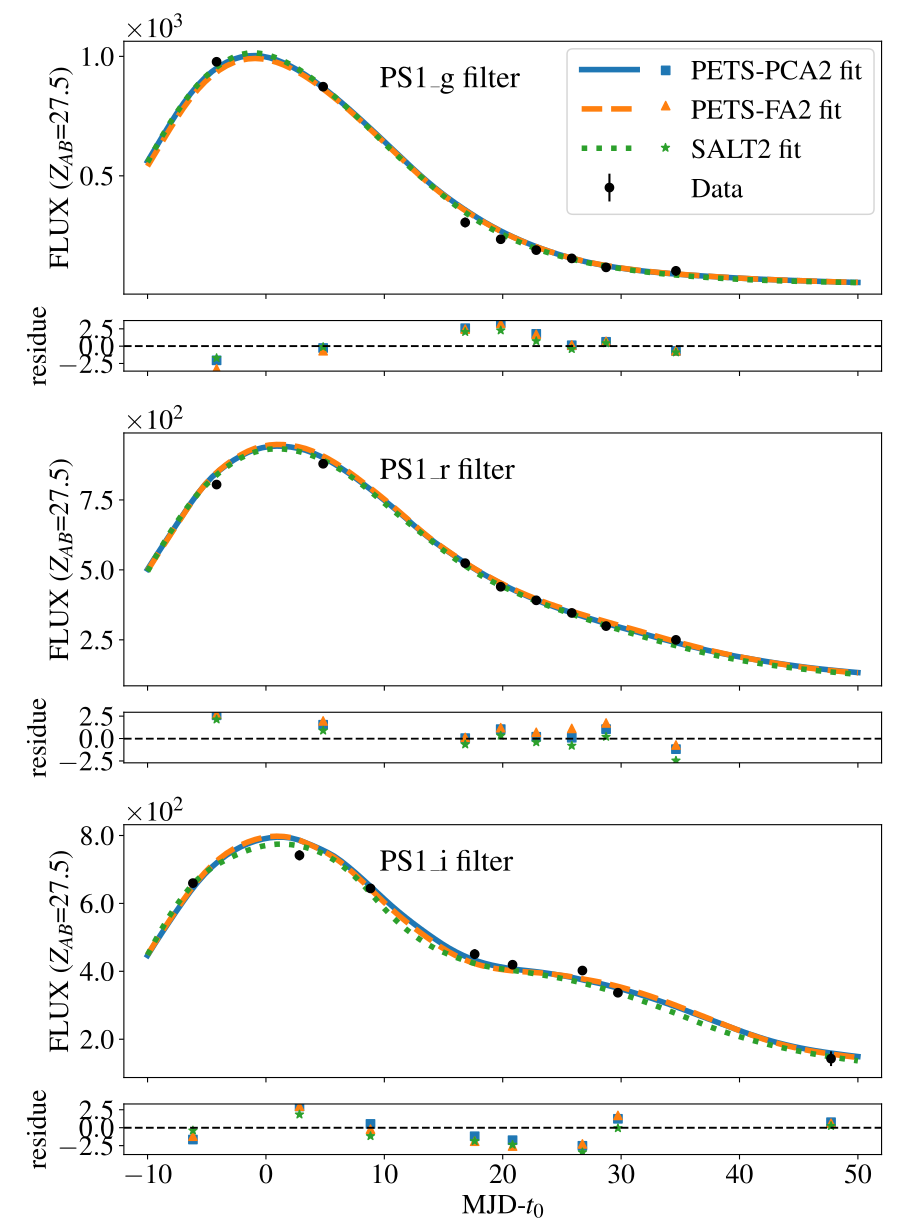} 
\end{minipage}
\caption{Comparison of SALT2 fits and PETS fits for SN 91 ($z=0.1526$) and SN 120143 ($z=0.1735$) from Pan-STARRS1 Surveys (PS1). Here the residue is defined as the difference from the curve to the original data divided by the data uncertainty.}
\label{fig:comparisonfits}
\end{figure}

\begin{table}
	\centering
	\caption{Cuts applied to the light curve fit parameters for Pantheon sample with PCA-PETS.}
	\label{tab:cutsexp}
	\begin{tabular}{lcccccc}
		\hline
		Cuts & \# of pass cut SNe \\
		\hline
		fit & 1023 \\
		$x_1$ 95th percentile & 893 \\
		$x_2$ 95th percentile & 900 \\
		$\chi^2/\textnormal{ndof}\leq 20$ & 973 \\
		$\sigma_{x_1}$ 90th percentile & 905 \\
		$\sigma_{x_2}$ 90th percentile & 910 \\
		\hline
         Total \# pass cut SNe & 746
	\end{tabular}
\end{table}
Table~\ref{tab:cutsexp_survey} shows the number of SNe passing the quality cuts per survey and in Fig.~\ref{fig:redshiftdist_pca} we have the histograms of redshift distributions for each survey. This same subsample is selected from SALT2 fits which were also performed over the same light curves using \texttt{SNCosmo} built-in "salt2 v. 2.4". In Fig.~\ref{fig:comparisonfits} we see a comparison of PCA-PETS and SALT2 fits for an arbitrary SNIa from Pan-STARRS1 survey. The main limitation for PCA-PETS is again verified to be the estimation of the flux at $p=0$ simultaneously for different filters.

\begin{table}
	\centering
	\caption{Accepted SNe per survey when applying the cuts for Pantheon subsample using PCA-PETS.}
	\label{tab:cutsexp_survey}
	\begin{tabular}{lcccccc} 
		\hline
		Survey & Total \# of SNe & \# of pass cut SNe \\
		\hline
		LOWZ & 172 & 136\\
		SDSS & 315 & 249 \\
		SNLS & 233 & 127 \\
		PS1 & 273 & 218\\
		HST & 26 & 16 \\
		\hline
	\end{tabular}
\end{table}

\begin{figure}
\centering
\includegraphics[width=0.6\columnwidth]{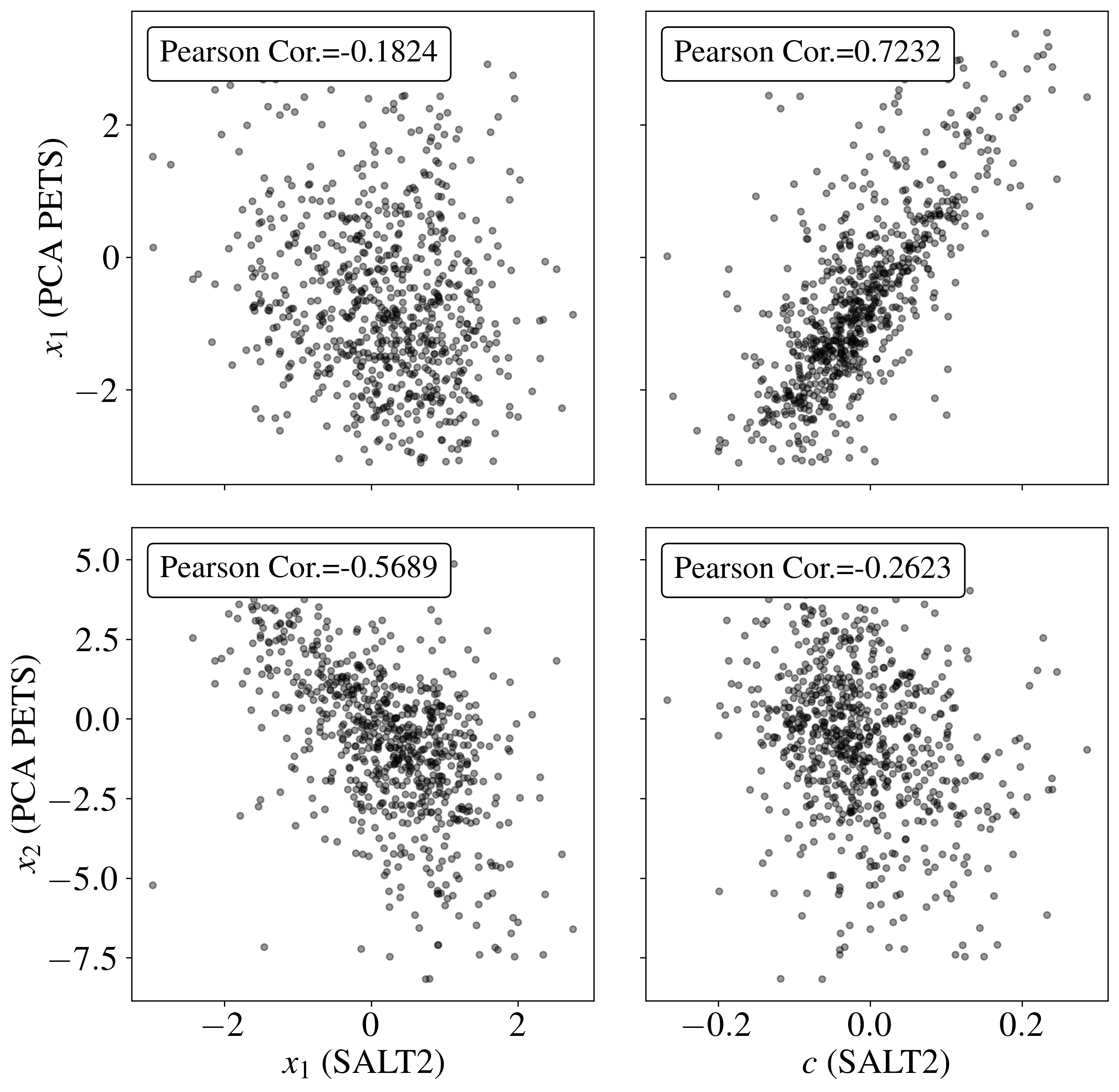} 
\caption{Scatter plots of SALT2 parameters and PCA-PETS parameters with the corresponding Pearson correlation coefficients for each pair.}
\label{fig:corr_salt_pca}
\end{figure}

In Fig.~\ref{fig:corr_salt_pca} we have a scatter plot for SALT2 parameters versus PCA-PETS parameters. To address their relationships we calculate the Pearson Correlation Coefficient. This coefficient measures the linear correlation between a pair of parameters. Values closer to unity indicate the measurement of one feature provides a good estimate of the second. On the other hand, values closer to zero indicate that when measuring one feature no information about the other is gained, regarding linear relationships. We see a positive correlation between the pairs ($c$ (SALT2), $x_1$ (PCA-PETS)). And negative correlations between the pairs ($x_1$ (SALT2), $x_1$ (PCA-PETS)), ($x_1$ (SALT2), $x_2$ (PCA-PETS)) and ($c$ (SALT2), $x_2$ (PCA-PETS)). The stronger the correlation, the stronger the indication of a similar origin for those features, i.e. the mechanism that drives $c$ (SALT2) for higher or lower values also affects $x_1$ (PETS) in the same direction. 

From the correlation coefficients shown in Fig.~\ref{fig:corr_salt_pca} we observe that for our PCA-PETS model, the most important correction is associated with a color-like phase-dependent component. And as expected $m_{B,\textnormal{SALT2}}^*$ and $m_{B,\textnormal{PCA-PETS}}^*$ are strongly correlated, with a correlation coefficient equal to 0.9992. The color index variations are mainly contained in our first surface $M_1$.

We can also analyze if the PCA-PETS parameters show redshift dependence. The $x_1$ distribution favors negative values but shows no strong apparent dependence with redshift. Regarding the $x_2$ distribution, for intermediate and high redshift there is no strong apparent dependence with redshift. For the low-z sample, we observe a deviation for higher values of $x_2$. However, this apparent dependence is due to the lower redshift region being populated mainly by one sample which favors host galaxies masses, measured through $\log_{10}(M_{stellar}/M_{\odot})$, higher than 10, hence reflecting a selection effect. (See Fig.~\ref{fig:param_dep_mb_pca} and Fig.~\ref{fig:hg_mass_dep_pca} in Appendix \ref{ap:pcacosm_cont})

\subsection{Fitting results for FA-PETS}
\label{sec:fittingresults_fa}

\begin{figure}
\includegraphics[width=\columnwidth]{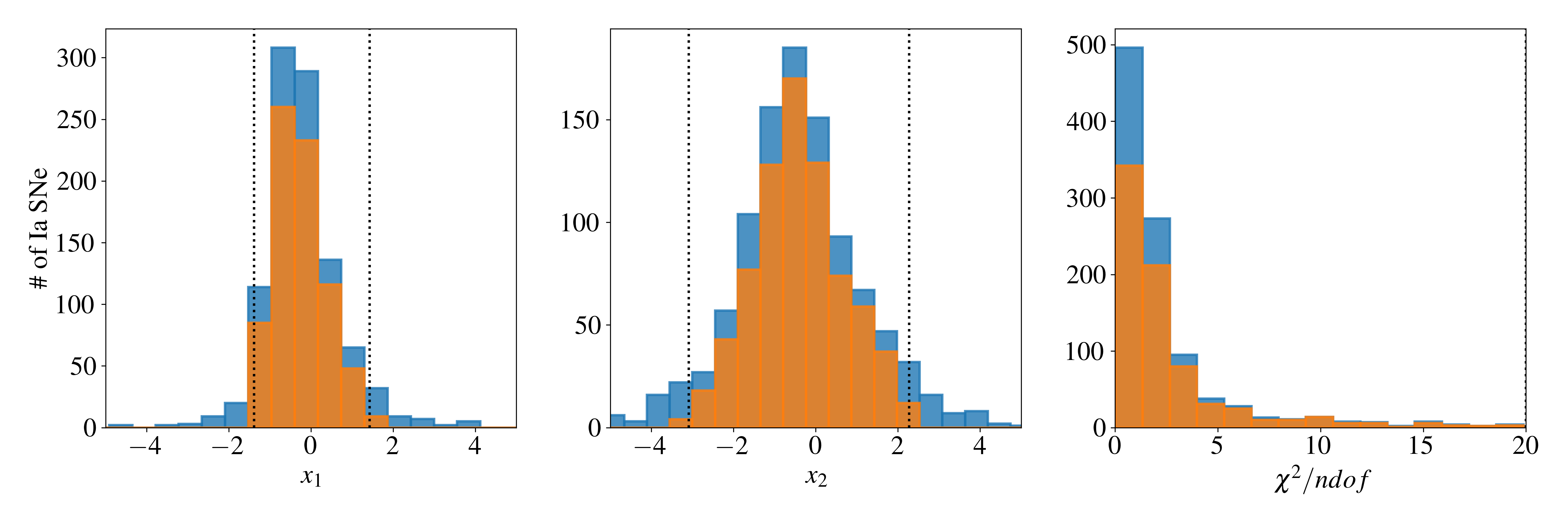} 
\caption{FA-PETS fit parameter distributions: in blue before the cuts and in orange after the cuts. The dashed line indicates the applied cuts.}
\label{fig:FA_param_dist}
\end{figure}

\begin{figure}
\centering
\includegraphics[width=0.5\columnwidth]{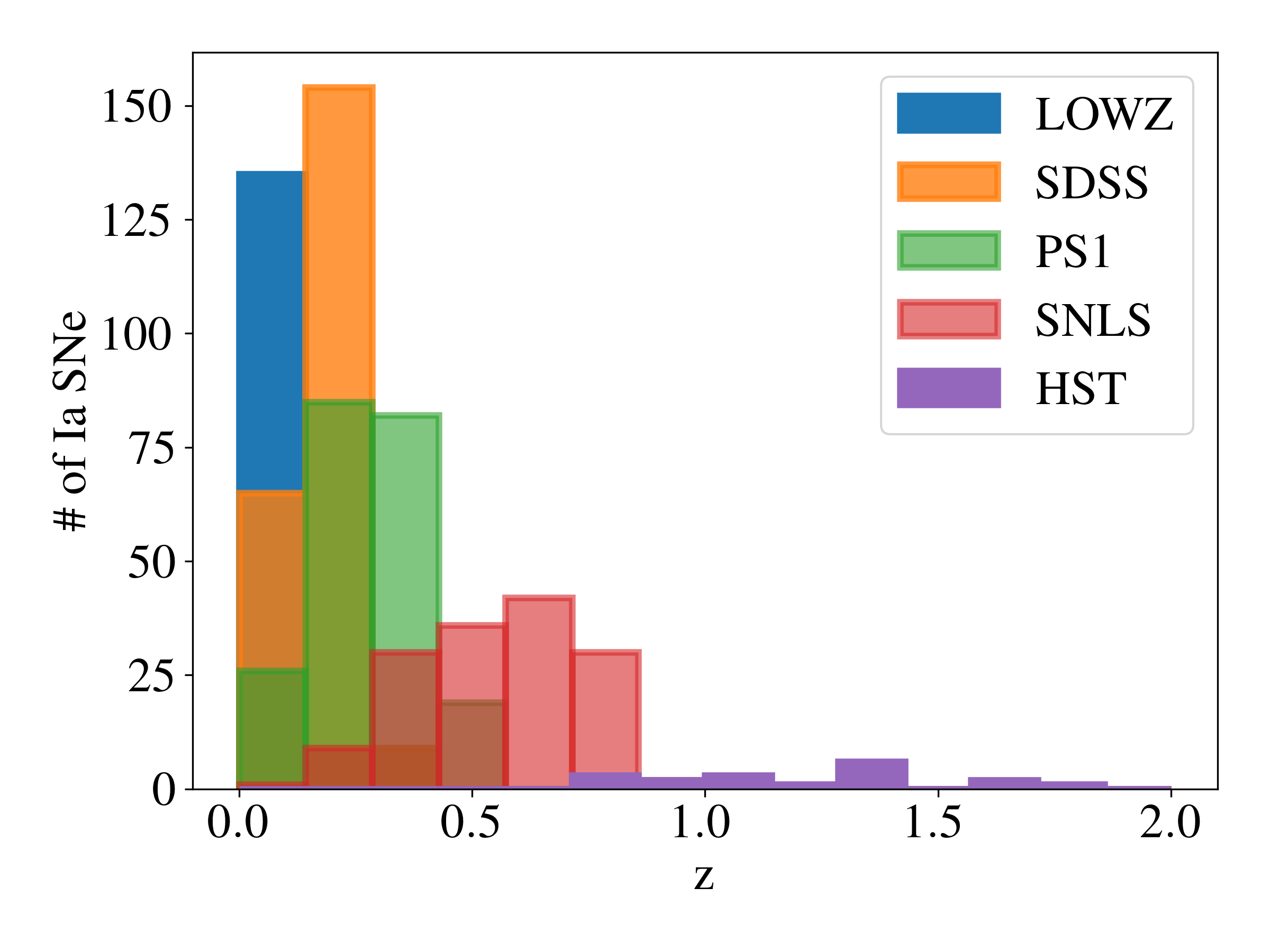} 
\caption{Redshift distribution of the supernovae sample used in the cosmological analysis with FA-PETS.}
\label{fig:redshiftdist_fa}
\end{figure}

We now analyze the fitting results when equation~(\ref{eq:restframeflux}) $M_i(p,\lambda)$ surfaces are the average SN SED and the first two FA common factors. Table~\ref{tab:cutsexpfa} shows the applied cuts over the Pantheon sample after light curve fitting along with the number of supernovas passing each cut. The same criteria defining PCA-PETS quality cuts were applied here. The fitting configuration was also maintained. Fig.~\ref{fig:FA_param_dist} shows in blue the original sample and in orange the reduced sample of 751 SNIa. Fig.~\ref{fig:redshiftdist_fa} shows the redshift distribution of FA-PETS SNIa cosmological sample.  

\begin{table}
	\centering
	\caption{Cuts applied to the light curve fit parameters for Pantheon sample with FA-PETS.}
	\label{tab:cutsexpfa}
	\begin{tabular}{lcccccc} 
		\hline
		Cuts & \# of pass cut SNe \\
		\hline
		fit & 1019 \\
		$x_1$ 95th percentile & 910 \\
		$x_2$ 95th percentile & 983 \\
		$\chi^2/\textnormal{ndof}\leq 20$ & 986 \\
		$\sigma_{x_1}$ 90th percentile & 908 \\
		$\sigma_{x_2}$ 90th percentile & 910 \\
		\hline
        Total \# of pass cut SNe & 751
	\end{tabular}
\end{table}

\begin{table}
	\centering
	\caption{Accepted SNe per survey when applying the cuts for Pantheon subsample using FA-PETS.}
	\label{tab:cutsexp_survey_fa}
	\begin{tabular}{lcccccc}
		\hline
		Survey & Total \# of SNe & \# of pass cut SNe \\
		\hline
		LOWZ & 172 & 143\\
		SDSS & 315 & 228 \\
		SNLS & 233 & 150 \\
		PS1 & 273 & 212 \\
		HST & 26 & 18 \\
		\hline
	\end{tabular}
\end{table}

A comparison of FA-PETS and SALT2 fits for an arbitrary SNIa from Pan-STARRS1 survey can be seen in Fig.~\ref{fig:comparisonfits}. We verify FA-PETS also outperforms PCA-PETS in light-curve fitting. 

\begin{figure}
\centering
\includegraphics[width=0.6\columnwidth]{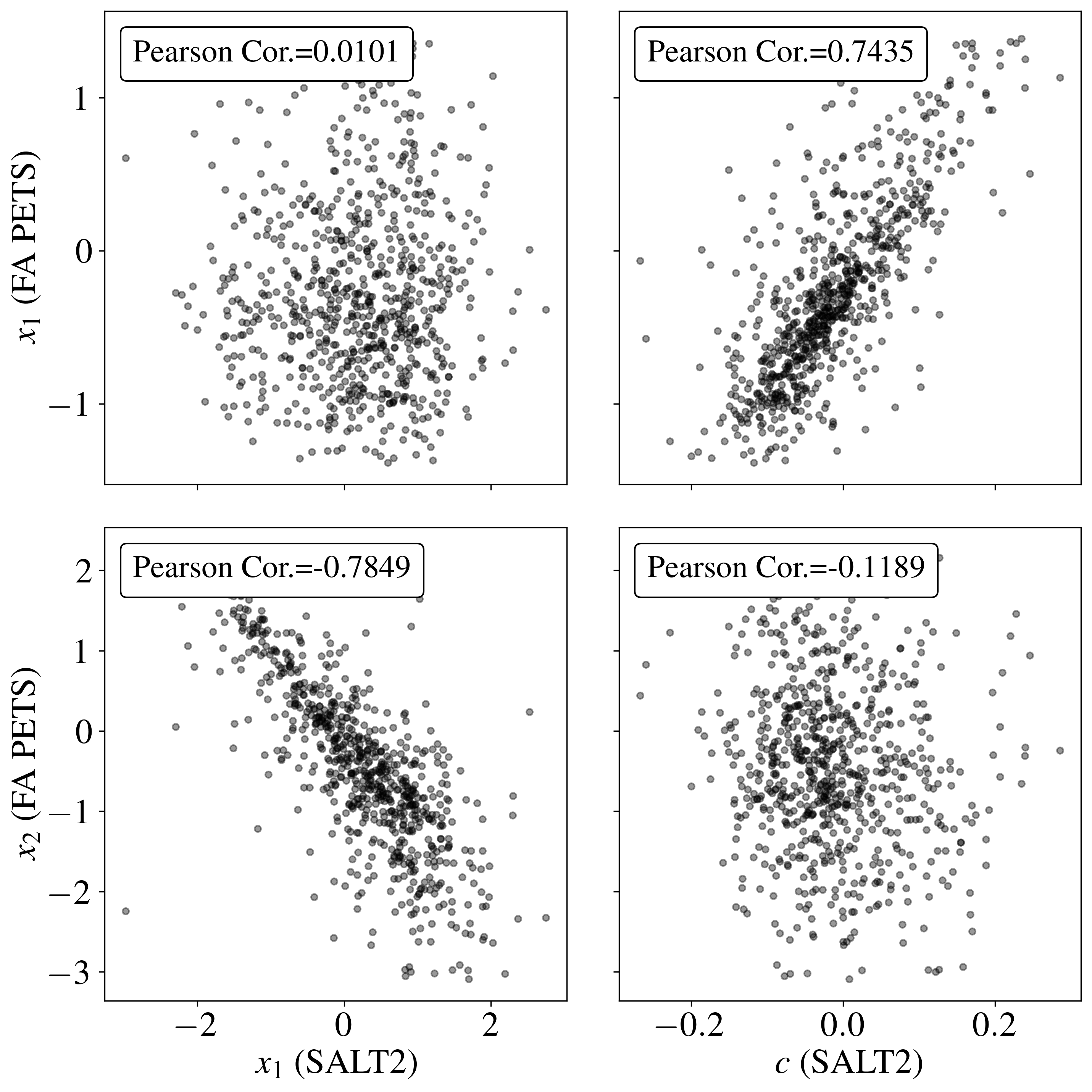} 
\caption{Scatter plots of SALT2 parameters and FA-PETS parameters with the corresponding Pearson correlation coefficients for each pair.}
\label{fig:corr_salt_fa}
\end{figure}

We can also compare the scatter plots for FA-PETS and SALT2 pairs of parameters. Fig.~\ref{fig:corr_salt_fa} reveals greater Pearson correlation coefficients between the two models than seen for the PCA approach. The greatest correlations are seen for the pairs ($x_2$ (FA-PETS), $x_1$ (SALT2)) and ($x_1$ (FA-PETS), $c$ (SALT2)), which also imply that for FA-PETS the first-order correction is mainly due to color index variations instead of light curve shape variations. And as expected $m_{B,\textnormal{SALT2}}^*$ and $m_{B,\textnormal{FA-PETS}}^*$ are also strongly correlated, with a correlation coefficient equal to 0.9994.

Following the analysis performed over the PCA fitting results, we see no apparent correlation with redshift apart from $x_2$ tendency to positive values at low redshifts, which is also a consequence of the host galaxy masses distribution of the LOWZ sample. Both $x_1$ and $x_2$ parameters show weak correlations with host galaxy mass when dealing with PCA components while only $x_2$ shows this same weak correlation when considering FA components. These results reinforce that PCA-PETS and FA-PETS act on different aspects of SNeIa variability. (See Fig.~\ref{fig:param_dep_mb_fa} and Fig.~\ref{fig:hg_mass_dep_fa} in Appendix~\ref{ap:facosm_cont}).

\section{Cosmology results for PETS}
\label{sec:cosmoresults}
The theoretical distance modulus is obtained from the luminosity distance in units of Hubble distance today through
\begin{equation}
    \mu_{th}(z;\boldsymbol{\theta},h) = 5\log_{10}(\mathcal{D}_L(z;\boldsymbol{\theta}))+\mu_0(h),
    \label{eq:mutheo}
\end{equation}
with $\mu_0(h)=5\log_{10}(1000c/(\textnormal{km}/\textnormal{s}))-5\log_{10}h$. Here $h$ is a dimensionless quantity parametrizing the Hubble constant, $H_0=100 h$ km s$^{-1}$ Mpc$^{-1}$ and $\boldsymbol{\theta}$ stands for the remaining cosmological parameters for a specific cosmological model. The luminosity distance in units of Hubble distance today is defined as
\begin{equation}
\mathcal{D}_L(z;\boldsymbol{\theta}) :=
\begin{cases}
    \displaystyle\frac{1+z}{\sqrt{\Omega_{k0}}}\sinh \left( \sqrt{\Omega_{k0}}\int_{0}^{z}\frac{dz'}{E(z'; \boldsymbol{\theta})} \right)& \textnormal{if}\; \Omega_{k0} > 0,\\  
    (1+z) \displaystyle\int_{0}^{z'} \frac{dz'}{E(z'; \boldsymbol{\theta})} & \textnormal{if}\; \Omega_{k0} = 0,\\
    \displaystyle\frac{1+z}{\sqrt{-\Omega_{k0}}}\sin \left( \sqrt{-\Omega_{k0}}\int_{0}^{z} \frac{dz'}{E(z'; \boldsymbol{\theta}) } \right)&  \textnormal{if}\; \Omega_{k0} < 0,\\
\end{cases}
\label{eq:luminositydistance}
\end{equation}
where $E(z;\boldsymbol{\theta})$ is the Hubble parameter in units of the Hubble constant and $\Omega_{k0}$ is
the dimensionless curvature parameter measured today.

The cosmological model considered is $\Lambda$CDM, assuming a universe with curvature, cold dark matter, and cosmological constant, which
give us the following dimensionless Hubble parameter
\begin{equation}
    E_{\Lambda\textnormal{CDM}}(z;\boldsymbol{\theta}) = \sqrt{\Omega_{m0} (1+z)^3 + \Omega_{\Lambda 0} + \Omega_{k_0}(1+z)^2}.
\end{equation}
Recalling that $\Omega_{k_0} = 1-\Omega_{m_0}-\Omega_{\Lambda_0}$. Here $\boldsymbol{\theta} = (\Omega_{m0}, \Omega_{\Lambda 0})$, representing the cosmological parameters we are interested in constraining and $\Omega_{r0}$, the dimensionless radiation density parameter measured today, is neglected since it is about 5 orders of magnitudes smaller than the cosmological parameters $\boldsymbol{\theta}$.

\subsection{SALT2 distance modulus}
The SALT2 empirical distance modulus, that carries information from the derived light-curve fitting parameters, is given by
\begin{equation}
    \mu_{\textrm{SALT2}}=m_B^{corr}-M_B=m_B^*+\alpha x_1 -\beta c-M_B,
    \label{eq:distmodsalt2}
\end{equation}
where $m_B^*$ is the rest-frame B-band magnitude at maximum light, and according to \citet{Mosher2014} this quantity can be obtained from the light curve parameter $x_0$ using $m_B^*=-2.5\log_{10} x_0 +10.635$. $M_B$ is the SNIa absolute magnitude, the parameters $x_1$ and $c$ corrects the supernova apparent magnitude to account for the observed dispersion. Lastly, $\alpha$ and $\beta$ are nuisance parameters that need to be included in the parameter inference.

The statistical uncertainty matrix for this distance modulus can be portrayed as  
\begin{equation}
    \sigma^2_{\mu_{\textrm{SALT2}}}=\sigma^2_{m_B^*}+\alpha^2 \sigma_{x_1}^2 +\beta^2 \sigma_c^2 +2\alpha\sigma_{m_B^*,x_1}-2\beta\sigma_{m_B^*,c}-2\alpha\beta\sigma_{x_1,c}+\sigma^2_{\mu,z},
    \label{eq:sig2salt2}
\end{equation}
where $\sigma^2_{m_B^*}$, $\sigma_{x_1}^2$, $\sigma_c^2$, $\sigma_{m_B^*,x_1}$, $\sigma_{m_B^*,c}$ and $\sigma_{x_1,c}$ are elements of the covariance matrix for the light curve parameters. The last term propagates to the distance modulus an error due to the redshift measurement and due to peculiar velocities contamination. Following \citet{Kessler2009}, we adopt the distance-redshift relation for an empty universe leading to
\begin{equation}
    \sigma_{\mu,z}=\sigma_z\left(\frac{5}{\log 10}\right)\frac{1+z}{z(1+z/2)},
\end{equation}
where $\sigma^2_z=\sigma^2_{\textrm{spec}}+\sigma^2_{\textrm{pec}}$. The term $\sigma^2_{\textrm{spec}}$ represents the measurement uncertainty and $\sigma^2_{\textrm{pec}}$ is the contribution due to peculiar velocity uncertainties, estimated as $0.0012$ by \citet{kessler2009snana}. 

\subsection{PETS distance modulus}

Based on the empirical observed correlations, our distance modulus description relies on adding linear corrections to $m_B^*$ that are proportional to PETS light curve fit parameters:
\begin{equation}
    \mu_{\textrm{PETS}}=m_B^{corr}-M_B=m_B^*+\alpha x_1 +\beta x_2-M_B,
    \label{eq:distmodpets}
\end{equation}
with the following statistical uncertainty matrix
\begin{equation}
    \sigma^2_{\mu_{\textrm{PETS}}}=\sigma^2_{m_B^*}+\alpha^2 \sigma_{x_1}^2 +\beta^2 \sigma_c^2 +2\alpha\sigma_{m_B^*,x_1}+2\beta\sigma_{m_B^*,x_2}+2\alpha\beta\sigma_{x_1,x_2}+\sigma^2_{\mu,z}.
    \label{eq:sig2pets}
\end{equation}
Where $\sigma^2_{m_B^*}$, $\sigma_{x_1}^2$, $\sigma_{x_2}^2$, $\sigma_{m_B^*,x_1}$, $\sigma_{m_B^*,x_2}$ and $\sigma_{x_1,x_2}$ are obtained from the light curve fitting.

We also consider a correction due to different host galaxies masses, if $M_{stellar}>10^{10} M_{\odot}$ we replace in the aforementioned equations $M_B$ for $M_B+\Delta_M$.
 
The parameter inference is performed through a Monte Carlo Markov Chain (MCMC) using the Python package \texttt{emcee}, \citet{emcee}. Only with SNIa data we can not simultaneously constrain $M_B$ and $h$, then we define $\mathcal{M}(M_B,h)=M_B+\mu_0(h)$. We can write the $\chi^2$ for PETS model, neglecting correlations between different supernovae, as
\begin{equation}
    \chi^2_{\textrm{PETS}}(\theta,\delta,\mathcal{M})=\sum^N_i \frac{[\mu_{i,\textrm{PETS}}(z_i,\delta,M_B)-\mu_{th}(z_i;\boldsymbol{\theta},h)]^2}{\sigma_{i,\textrm{PETS}}^2(\delta)+\sigma_{\textrm{int}}^2}.
\end{equation}
An analogous expression is defined for SALT2. The input log-likelihood is
\begin{equation}
    \log L = -0.5\left[\chi^2_{\textrm{PETS}}(\boldsymbol{\theta},\mathcal{M},\boldsymbol{\delta}, \sigma_{\textrm{int}})+\sum_i^N \log{[\sigma^2_{i,\textrm{PETS}}(\boldsymbol{\theta},\boldsymbol{\delta})+\sigma_{\textrm{int}}^2]}\right],
    \label{eq:loglike}
\end{equation}
where $\boldsymbol{\delta}=(\alpha,\beta)$ are the model nuisance parameters, $\boldsymbol{\theta} = (\Omega_{m0}, \Omega_{\Lambda 0})$ are the cosmological parameters. And $\sigma_{\textrm{int}}$, the intrinsic scatter, is a free parameter that will store any remaining unmodeled coherent variability. 

When performing the cosmological inference via MCMC, $\Omega_{m0}$, $\Omega_{\Lambda 0}$, $\alpha$, $\beta$, $\sigma_{\textnormal{int}}$ and $\mathcal{M}$ parameters are free to vary with flat uninformative priors.

\subsection{Results for PCA-PETS}
\label{sec:cosmo_results_pca}

\begin{figure}
\centering\includegraphics[width=0.5\columnwidth]{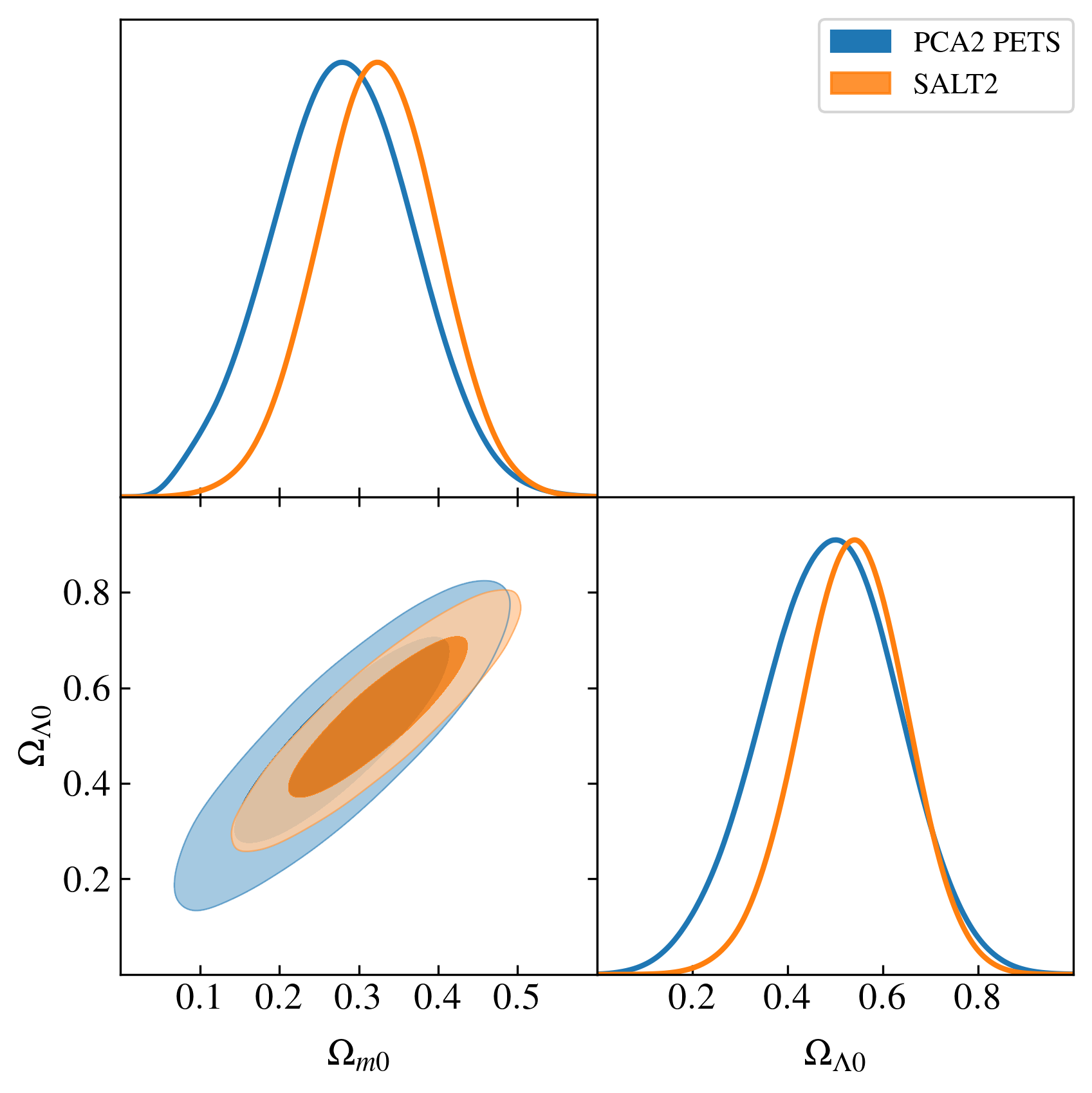} 
\caption{Comparison of marginalized cosmology results for SALT2 and PCA-PETS for $\Lambda$CDM model.}
\label{fig:triangleplots_pca}
\end{figure}

With light curve fitting results for PCA-PETS we perform a cosmological parameter inference using \texttt{emcee} sampling package \cite{emcee}, with the log-likelihood described in equation~(\ref{eq:loglike}) and flat prior over all parameters. The marginalized contours and one-dimensional marginalized distributions for the main cosmological parameters are shown in Fig.~\ref{fig:triangleplots_pca}. The constraining power of our PCA-PETS model is smaller than SALT2's but the marginalized one-dimensional distributions show good agreement inside the 68\% confidence region. 
Table~\ref{tab:lcdmtableresults} shows the MCMC results for $\Lambda$CDM model, here only including this subsample of SNIa as cosmological probes. 

\begin{table*}
	\centering
	\caption{Result of marginalized parameters for $\Lambda$CDM cosmological model for our PCA-PETS and FA-PETS models as well as for SALT2 without distance bias correction.\newline}
	\label{tab:lcdmtableresults}
	\resizebox{\columnwidth}{!}{\begin{tabular}{lccccccc} 
		\hline
		Model & $\Omega_{m0}$ & $\Omega_{\Lambda0}$ & $\mathcal{M}$ & $\Delta_M$ & $\alpha$ & $\beta$ & $\sigma_{\textrm{int}}$ \\
		\hline
		PCA-PETS & 0.28 $\pm$ 0.09 & 0.49 $\pm$ 0.14 & 24.11 $\pm$  0.02 & -0.033 $\pm$ 0.015 & -0.163 $\pm$ 0.006 & -0.038 $\pm$ 0.003 & 0.158 $\pm$ 0.006\\
		\\
		SALT2 (PCA subsample) & 0.32 $\pm$ 0.07 & 0.54 $\pm$ 0.11 & 24.14 $\pm$  0.01 & -0.048 $\pm$ 0.010 & 0.130 $\pm$ 0.006 & 2.73 $\pm$ 0.06 & 0.097 $\pm$ 0.005 \\
		\\
		FA-PETS & 0.35$_{-0.09}^{+0.08}$ & 0.54 $_{-0.14}^{+0.13}$ & 24.12 $\pm$  0.02 & -0.036 $\pm$ 0.014 & -0.334 $\pm$ 0.013 & -0.109 $\pm$ 0.007 & 0.145 $\pm$ 0.006\\
		\\
		SALT2 (FA subsample) & 0.30 $\pm$ 0.07 & 0.49 $\pm$ 0.11 & 24.15 $\pm$0.01  & -0.051 $\pm$ 0.011 & 0.131 $\pm$ 0.006 & 2.69 $\pm$ 0.06 & 0.099 $\pm$ 0.004 \\
		\hline
	\end{tabular}}
\end{table*}

The intrinsic scatter, $\sigma_{\textrm{int}}$, is a term added in quadrature to the distance modulus uncertainty of each supernova and is meant to capture any behavior unaccounted by the model. PCA-PETS returns an intrinsic scatter value 63\% higher than SALT2. When adding a model covariance, this value is expected to decrease.

The histogram of the HD residues has a mean of -0.002 mag, indicating no visible rigid translation of the distribution, which would indicate a bias or an uncounted systematic error. It also shows a dispersion of 0.20. On top of that, except for $x_1$ lower boundary, all fit parameters show normal dispersion around the null residue line (See Fig.~\ref{fig:HD_pca} and Fig.~\ref{fig:HD_residue_dependences_pca} in Appendix \ref{ap:pcacosm_cont}). 

The dependencies seen here for light-curve fit parameters $x_1$, $x_2$, and HD residues with host galaxy mass resemble that observed for SALT2' $x_1$ parameter reported in \citet{Scolnic2018}, with opposite sign. The binned scatter points show a preference for slightly lower values of $x_1$ for lower host galaxies' mass. While for $x_2$, the higher positive values are observed for more massive host galaxies. Regarding the HD residues, it again does not exhibit any dependence (See Fig.~\ref{fig:hg_mass_dep_pca} in Appendix \ref{ap:pcacosm_cont}).

Allowing the nuisance parameters and the intrinsic scatter to evolve with redshift we solve a binned MCMC simulation for all cosmological parameters fixed at their best fit values and find evidence for $\alpha$ evolution with redshift. This nuisance parameter multiplies the color phase-dependent component in our PCA-PETS. This evolution can be associated with SNIa evolution and/or with selection effects. (See Fig.~\ref{fig:param_evol_w_z_pca} in Appendix \ref{ap:pcacosm_cont})

\begin{figure}
\centering\includegraphics[width=0.5\columnwidth]{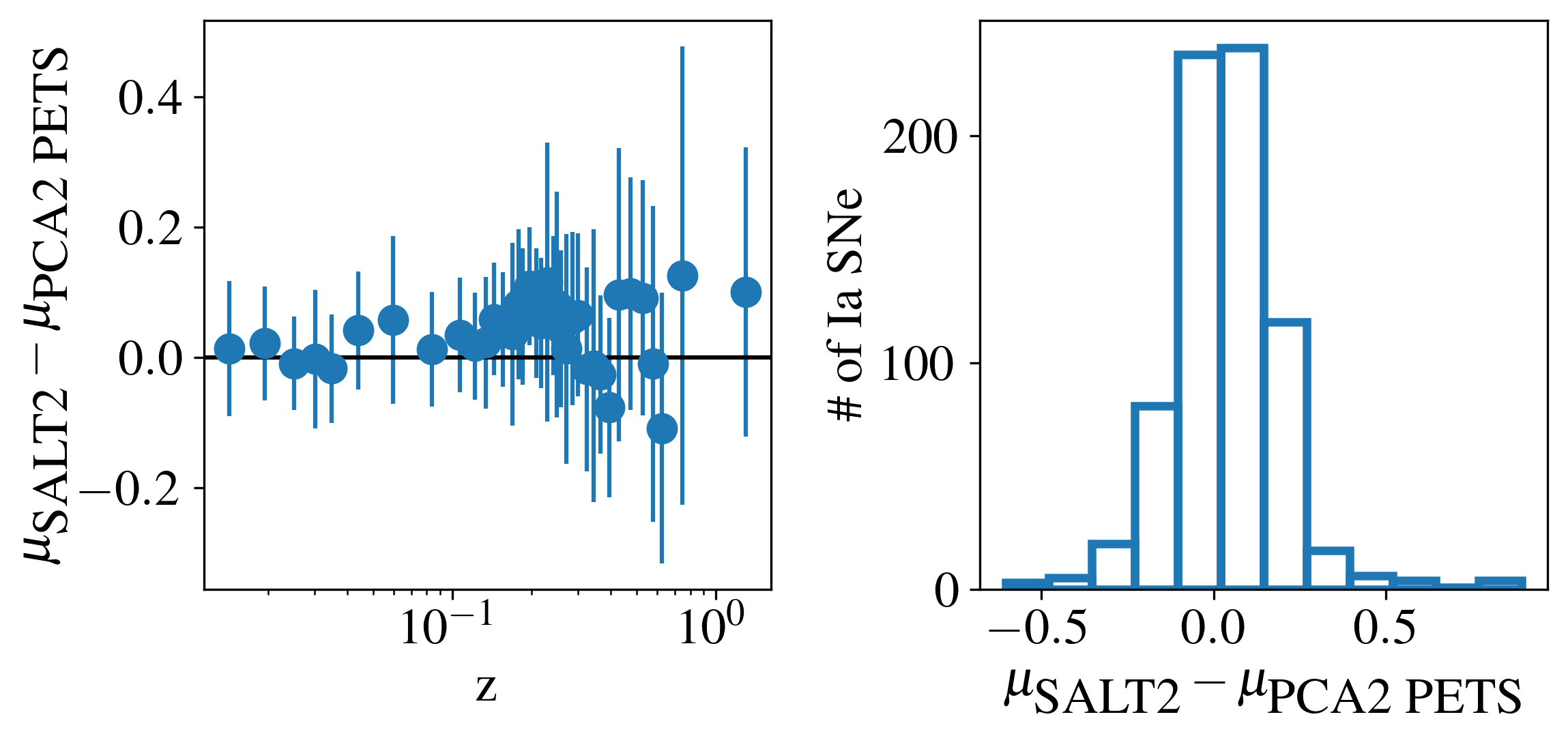} 
\caption{The left panel shows the dependence with redshift of the difference $\mu_{\textrm{SALT2}}-\mu_{\textrm{PCA-PETS}}$. On the right panel is the histogram for those differences for all redshifts.}
\label{fig:dist_mod_dif_salt2_pca}
\end{figure}

We see no clear redshift dependence for the difference $\mu_{SALT2}-\mu_{\textrm{PETS}}$, as shown in Fig.~\ref{fig:dist_mod_dif_salt2_pca}. Considering a Kolmogorov–Smirnov test with the null hypothesis that both distance modulus are realizations of the same distribution, we find the null hypothesis can not be rejected with a p-value equal to 0.99. Recalling no distance bias corrections was included for either model. 

\subsection{Results for FA-PETS}
\label{sec:cosmo_results_fa}

We now analyze the cosmological results when considering the FA-PETS model. As seen previously, the components retrieved from the FA method concentrate on explaining the off-diagonal elements of the sample covariance matrix and show less explained variability when fitting the SNIa SEDs in contrast to PCA-PETS. 

\begin{figure}
\centering\includegraphics[width=0.5\columnwidth]{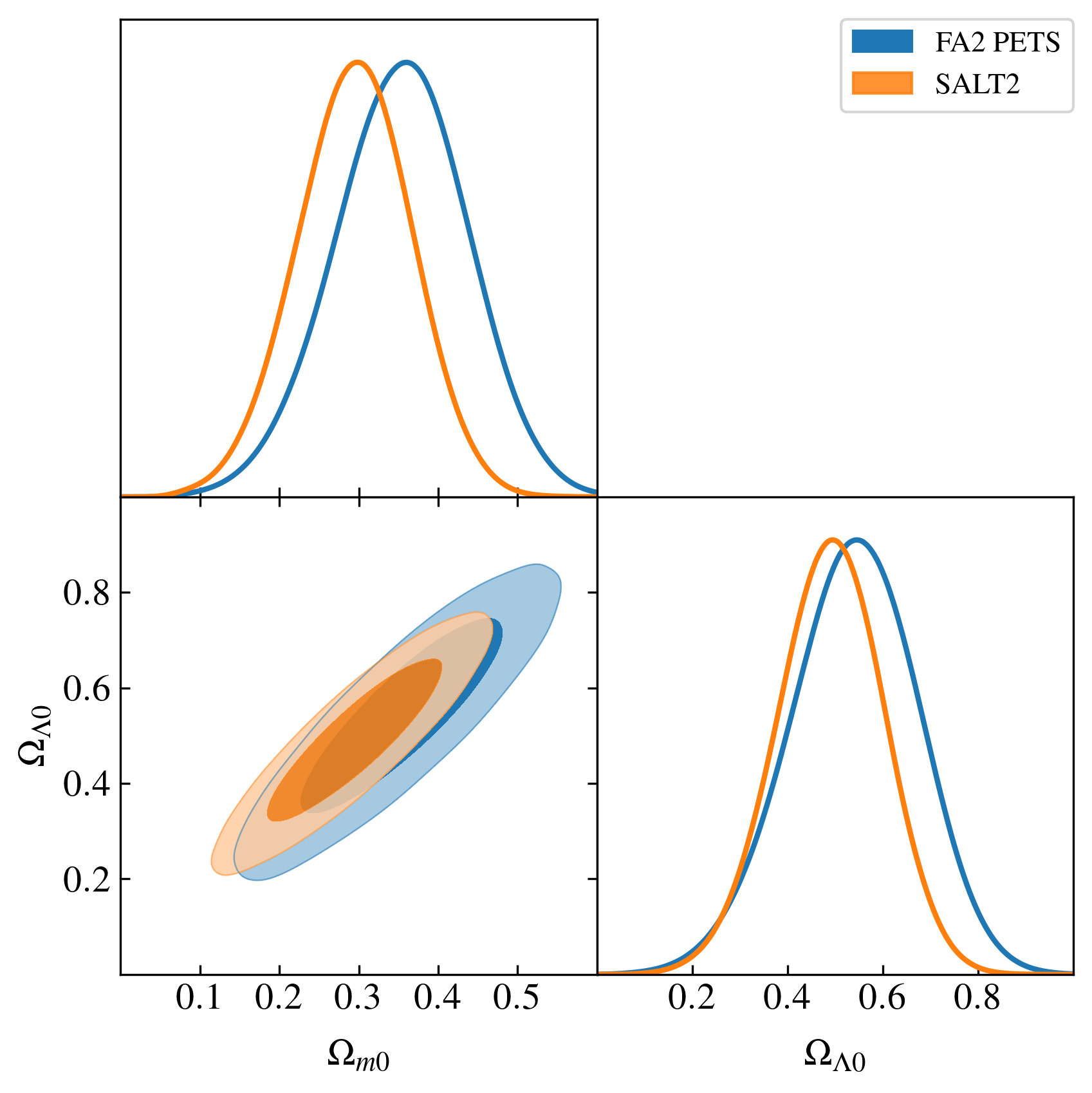} 
\caption{Comparison of marginalized cosmology results for SALT2 and FA-PETS for $\Lambda$CDM model.}
\label{fig:triangleplots_fa}
\end{figure}

Fig. \ref{fig:triangleplots_fa} shows the confidence levels in the ($\Omega_{m0}$, $\Omega_{\Lambda 0}$) plane and the given parameters posteriors. In addition to the confidence plot, the best fit parameters are given in Table~\ref{tab:lcdmtableresults} for both FA-PETS and SALT2 considering the same subsample.  For FA-PETS constraints in Fig. \ref{fig:triangleplots_fa} it is possible to detect a rigid translation in parameter space towards higher $\Omega_{m0}$ and $\Omega_{\Lambda 0}$ relative to SALT2.

For $x_1$ there is no apparent correlation with host galaxy mass while for $x_2$ there is a similar behavior previously seen for PCA-PETS, higher values of $x_2$ are more common for more massive galaxies. Regarding the HD residues, no host galaxy mass dependence is observed for FA-PETS. (See  Fig.~\ref{fig:param_dep_mb_fa} and Fig.~\ref{fig:hg_mass_dep_fa} in Appendix~\ref{ap:facosm_cont})

HD residues do not show dependencies with redshift nor with fit parameters. FA-PETS also predicts compatible results of $\Omega_{m0}$ and $\Omega_{\Lambda 0}$ parameters with SALT2 predictions within 68\% confidence level, with a higher estimate of coherent intrinsic scatter, $\sigma_{\textnormal{int}}$. And lastly, no redshift evolution is observed for the difference $\mu_{\textnormal{FA-PETS}}-\mu_{\textnormal{SALT2}}$ as seen in Fig.~\ref{fig:dist_mod_dif_salt2_fa}. (See  Fig.~\ref{fig:HD_fa} and Fig.~\ref{fig:HD_residue_dependences_fa}) in Appendix~\ref{ap:facosm_cont}).

Performing the redshift binned MCMC analysis with cosmological parameters fixed at their best fit values we encounter a similar redshift evolution previously seen for $\alpha$ in the context of PCA-PETS. (See Fig.~\ref{fig:param_evol_w_z_fa} in Appendix~\ref{ap:facosm_cont}).

\begin{figure}
\centering\includegraphics[width=0.5\columnwidth]{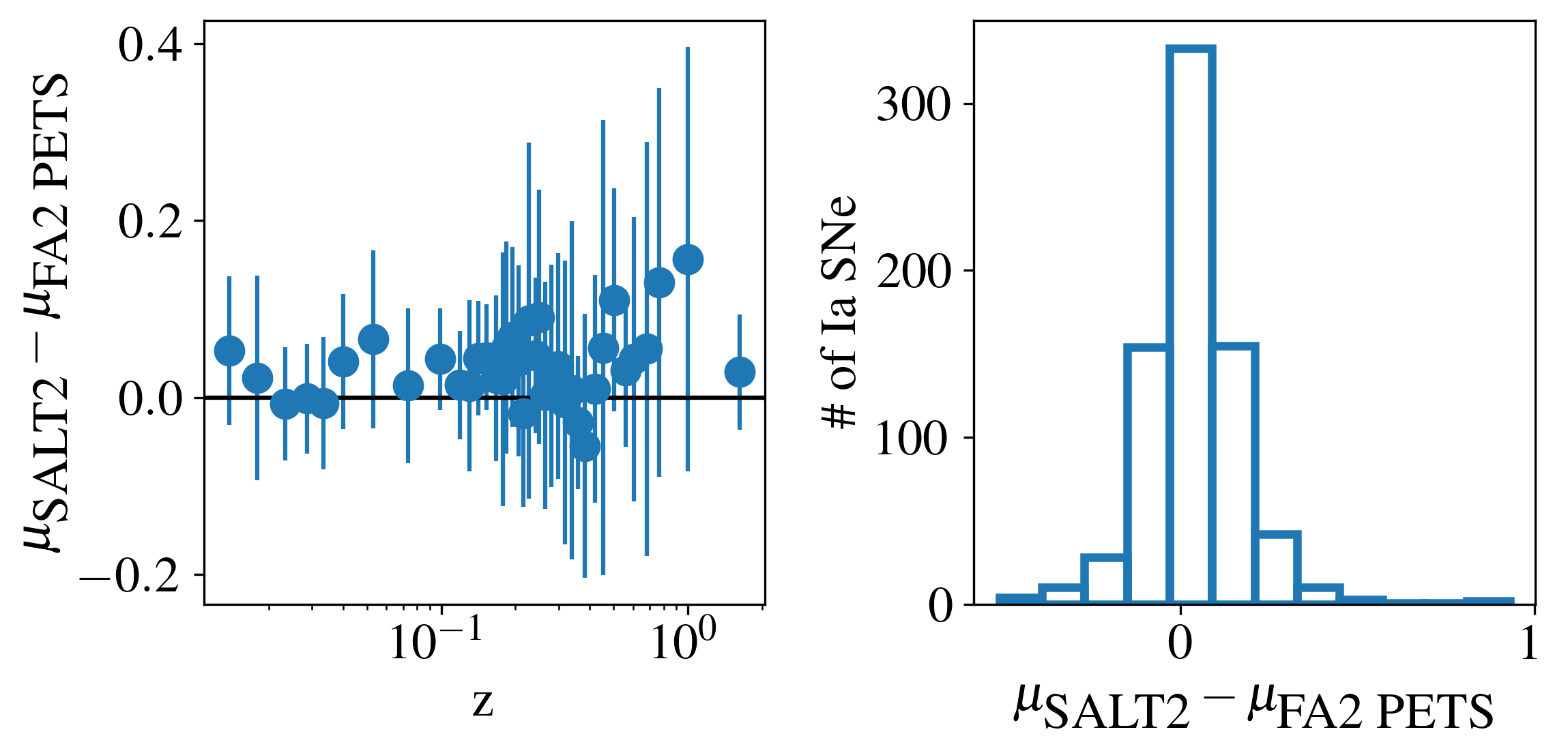} 
\caption{The left panel shows the dependence with redshift of the difference $\mu_{\textrm{SALT2}}-\mu_{\textrm{FA2 PETS}}$. On the right panel is the histogram for those differences for all redshift.}
\label{fig:dist_mod_dif_salt2_fa}
\end{figure}

\section{Conclusion}
\label{sec:conclusions}

In this paper, we investigated two different models describing the rest-frame flux of SNIa as simple 3-component linear expansions, called PCA-PETS and FA-PETS. 
We obtained the model components through Principal Component Analysis and Factor Analysis after a robust reconstruction of the reddened training sample SEDs using 1D Gaussian Process Regression for each plane of constant wavelength with \cite{Hsiao2007} template as the mean. This reconstruction provided a SED resolution of 10\AA\, $\times$ 1 day. 

The main idea was to take advantage of SNeIa similarities and adopt an alternative description using decomposition methods over a training set of SEDs. 
Unlike the usual rest-frame flux description that includes an exponential term to model two different color-related effects, reddening due to dust and intrinsic color variations, here a simple expansion model was applied, replacing the color-law term with a data-driven component. 
We verified this pure expansion is a reasonable assumption when constructing a light curve fitter and both PCA and FA methods showed within limitations reliable light-curve modeling results but with FA constantly outperforming PCA. 

Our analysis of PCA-PETS and FA-PETS cosmological results showed inferences of cosmological parameters within 68\% agreement with the most widely used fitter SALT2 results, with our models predicting more conservative statistical uncertainties over the cosmological parameters.
The cosmological constraints for PCA-PETS favor slightly lower values of $\Omega_{m0}$ and of $\Omega_{\Lambda 0}$ when compared to SALT2. And the opposite is observed for FA-PETS.
We verified from fit parameters correlations with SALT2 and SNEMO2 that the main SNIa spectra feature retrieved from the decomposition methods describes color-index variation. 
Thus this color variability can be accessed through a linear 3-component model. 

The Hubble Diagram residues showed no evolution either with redshift or with host galaxy mass for both PCA and FA models. The residues with respect to SALT2 distance modulus predictions show no redshift dependence. 
In future works, we plan on extending the model coverage to include the UV region, a crucial feature that will augment the supernova cosmology sample and is currently the main limitation of our model.

We plan on creating a model covariance to estimate more precisely the unaccounted intrinsic uncertainty $\sigma_\textnormal{int}$. 
We also plan on including higher redshift SNIa data, when available, to reassess the nuisance parameters evolution with redshift and apply the effects of distance bias selection to our analysis.

Overall, both the PCA and FA-PETS performances will heavily depend on how representative the training set is. 
These methods undergo an approximation when keeping only the first two terms of the rest-frame flux expansion, as other empirical descriptions of SNeIa rest-frame fluxes. 
The main differences reside in what kind of information is being neglected. To avoid correlating dust extinction reddening with intrinsic luminosity, we decided not to separate the reddening effect from our SEDs, but at the same time, this raised questions about whether important stretch and color information is being neglected. 
We argue that the fitting and constraining power of both PCA-PETS and FA-PETS are good indicators that this information is not being completely neglected but would instead be more diluted over the model components and now the color variations can be represented by a phase-dependent component.

It is important to note we only considered a coherent intrinsic scatter in the cosmological analysis, which is also a limitation of our analysis. Other intrinsic scatter could be explored, giving further insight into whether the missing information was wavelength-dependent. 
In this scenario, the best intrinsic scatter modeling may not be wavelength-independent for a model that does not correlate extinction and intrinsic luminosity variations.

We verify a clear evolution with redshift for $\alpha$ nuisance parameter. 
This parameter multiplies the $x_1$ fit parameter in our distance modulus definition, which is strongly correlated with SALT2 color index variation. As $x_1$ seems to describe the majority of color index variation it is reasonable that $\alpha$ shows a redshift evolution as does SALT2 $\beta$ parameter. 
This evolution can be related to selection effects, which are not covered in this stage of PETS development.
Incorporating the model covariance and extending the model to the UV region will allow us to reassess these dependencies.

Mapping the systematics is essential to the current cosmological precision era, thus supporting improvements in current empirical models.  
FA-PETS can be further explored if we consider accommodating measurement uncertainties and common factor rotations. These rotations can provide a different insight into the hidden variables since oblique rotations allow for correlated factors.

\newpage
\section*{Acknowledgements}
We thank Arthur L. da Fonseca for the helpful discussions. We would like to express our gratitude to Martin Makler for his generosity and support throughout this collaboration. We would also like to thank Kyle Barbary and Kyle Boone for their helpful contributions towards resolving SNCOSMO GitHub issues, which significantly eased the fitting analysis process. We thank anonymous reviewers for their helpful comments and suggestions.

This work has also been supported by the Brazilian funding agencies CAPES and CNPq. CSN thanks the support of CNPq for the PhD scholarship no. 155994/2019-0. JPCF thanks the Brazilian funding agencies CAPES for the MS scholarship no. 88887.336370/2019-00 and CNPq for the PhD scholarship no. 140210/2021-0. RRRR acknowledges CNPq (grant no. 309868/2021-1).

\section*{Data Availability}

The SNIa spectra used in this paper to construct the PETS models are from Nearby Supernova Factory Data Release 9 publicly available at \url{https://snfactory.lbl.gov/}. The light curves used for cosmology are from \texttt{SNANA}, available for download in \url{https://zenodo.org/record/4015325}. Our codes for training, light curve fitting and cosmology will be available after publication at 
\url{https://github.com/CassiaNascimento/PETS_model_for_SN_Ia_LC_fitting}.

\section*{SOFTWARE CITATIONS}
This work uses the following software packages:
\begin{itemize}
  \item \href{https://github.com/astropy/astropy}{Astropy} (\citet{astropy_colab_1, astropy_colab_2}).
  \item \href{https://github.com/dfm/emcee}{Emcee} (\citet{emcee}).
  \item \href{https://github.com/cmbant/getdist}{GetDist} (\citet{getdist}).
  \item \href{https://github.com/matplotlib/matplotlib}{Matplotlib} (\citet{hunter_2007}).
  \item \href{https://github.com/numpy/numpy}{Numpy} (\citet{2020NumPy-Array}).
  \item \href{https://github.com/python-pillow/Pillow}{Pillow} (\citet{pillow}).
  \item \href{https://www.python.org/}{Python} (\citet{python}).
  \item \href{https://github.com/sigma-py/quadpy}{Quadpy} (\citet{quadpy}).
  \item \href{https://github.com/scipy/scipy}{Scipy} (\citet{scipy}).
  \item \href{https://github.com/RickKessler/SNANA}{SNANA} (\citet{kessler2009snana}).
 \item \href{https://github.com/sncosmo/sncosmo}{SNCosmo} (\citet{barbary_kyle_2022_7117347}).
 \item \href{https://github.com/scikit-learn/scikit-learn}{Scikit-learn} (\citet{sklearn_api, scikit-learn}).

\end{itemize}

\printbibliography

\appendix

\section{Further cosmology results for PCA-PETS}
\label{ap:pcacosm_cont}
\begin{figure}
\centering\includegraphics[width=0.5\columnwidth]{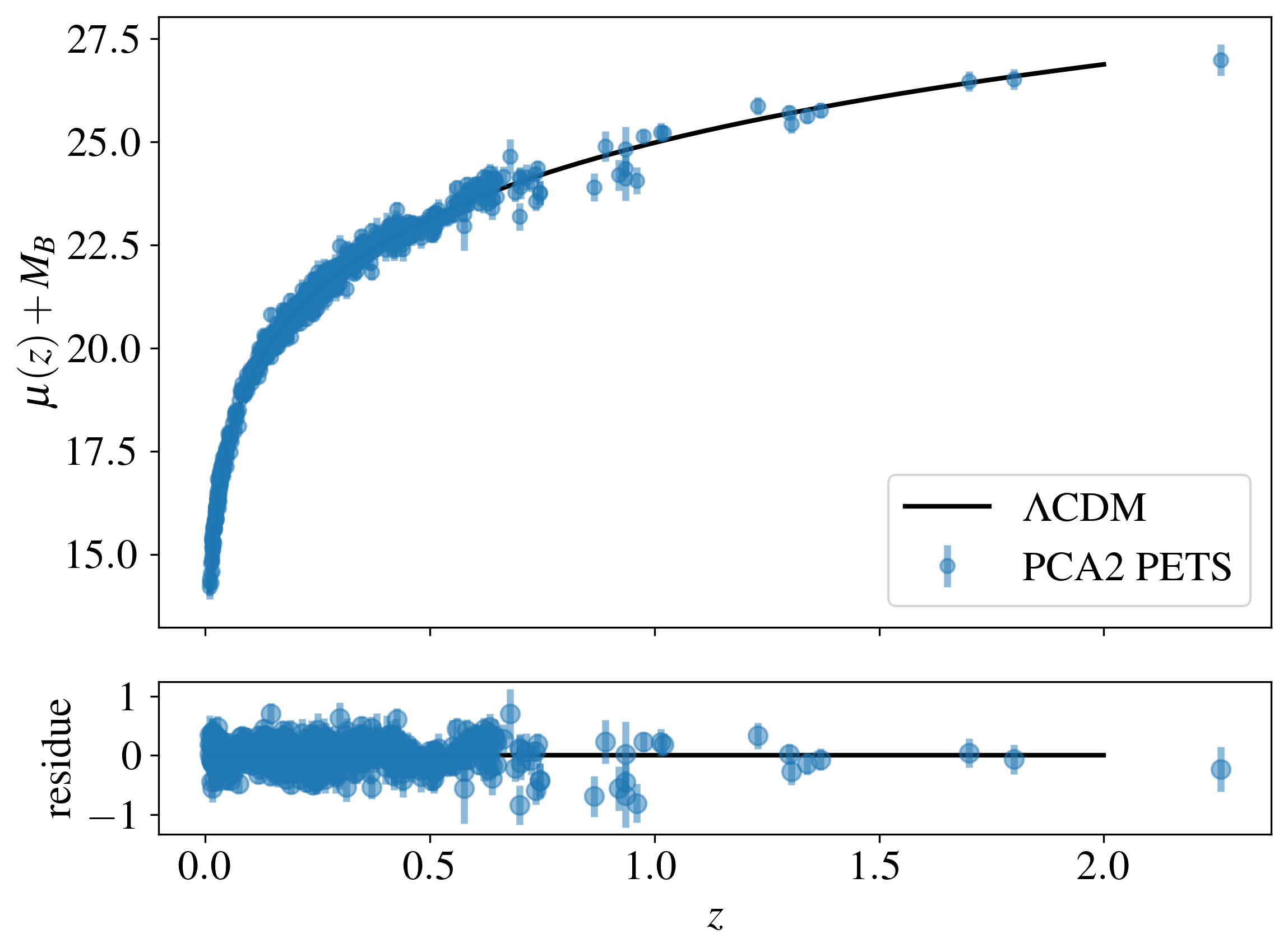} 
\caption{The upper panel shows the Hubble Diagram for the PCA-PETS model. In the bottom panel is the HD residue as a function of redshift.}
\label{fig:HD_pca}
\end{figure}

In the top panel of Fig.~\ref{fig:HD_pca} we show the Hubble Diagram (HD) for our Pantheon subsample, corrected for intrinsic dispersion using PCA-PETS. The black solid line is the theoretical distance modulus with parameters fixed at the best fit displayed in Table~\ref{tab:lcdmtableresults}, and the blue scatter points are drawn from equation~(\ref{eq:distmodpets}) at each supernova redshift, with statistical errors given by the square root of equation~(\ref{eq:sig2pets}) with parameters fixed at their best fits. In the bottom panel we have the residues for this HD, here defined as $\mu_{\textrm{PETS}}-\mu_{th}$. 

\begin{figure}
\centering
\includegraphics[width=0.7\columnwidth]{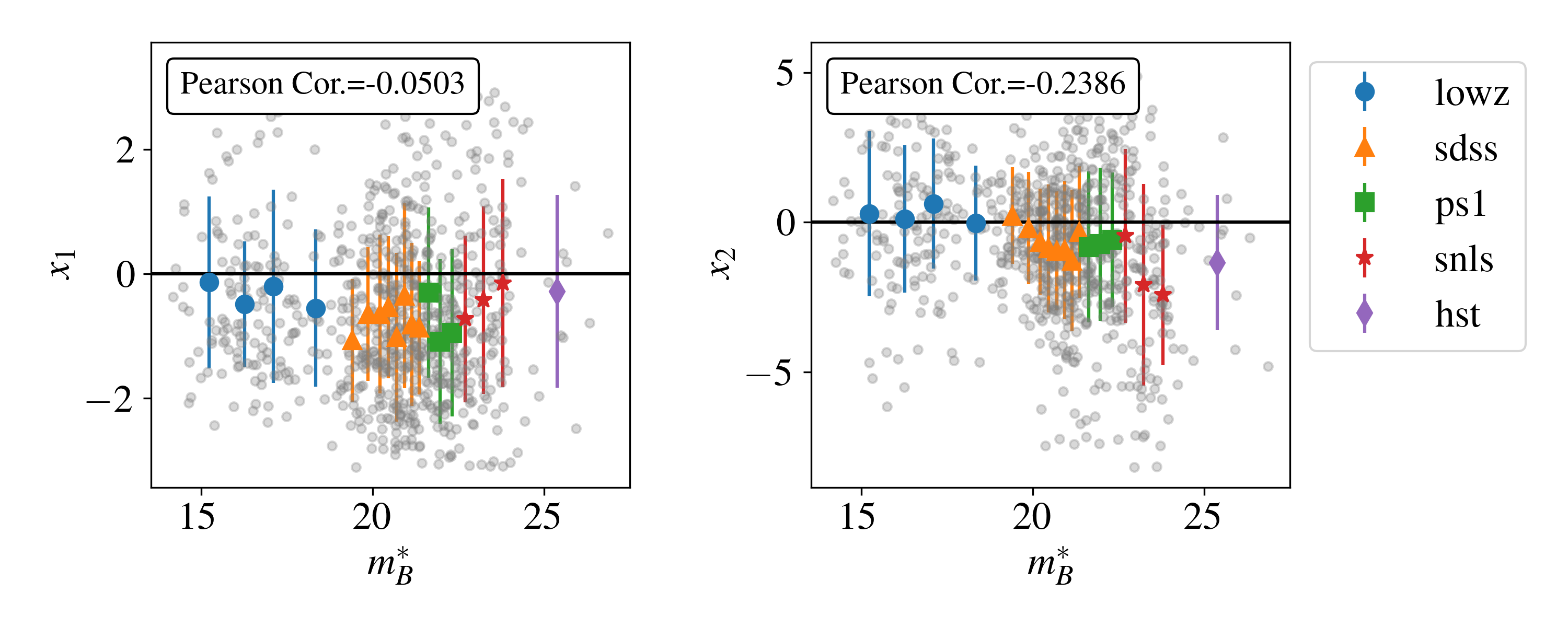} 
\caption{Scatter plot of PCA-PETS fit parameters as a function $m_B^*$, a quantity that correlates linearly with $\log_{10}(z)$. The binned scatter points with corresponding standard deviations are shown colored according to the most frequent survey in each bin.}
\label{fig:param_dep_mb_pca}
\end{figure}

\begin{figure}
\centering\includegraphics[width=0.5\columnwidth]{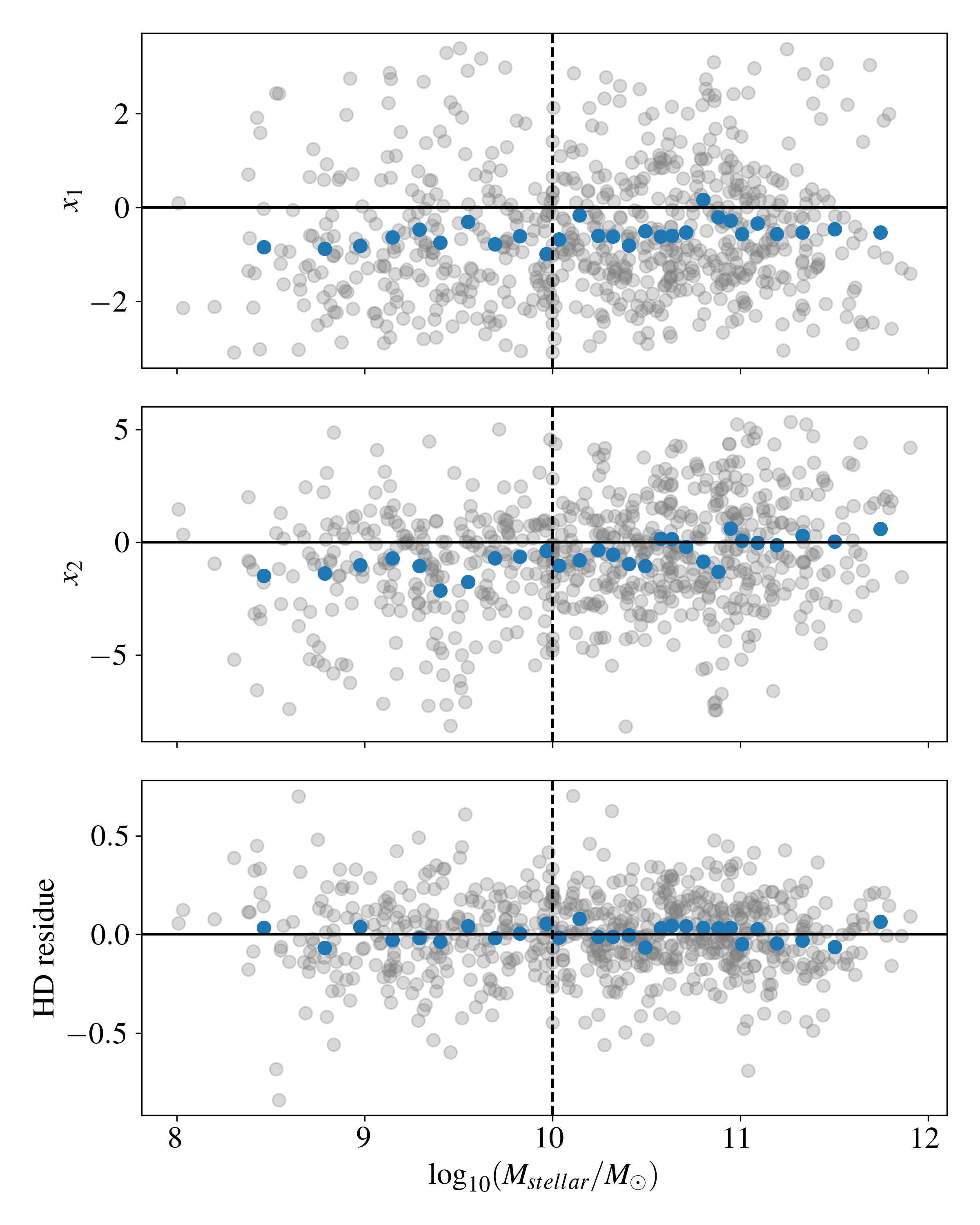} 
\caption{Fit parameters and HD residue dependence with host galaxy mass for PCA-PETS fits.}
\label{fig:hg_mass_dep_pca}
\end{figure}

\begin{figure}
\centering\includegraphics[width=\columnwidth]{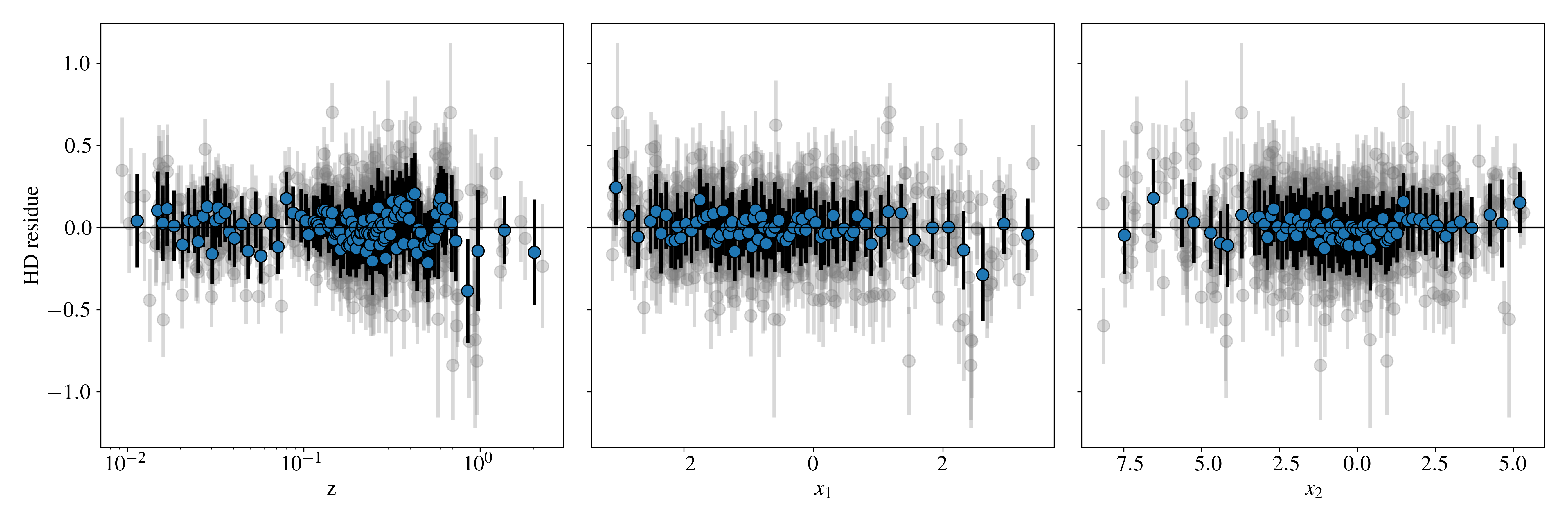} 
\caption{Hubble Diagram residues dependence with redshift and fit parameters, in magnitude, for PCA-PETS. The blue scatter points are binned HD residues of the underlying gray scatter points.}
\label{fig:HD_residue_dependences_pca}
\end{figure}

In Fig.~\ref{fig:HD_residue_dependences_pca} the HD residues dependence with redshift and fit parameters, $m_B^*$, $x_1$ and $x_ 2$, are displayed for the PCA-PETS model. Each gray scatter point in the background represents a SNIa from the underlying sub-sample, the binned results shown in blue are the mean and standard deviation of HD residues for the set contained in each bin.

\begin{figure}
\centering\includegraphics[width=0.5\columnwidth]{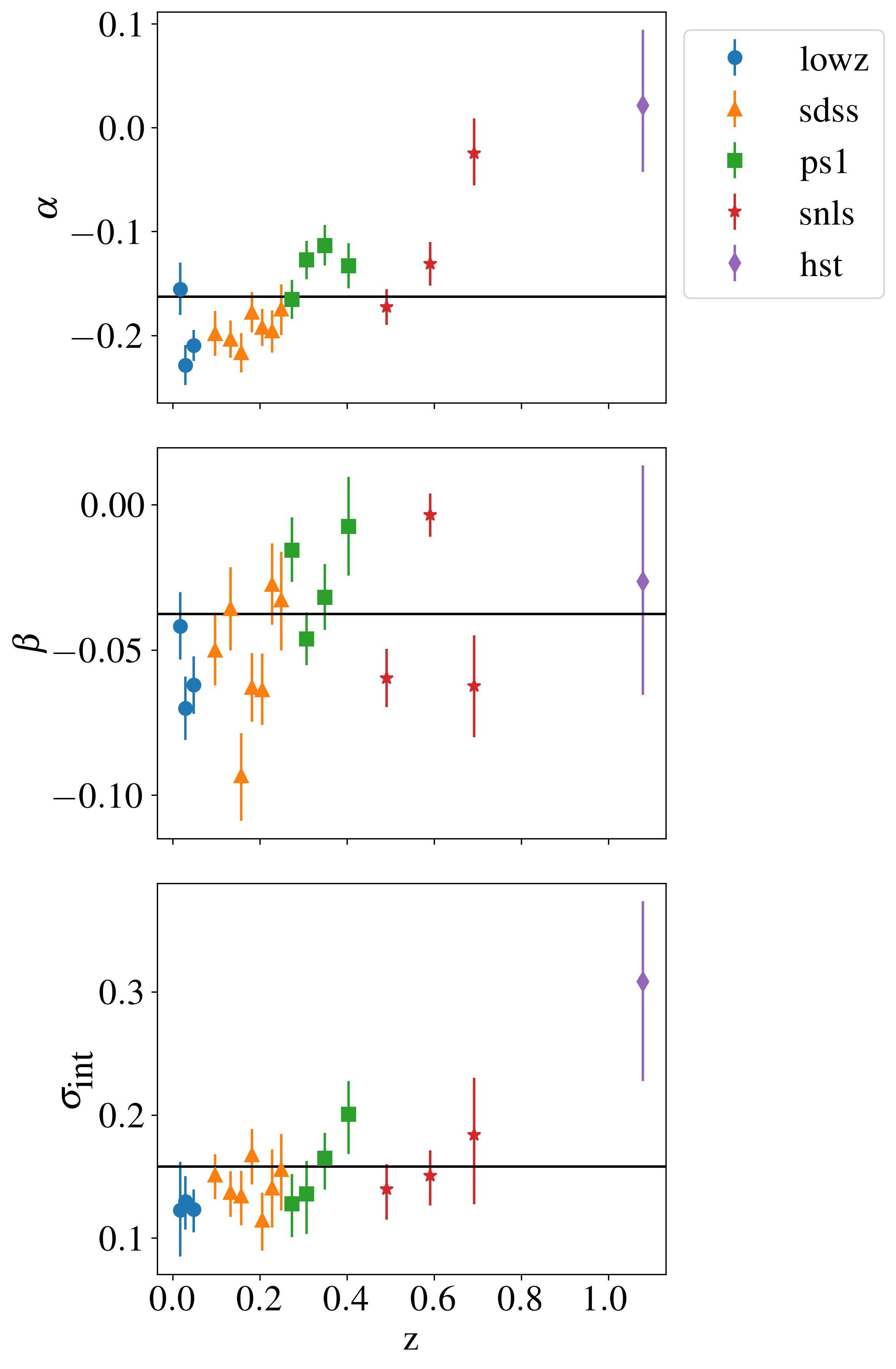} 
\caption{Evolution of PCA-PETS fit parameters with redshift.}
\label{fig:param_evol_w_z_pca}
\end{figure}

In Fig.~\ref{fig:param_evol_w_z_pca}, the binned results are calculated for groups of similar number of supernovas, around 40 per group. All three parameters tend to have higher values with increasing redshift. As the error bars shown here are only statistical contributions, including systematic terms would reduce this apparent evolution primarily for the intrinsic scatter. 

\begin{figure}
\centering\includegraphics[width=0.7\columnwidth]{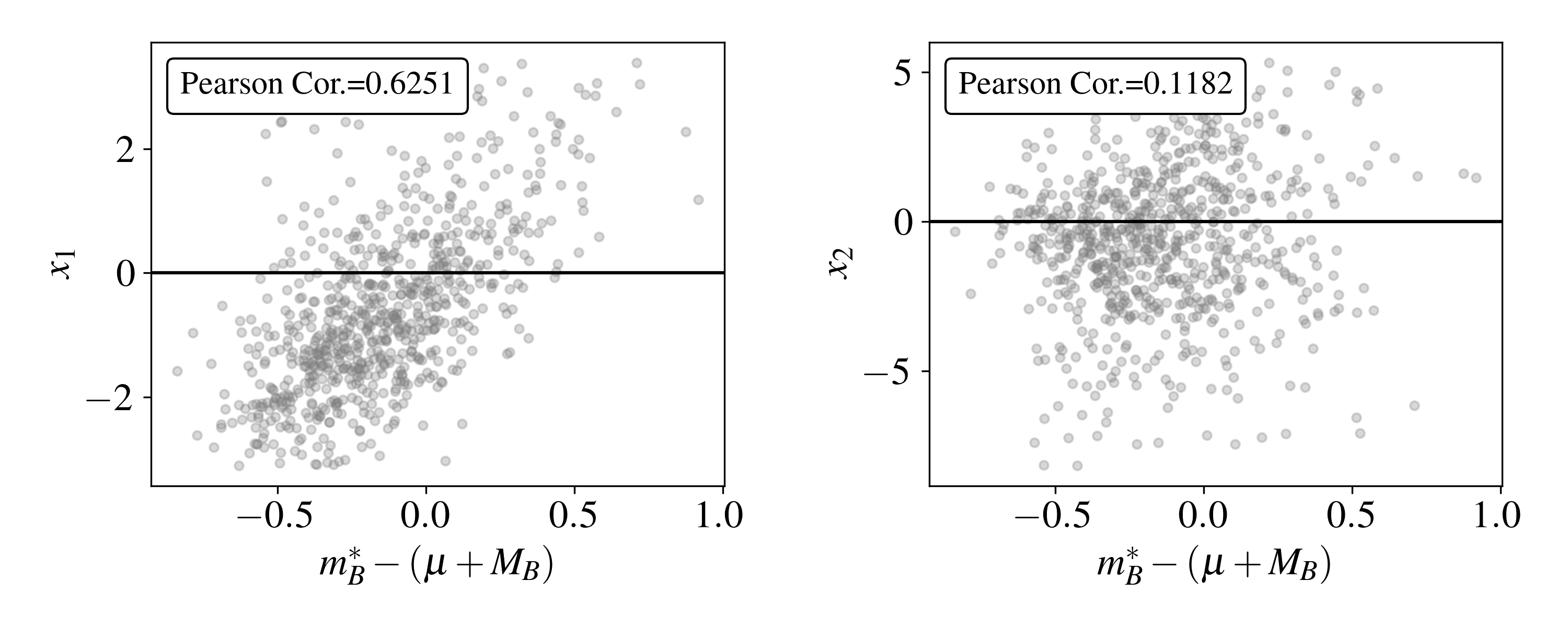} 
\caption{Scatter plots of fitted PCA-PETS parameters for the Pantheon sample with their correspondent Pearson correlation. \textit{Left:} $x_1$ versus $m^{\ast}_B-(\mu +M_B)$. \textit{Right:} $x_2$ versus $m^{\ast}_B-(\mu +M_B)$.}
\label{fig:distmodcorr_pca}
\end{figure}

In Fig.~\ref{fig:distmodcorr_pca}, we present the correlation between PCA-PETS fit parameters $x_1$ and $x_2$ with $m^{\ast}_B-(\mu +M_B)$. The linear correlations observed support the use of these parameters as linear corrections to our distance modulus.

\section{Further cosmology results for FA-PETS}
\label{ap:facosm_cont}

\begin{figure}
\centering
\includegraphics[width=0.7\columnwidth]{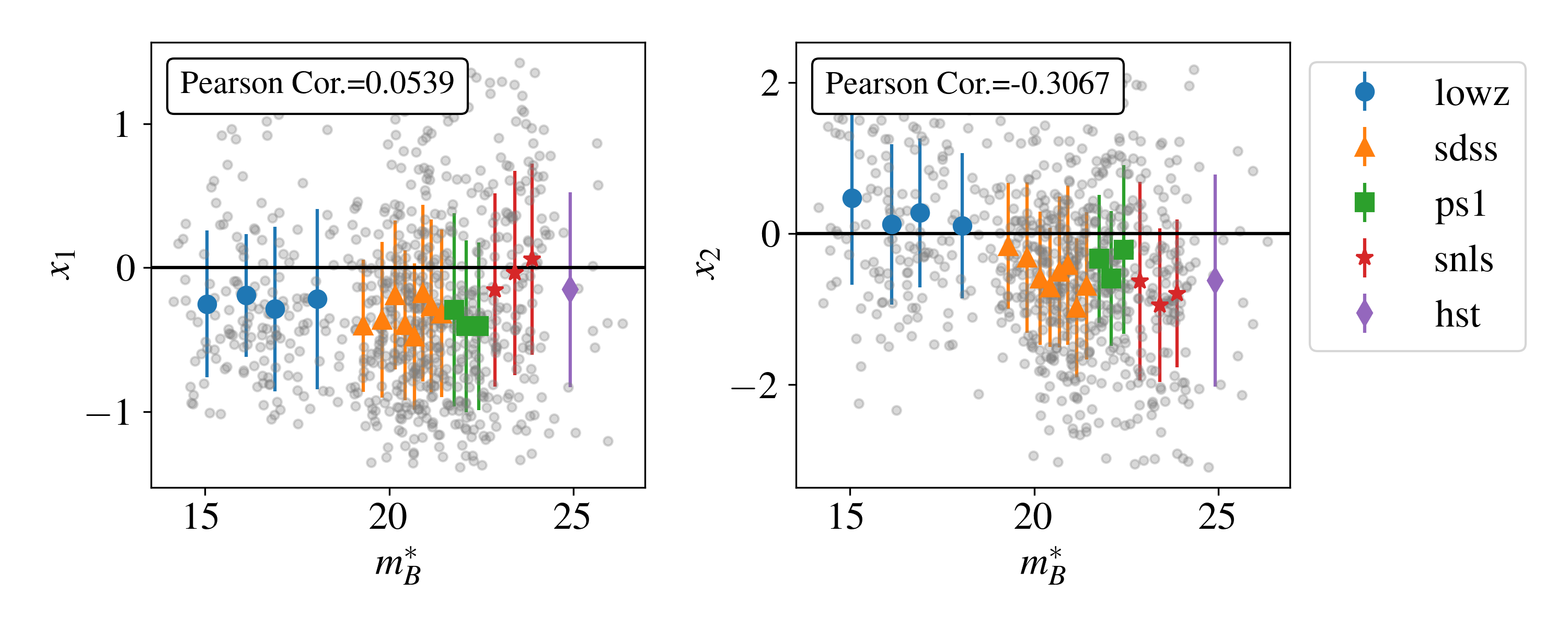} 
\caption{Scatter plot of FA-PETS fit parameters as a function $m_B^*$, a quantity that correlates linearly with $\log_{10}(z)$. The binned scatter points with corresponding standard deviations are shown colored according to the most frequent survey in each bin..}
\label{fig:param_dep_mb_fa}
\end{figure}

\begin{figure}
\centering\includegraphics[width=0.5\columnwidth]{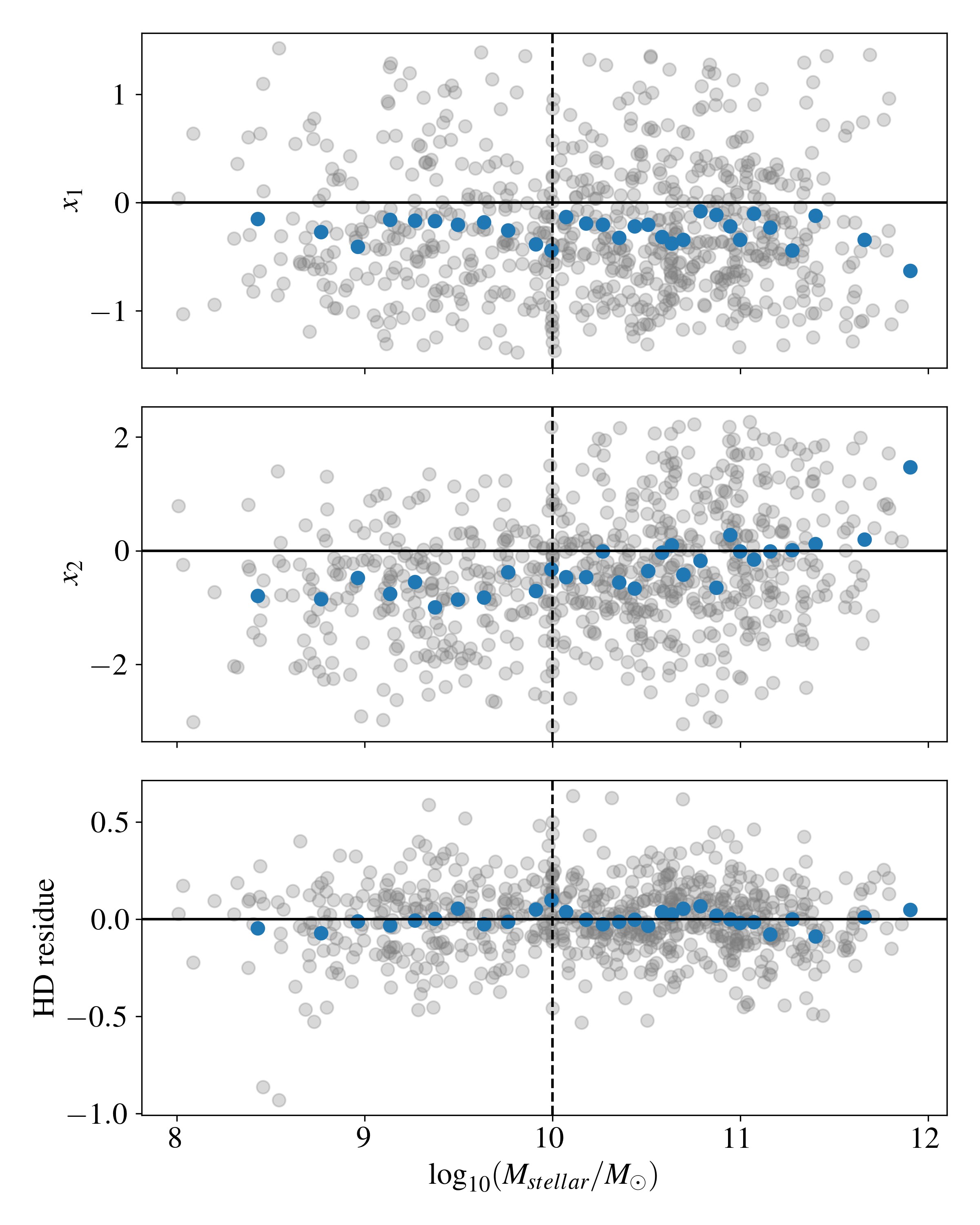} 
\caption{Fit parameters and HD residue dependence with host galaxy mass for FA-PETS fits.}
\label{fig:hg_mass_dep_fa}
\end{figure}

\begin{figure}
\centering\includegraphics[width=0.5\columnwidth]{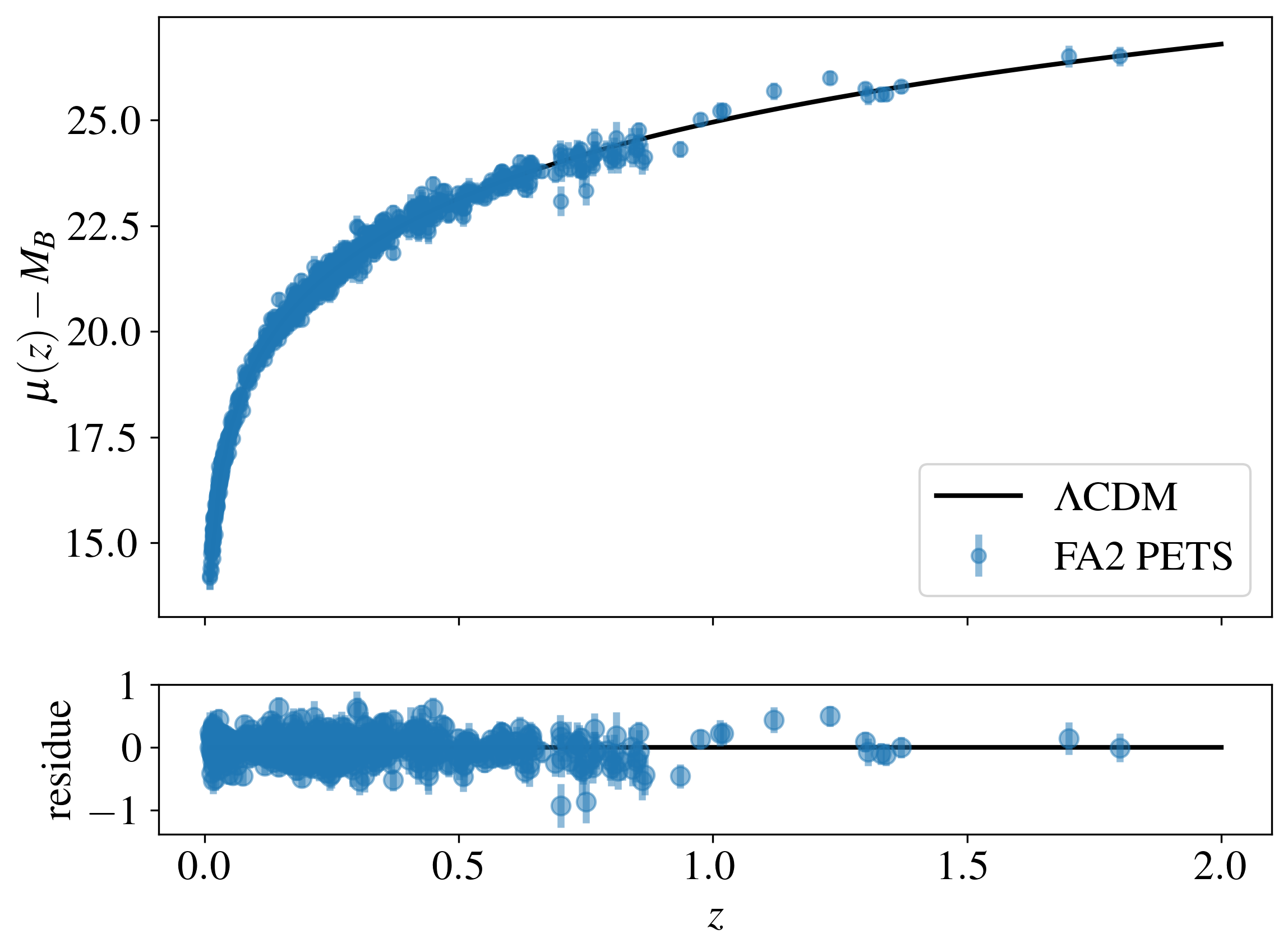} 
\caption{The upper panel shows the Hubble Diagram for FA-PETS model. In the bottom panel is the HD residue as a function of redshift.}
\label{fig:HD_fa}
\end{figure}

\begin{figure}
\centering\includegraphics[width=\columnwidth]{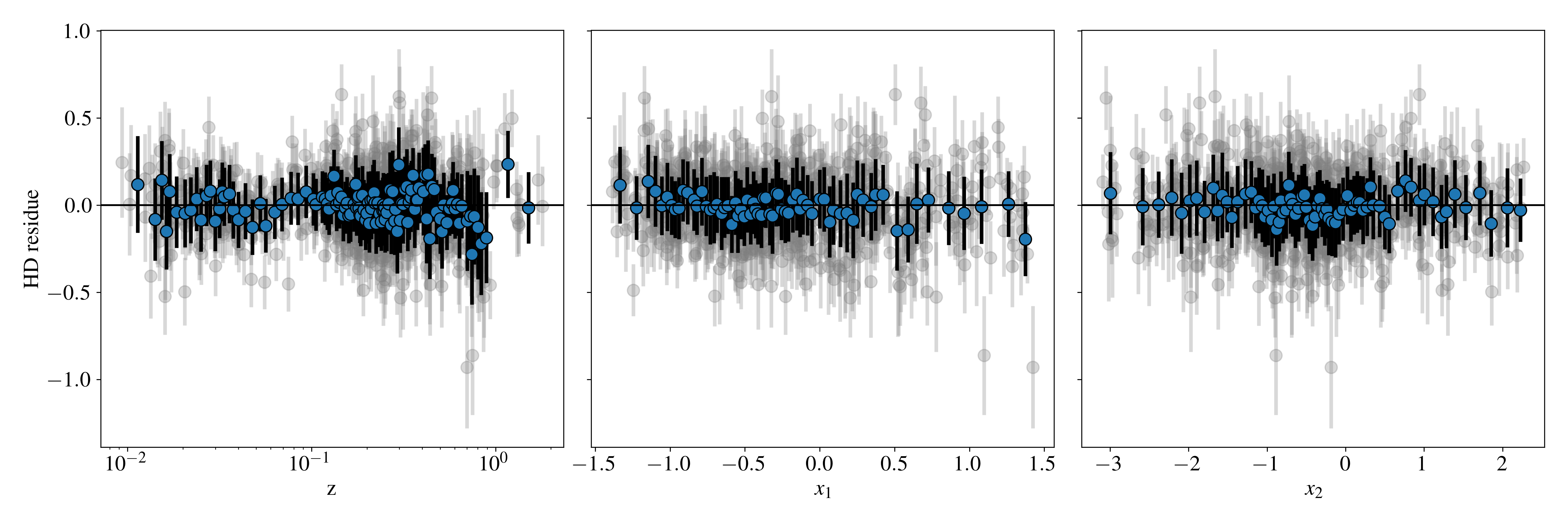} 
\caption{Hubble Diagram residues dependence with redshift and fit parameters, in  magnitude, for FA-PETS. The blue scatter points are binned HD residues of the underlying gray scatter points.}
\label{fig:HD_residue_dependences_fa}
\end{figure}

\begin{figure}
\centering\includegraphics[width=0.5\columnwidth]{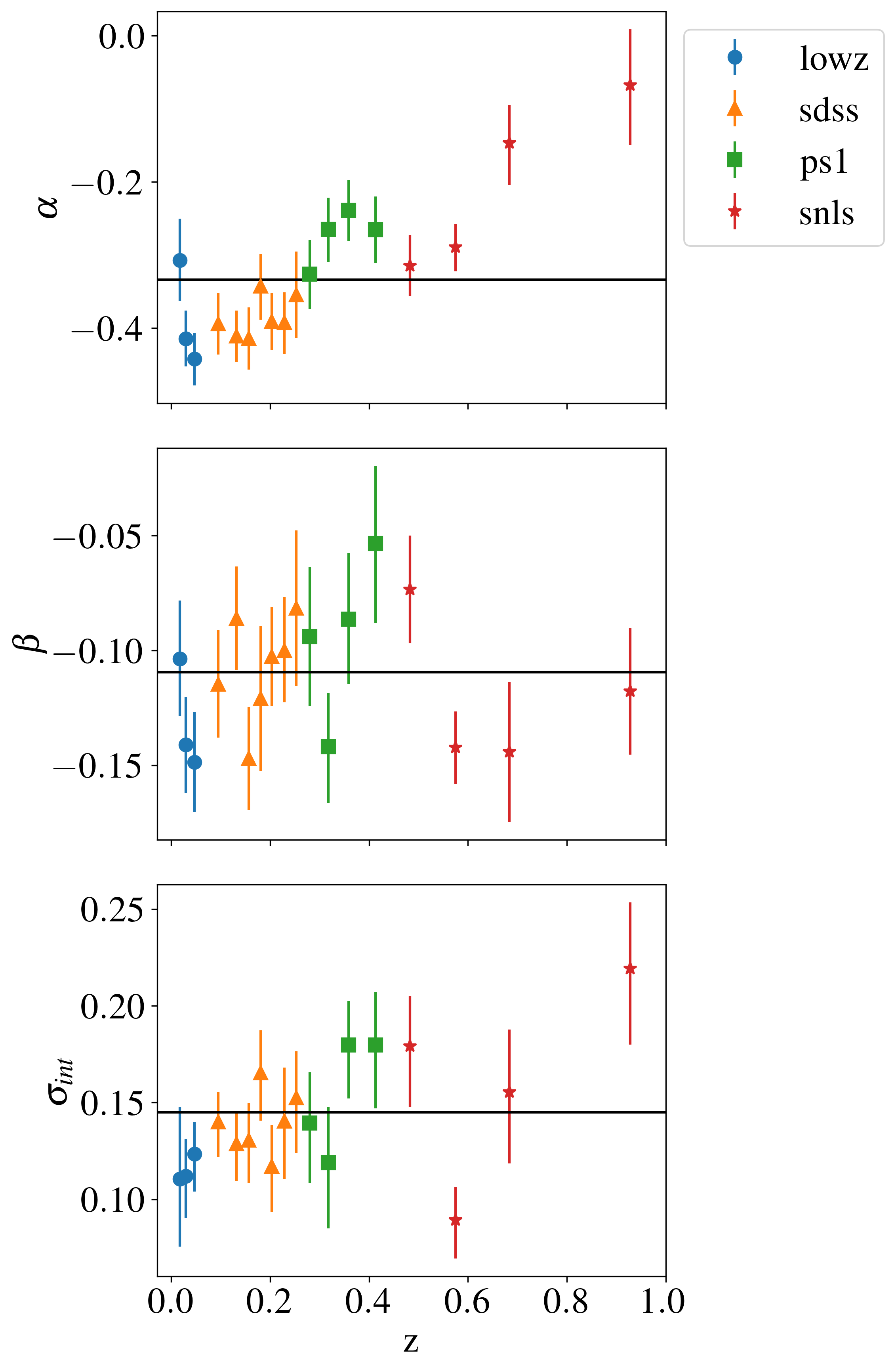} 
\caption{Evolution of FA-PETS fit parameters with redshift.}
\label{fig:param_evol_w_z_fa}
\end{figure}

\begin{figure}
\centering\includegraphics[width=0.7\columnwidth]{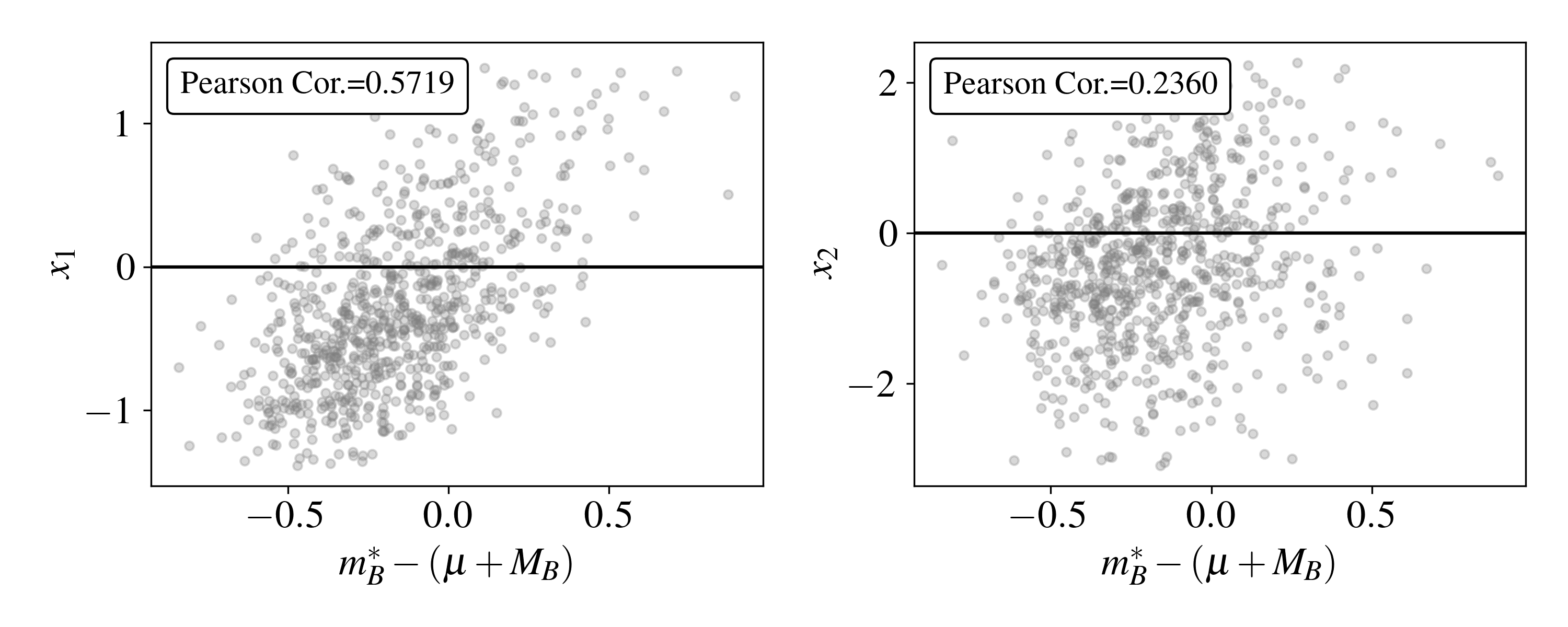} 
\caption{Scatter plots of fitted FA-PETS parameters for the Pantheon sample with their correspondent Pearson correlation. \textit{Left:} $x_1$ versus $m^{\ast}_B-(\mu +M_B)$. \textit{Right:} $x_2$ versus $m^{\ast}_B-(\mu +M_B)$.}
\label{fig:dismodcorr_fa}
\end{figure}

Fig.~\ref{fig:HD_fa} shows the Hubble Diagram for the Pantheon subsample and the corresponding residues with respect to the black solid line for FA-PETS. This black solid line is the theoretical distance modulus prediction with cosmological parameters fixed at the best fit parameters reported in Table~\ref{tab:lcdmtableresults}. The HD residues dependencies with redshift and fit parameters are shown on Fig.~\ref{fig:HD_residue_dependences_fa}.  Fig.~\ref{fig:param_evol_w_z_fa} shows FA-PETS fit parameters evolution with redshift for a binned cosmology analysis. Lastly, Fig.~\ref{fig:dismodcorr_fa} shows the observed linear correlations that also support the use of the fit parameters as linear corrections to our distance modulus.

\end{document}